\documentclass[10pt]{amsart}
\usepackage{amsmath,amssymb}
\usepackage[dvips]{epsfig}

\newtheorem{theorem}{Theorem}[section]
\newtheorem{lemma}[theorem]{Lemma}
\newtheorem{corollary}[theorem]{Corollary}
\newtheorem{proposition}[theorem]{Proposition}
\newtheorem{conjecture}[theorem]{Conjecture}
\allowdisplaybreaks \numberwithin{equation}{section}

\newcommand{\PP}{{\mathbb P}}
\def\Im{\mathop{\rm Im}\nolimits}
\def\Re{\mathop{\rm Re}\nolimits}
\def\div{\mathop{\rm Div}\nolimits}
\def\Jac{\mathop{\rm Jac}\nolimits}
\def\diag{\mathop{\rm Diag}\nolimits}
\def\dim{\mathop{\rm Dim}\nolimits}

\begin{document}

\title[Monopoles, Curves and Ramanujan]
{Monopoles, Curves and Ramanujan}
\author{H.W. Braden}
\address{School of Mathematics, Edinburgh University, Edinburgh.}
\email{hwb@ed.ac.uk}
\author{V.Z. Enolski}
\address{Department of Mathematics and Statistics, Concordia
University, Montreal.
\newline
On leave: Institute of Magnetism, National Academy of Sciences of
Ukraine.} \email{vze@ma.hw.ac.uk}
\begin{abstract}
We develop the Ercolani-Sinha construction of $SU(2)$ monopoles and
make this effective for (a five parameter family of centred) charge
3 monopoles. In particular we show how to solve the transcendental
constraints arising on the spectral curve. For a class of symmetric
curves the transcendental constraints become a number theoretic
problem and a recently proven identity of Ramanujan provides a
solution.
\end{abstract}

 \maketitle

\tableofcontents

\section{Introduction}
This article deals with the construction of magnetic monopoles,
algebraic curves subject to certain transcendental constraints and
number theory. Magnetic monopoles, or the topological soliton
solutions of Yang-Mills-Higgs gauge theories in three space
dimensions, have been objects of fascination for over a quarter of
a century. BPS monopoles in particular have been the focus of much
research (see \cite{ms04} for a recent review). These monopoles
arise as a limit in which the Higgs potential is removed and
satisfy a first order Bogomolny equation
$$B_i=\frac{1}{2}\sum_{j,k=1}\sp3\epsilon_{ijk}F\sp{jk}=D_i\Phi$$
(together with certain boundary conditions, the remnant of the Higgs
potential). Here $F_{ij}$ is the field strength associated to a
gauge field $A$, and $\Phi$ is the Higgs field. These equations may
be viewed as a dimensional reduction of the four dimensional
self-dual equations upon setting all functions independent of $x_4$
and identifying $\Phi=A_4$. Just as Ward's twistor transform relates
instanton solutions in $\mathbb{R}\sp4$ to certain holomorphic
vector bundles over the twistor space $\mathbb{CP}\sp3$, Hitchin
showed \cite{hitchin82} that the dimensional reduction leading to
BPS monopoles could be made at the twistor level as well.
Mini-twistor space is a two dimensional complex manifold isomorphic
to T$\mathbb{P}\sp1$, and BPS monopoles may be identified with
certain bundles over this space. In particular a curve
$\mathcal{C}\subset$ T$\mathbb{P}\sp1$, the spectral curve, arises
in this construction and, subject to certain nonsingularity
conditions, Hitchin was able to prove all monopoles could be
obtained by this approach \cite{hitchin83}. Nahm also gave a
transform of the ADHM instanton construction to produce BPS
monopoles \cite{nahm82}. The resulting Nahm's equations have Lax
form and the corresponding spectral curve is again $\mathcal{C}$.

These curves $\mathcal{C}$ are the curves of the title. Hitchin's
construction (as we shall soon recall) places various
transcendental constraints upon them and the outstanding and
difficult problem is to construct curves satisfying these
constraints. Now, given an appropriate curve, Ercolani and Sinha
showed in their seminal paper \cite{es89} how one could solve (a
gauge transform of) the Nahm equations in terms of a
Baker-Akhiezer function for the curve $\mathcal{C}$. The authors
have extended this work \cite{be06} and, given a curve, one can
solve both the Nahm data and reconstruct the desired monopole in
terms of standard integrable systems constructions. Our purpose
here is not to focus on this integrable system side of the story
but on the curves $\mathcal{C}$ and we shall only cite various
formulae as needed to illustrate the type of detail needed of a
curve in order to implement the construction. The number theory
(and in particular `Ramanujan' of the title) appear when we try to
implement the transcendental constraints on the curve and
construct the required quantities associated with the curve. This
is the story we now recount.

An outline of our article is as follows. In section 2 we recall
the relevant aspects of the Hitchin and Ercolani-Sinha
constructions. This recounts the constraints of Hitchin and we
highlight the ingredients needed to make effective the integrable
system construction and show how these reduce to evaluating
quantities intrinsic to the curve. The first hurdle in
implementing the construction is to analytically determine the
period matrix for $\mathcal{C}$ and then understand the theta
divisor. In section 3 we will introduce a class of (genus 4)
curves for which we can do this. They are of the form
\begin{equation}
\eta^3+\hat\chi(\zeta-\lambda_1)(\zeta-\lambda_2)(\zeta-\lambda_3)
(\zeta-\lambda_4)(\zeta-\lambda_5)(\zeta-\lambda_6)=0,
\label{welstein99}
\end{equation}
where $\lambda_i$, $i=1,\ldots,6$ are distinct complex numbers.
(For appropriate $\lambda_i$ this yields a charge 3 monopole.)
This class of curves was studied by Wellstein over one hundred
years ago \cite{wel99} and more recently by Matsumoto
\cite{matsu00}. Here we will introduce our homology basis and
define branch points in terms of $\theta$-constants following
\cite{matsu00}.

Corresponding to (some of) Hitchin's nonsingularity conditions
Ercolani and Sinha obtain restrictions on the allowed period
matrices for the spectral curve. Equivalent formulations of these
conditions were given in \cite{hmr99}. The Ercolani-Sinha conditions
are transcendental constraints and to solve these is the next
(perhaps \emph{the}) major hurdle to overcome in the construction.
In section 4 we do this for our curves. At this stage we have
replaced the constraints by relations between various hypergeometric
integrals. To simplify matters for the present paper we next demand
more symmetry and consider in section 5 the genus 4 curves
\begin{equation}
\eta^3+\chi(\zeta^6+b \zeta^3-1)=0\label{bren03}
\end{equation}
where $b$ is a certain real parameter. This restriction has the
effect of reducing the number of hypergeometric integrals to be
calculated to two. Interestingly the relations we demand of these
integrals are assertions of Ramanujan only recently proven. We
will denote curves of the form (\ref{bren03}) as \emph{symmetric
monopole} curves (though in fact they may not satisfy all of
Hitchin's nonsingularity conditions). The tetrahedrally symmetric
charge 3 monopole is of this form.

The curve (\ref{bren03}) covers a hyperelliptic curve of genus two
and two elliptic curves. We discuss these coverings. Using
Weierstrass-Poincar\'e reduction theory we are able to express the
theta function behaviour of these symmetric monopoles in terms of
elliptic functions and fairly comprehensive results may be
obtained. Finally, in section 6, we shall consider the curve
(\ref{bren03}) associated with tetrahedrally symmetric 3-monopole
when  the above parameter $b=2\sqrt{5}$. This genus 4 curve covers
4 elliptic curves and all entries to the period matrices are
expressible in terms of elliptic moduli. The analytical means
which we are using for our analysis involve Thomae-type formulae,
Weierstrass-Poincar\'e reduction theory, multivariable
hypergeometric function and higher hypergeometric equalities of
Goursat. Our conclusions in section 7 will highlight various of
our results.

\emph{This article is based upon the second part of the preprint
\cite{be06}. New theoretical results and the unwieldy length of that
paper have led us to separate general constructions from the
exploration of the transcendental constraints upon which we now
report.}

\section{The Monopole Spectral Curve}

In this section we shall recount the transcendental constraints
placed upon the spectral curve coming from Hitchin's construction.
We shall then describe those quantities related to the curve needed
in the Ercolani-Sinha construction and its extensions.

\subsection{Hitchin Data}

Using twistor methods Hitchin \cite{hitchin83} has shown that each
static $SU(2)$ Yang-Mills-Higgs monopole in the BPS limit with
magnetic charge $n$ is equivalent to a spectral curve of a
restricted form. If $\zeta$ is the inhomogeneous coordinate on the
Riemann sphere, and $(\zeta,\eta)$ are the standard local
coordinates on $T\PP\sp1$ (defined by
$(\zeta,\eta)\rightarrow\eta\frac{d}{d\zeta}$), the spectral curve
is an algebraic curve $\mathcal{C} \subset T\PP\sp1$ which has the
form
\begin{equation}
P(\eta,\zeta)=\eta^n+\eta^{n-1} a_1(\zeta)+\ldots+\eta^r
a_{n-r}(\zeta)+ \ldots+\eta\,
a_{n-1}(\zeta)+a_n(\zeta)=0.\label{spectcurve}
\end{equation}
Here $a_r(\zeta)$  (for $1\leq r\leq n$) is a polynomial in
$\zeta$ of maximum degree $2r$.

The Hitchin data constrains the curve $\mathcal{C}$ explicitly in
terms of the polynomial $P(\eta,\zeta)$ and implicitly in terms of
the behaviour of various line bundles on $\mathcal{C}$. If the
homogeneous coordinates of $ \PP\sp1$ are $[\zeta_0,\zeta_1]$ we
consider the standard covering of this by the open sets
$U_0=\{[\zeta_0,\zeta_1]\,|\,\zeta_0\ne0\}$ and
$U_1=\{[\zeta_0,\zeta_1]\,|\,\zeta_1\ne0\}$, with
$\zeta=\zeta_1/\zeta_0$ the usual coordinate on $U_0$. We will
denote by $\widehat U_{0,1}$ the pre-images of these sets under the
projection map $\pi:T\PP\sp1\rightarrow\PP\sp1$. Let
 $L^{\lambda}$ denote the holomorphic line bundle on
$T\PP\sp1$ defined by the transition function
$g_{01}=\rm{exp}(-\lambda\eta/\zeta)$ on $\widehat U_{0}\cap \widehat
U_{1}$, and let $L^{\lambda}(m)\equiv
L^{\lambda}\otimes\pi\sp*\mathcal{O}(m)$ be similarly defined in
terms of the transition function
$g_{01}=\zeta^m\exp{(-\lambda\eta/\zeta)}$. A holomorphic section
of such line bundles is given in terms of holomorphic functions
$f_\alpha$ on $\widehat U_\alpha$ satisfying
$f_\alpha=g_{\alpha\beta}f_\beta$. We denote line bundles on
$\mathcal{C}$ in the same way, where now we have holomorphic
functions $f_\alpha$ defined on $\mathcal{C}\cap\widehat U_\alpha$.

The Hitchin data constrains the curve to satisfy:\\

\textbf{H1}. Reality conditions
\begin{equation}\label{spectcurvereal}
a_r(\zeta)=(-1)^r\zeta^{2r}\overline{a_r(-\frac{1}{\overline{\zeta}})}
.\end{equation}

\textbf{H2}. $L^2$ is trivial on $\mathcal{C}$ and $L(n-1)$ is
real.\\

\textbf{H3}. $H^0(\mathcal{C},L^{\lambda}(n-2))=0$ for $\lambda\in(0,2)$.\\

The reality conditions express the requirement that $\mathcal{C}$ is
real with respect to the standard real structure on $T\PP\sp1$
\begin{equation}
\tau:(\zeta,\eta)\mapsto(-\frac{1}{\bar{\zeta}},
-\frac{\bar{\eta}}{\bar{\zeta}^2}).
\end{equation}
This is the anti-holomorphic involution defined by reversing the
orientation of the lines in ${\mathbb R}\sp3$. A consequence of
the reality condition is that we may parameterise $a_r(\zeta)$ as
follows,
\begin{equation}\label{arpar}
a_r(\zeta)=\sum_{k=0}\sp{2r}a_{r k}\,\zeta\sp{k}= \chi_r
\left[\prod_{l=1}\sp{r}\left(
\frac{\overline{\alpha}_l}{\alpha_l}\right)\sp{1/2}\right]
\prod_{k=1}\sp{r}(\zeta-\alpha_k)(\zeta+\frac{1}{\overline{\alpha}_k}),\qquad
\alpha_r\in \mathbb{C},\ \chi_r\in\mathbb{R}.
\end{equation}
Thus each $a_r(\zeta)$ contributes $2r+1$ (real) parameters.
Certainly this constraint may be readily implemented. The remaining
two constraints are however transcendental. The triviality of $L^2$
on $\mathcal{C}$ means that there exists a nowhere-vanishing
holomorphic section. In terms of our open sets $\widehat U_{0,1}$ we
then have two, nowhere-vanishing holomorphic functions, $f_0$ on
$\widehat U_0\cap\mathcal{C}$ and $f_1$ on $\widehat U_1\cap\mathcal{C}$,
such that on $\widehat U_{0}\cap \widehat U_{1}\cap\mathcal{C}$
\begin{equation}
 f_{0}(\eta,\zeta)=\mathrm{exp} \left\{
-2\frac{\eta}{\zeta} \right\} f_1(\eta,\zeta). \label{triv3}
\end{equation}
Ercolani and Sinha utilize this to give an alternate
characterisation we will present in due course.

For a generic $n$-monopole the spectral curve is irreducible and
has genus $g_\mathcal{C}=(n-1)^2$. This may be calculated as
follows. For fixed $\zeta$ the $n$ roots of $P(\eta,\zeta)=0$
yield an $n$-fold covering of the Riemann sphere. The branch
points of this covering are given by
$$0= {\text{ Resultant}_\eta}(P(\eta,\zeta),\partial_\eta P(\eta,\zeta))=
\prod_{i=1}\sp{n}\partial_\eta P(\eta_i,\zeta), \quad{\text{
where}}\ P(\eta_i,\zeta)=0.$$ This expression is of degree
$n\times\text{deg}\,a_{n-1}=n(2n-2)$ in $\zeta$ and so by the
Riemann-Hurwitz theorem we have that
$$2 g_\mathcal{C}-2=2n(g_{\mathbb{P}\sp1}-1)+n(2n-2)=2(n-1)\sp2-2,$$
whence the genus as stated.

The $n=1$ monopole spectral curve is given by
$$\eta= (x_1+i x_2)-2 x_3 \zeta-(x_1-ix_2)\zeta\sp2,$$
where $x=(x_1,x_2,x_3)$ is any point in ${\mathbb R}\sp3$. In
general the three independent real coefficients of $a_1(\zeta)$
may be interpreted as the centre of the monopole in ${\mathbb
R}\sp3$. Strongly centred monopoles have the origin as center and
hence $a_1(\zeta)=0$. The group $SO(3)$ of rotations of ${\mathbb
R}\sp3$ induces an action on $T\PP\sp1$ via the corresponding
$PSU(2)$ transformations. If
$$\begin{pmatrix} p&q\\-\bar q&\bar p\end{pmatrix}\in PSU(2),
\qquad |p|\sp2+|q|\sp2=1,
$$
the transformation on $T\PP\sp1$ given by
$$\zeta\rightarrow\dfrac{\bar p\, \zeta-\bar q}{q\, \zeta+p},
\qquad \eta\rightarrow \dfrac{\eta}{(q\, \zeta+p)\sp2}
$$
corresponds to a rotation by $\theta$ around ${\mathbf n}\in
S\sp2$, where $n_1\sin{(\theta/2)}=\Im q$,
$n_2\sin{(\theta/2)}=-\Re q$, $n_3\sin{(\theta/2)}=- \Im q$, and
$\cos{(\theta/2)}=\Re p$. (Here the $\eta$ transformation is given
by the derivative of the $\zeta$ transformation.) The $SO(3)$
action commutes with the real structure $\tau$. Although a general
M\"obius transformation does not change the period matrix of a
curve $\mathcal{C}$ only the subgroup $PSU(2)< PSL(2,\mathbb{C})$
preserves the desired reality properties . We have that
$$\alpha_k\rightarrow {\tilde\alpha}_k \equiv\frac{p\alpha_k+{\bar
q}}{{\bar p}-\alpha_k q},\qquad \chi_r \rightarrow
{\tilde\chi}_r\equiv \chi_r \prod_{k=1}\sp{r}\left[\frac{({\bar
p}-\alpha_k q)({ p}-{\bar\alpha_k}{\bar q})({\bar\alpha_k}{\bar
p}+q)(\alpha_k{ p}+ {\bar
q})}{{\alpha_k}{\bar\alpha_k}}\right]\sp{1/2},$$ and
$$
a_r\rightarrow \frac{{\tilde a}_r}{(q\, \zeta+p)\sp{2r}}\equiv
\frac{{\tilde\chi}_r}{(q\, \zeta+p)\sp{2r}}
\left[\prod_{l=1}\sp{r}\left(
\frac{\overline{\tilde\alpha}_l}{\tilde\alpha_l}\right)\sp{1/2}\right]
\prod_{k=1}\sp{r}(\zeta-{\tilde\alpha}_k)(\zeta+\frac{1}{\overline{\tilde\alpha}_k}).
$$
In particular the form of the curve does not change: that is, if
$a_r=0$ then so also ${\tilde a}_r=0$. It is perhaps worth
emphasising that the reality conditions are an extrinsic feature
of the curve (encoding the space-time aspect of the problem)
whereas the intrinsic properties of the curve are invariant under
birational transformations or the full M\"obius group. Such
extrinsic aspects are not a part of the usual integrable system
story.

\subsection{Quantities associated to the curve needed for reconstruction.}
In order to implement the reconstruction of the Nahm data and
monopole associated to a curve $\mathcal{C}$ various other
quantities are also required. We shall now describe these and in
so doing express Ercolani and Sinhas's alternate characterisation
of one of Hitchin's transcendental constraints.

The spectral curve (\ref{spectcurve}) is an $n$-sheeted cover of
$\mathbb{P}\sp1$. By a rotation if necessary it is always possible
to choose $n$ distinct preimages $\{\infty_j\}$ of $\zeta=\infty$
and consequently the roots of ${P(\eta,\zeta)}/{\zeta\sp{2n}}$ near
$\zeta=\infty$ behave as
\begin{equation}\label{rhodef}\frac{P(\eta,\zeta)}{\zeta\sp{2n}}\sim \prod_{j=1}\sp{n}\left(
\frac{\eta}{\zeta\sp2}-\rho_j\right),\end{equation} where the
$\rho_j$ may be assumed distinct. As a consequence we see that at
$\infty_j$ we have
\begin{equation}\label{infj}
\frac{\eta}{\zeta}\sim \rho_j\,\zeta,\qquad
{d}\left(\frac{\eta}{\zeta}\right)\sim \rho_j\,
{d}\zeta=\left(-\frac{\rho_j}{t^2} +O(1)\right){d}t,
\end{equation}
where $t=1/\zeta$ is a local coordinate. Also from (\ref{arpar}) we
have that at $\zeta=0$
\begin{equation}\label{rhobdef}
P(\eta,0)=\prod_{j=1}\sp{n}\left(
\eta+\overline{\rho_j}\right).\end{equation} In view of (\ref{infj})
there exists a differential $\gamma_\infty$, a sum of differentials
of the second kind, such that
\begin{align}
&\gamma_{\infty}(P)=
\left(\frac{\rho_l}{t^2}+O(1)\right){d}t,\quad \text{as}\quad
P\rightarrow\infty_l,\label{gaminf}\\
&\oint\limits_{\mathfrak{a}_k} \gamma_{\infty}(P) = 0,\quad \forall
k=1,\ldots,g.\label{gamnorm}
\end{align}
Here $\{\mathfrak{a}_k,\mathfrak{b}_k\}$  are a canonical homology
basis for $\mathcal{C}$. On the integrable systems side of the
story this differential yields the flow on the Jacobian
corresponding to the time evolution of the system. This direction
is given by
\begin{equation}\label{wvec}
U_k=\frac{1}{2\pi\imath}\oint_{\mathfrak{b}_k}\gamma_{\infty}(P).
\end{equation}
In practice one does not need $\gamma_{\infty}(P)$ explicitly but
(\ref{wvec}) and
\begin{equation}
\nu_{j}=\lim_{P\rightarrow \infty_j}
\left[\int\limits_{P_0}^P\gamma_{\infty}(P)+\frac{\eta}{\zeta}
\right].\label{etacondition}
\end{equation}

At this stage we have not imposed the Hitchin constraints H2, H3 on
our curve $\mathcal{C}$. Ercolani and Sinha  now make the following
connection between (\ref{wvec}) and H2. Consider the logarithmic
derivative of (\ref{triv3}) representing the triviality of $L^2$ on
$\mathcal{C}$,
\begin{equation}
d\mathrm{log}\,f_{0}={d}\left(
-2\frac{\eta}{\zeta}\right) +d\mathrm{log}
f_{1}\label{derivative}.
\end{equation}
(Hurtubise considered a similar construction in the $n=2$ case
\cite{hurt}.) Now in order to avoid essential singularities in
$f_{0,1}$ we have from (\ref{rhodef}, \ref{rhobdef}) that
\begin{align}
d\mathrm{log}\,f_{1}(P)
&=\left(-\frac{2\eta_j(0)}{\zeta^2}
+O(1)\right){d}\zeta=\left(\frac{2{\bar \rho}_j(0)}{\zeta^2}
+O(1)\right){d}\zeta,\quad \text{at}\quad P\rightarrow 0_j,\\
d\mathrm{log}\,f_{0}(P) &=\left(\frac{2\rho_j}{t^2}
+O(1)\right){d}t,\quad \text{at}\quad P\rightarrow
\infty_j.\label{foex}
\end{align}
Because $f_{0}$  is a function on $U_0=\mathcal{C}\setminus
\{P_j\}_{j=1}\sp{n}$, then
\begin{equation}
\mathrm{exp}\oint\limits_{\lambda}d\mathrm{log}\,f_{0}=1,
\end{equation}
for all cycles $\lambda$ from $H_1(\mathbb{Z},\mathcal{C})$. A
similar result follows for $f_1$ and upon noting
(\ref{derivative}) we may define
\begin{align}
m_j =
-\frac{1}{2\pi\imath}\oint\limits_{\mathfrak{a}_j}\,d\mathrm{log}
\,
f_{0}=-\frac{1}{2\pi\imath}\oint\limits_{\mathfrak{a}_j}d
\mathrm{log} \, f_{1}\label{denotem},\\
n_j =
\frac{1}{2\pi\imath}\oint\limits_{\mathfrak{b}_j}d\mathrm{log}
\, f_{0}=\frac{1}{2\pi\imath}\oint\limits_{\mathfrak{b}_j}
d\mathrm{log} \, f_{1}.\label{denoten}
\end{align}
Further, in view of (\ref{gaminf}) and (\ref{foex}), we may write
\begin{equation}\label{gamfo}
\gamma_\infty(P)=\frac{1}{2}\,d\mathrm{log} \,
f_{0}(P)+\imath \pi\,\sum_{j=1}^g m_j\, v_j(P),
\end{equation}
where $v_j$ are canonically $\mathfrak{a}$-normalized holomorphic
differentials. Integrating $\gamma_\infty$ around
$\mathfrak{b}$-cycles leads to the Ercolani-Sinha constraints
\begin{equation*}
\oint\limits_{\mathfrak{b}_k}\gamma_\infty=\imath\pi  n_k
+\imath\pi\sum_{l=1}^g\tau_{kl} m_l,\label{winding}
\end{equation*}
which are necessary and sufficient conditions for $L\sp2$ to be
trivial when restricted to $ \mathcal{C}$. Thus
\begin{equation}\label{ESC1}
\boldsymbol{U}=\frac12 \boldsymbol{n}+\frac12\tau\boldsymbol{m}.
\end{equation}
Therefore the vector $2\boldsymbol{U}\in\Lambda $, the period
lattice for the curve $\mathcal{C}$, and so the ``winding-vector"
vector $\mathcal{}$ is a half-period. Note that $\boldsymbol{U}\ne
0$ or otherwise $\gamma_\infty$ would be holomorphic contrary to
our choice.

Using the bilinear relations we have that, for any holomorphic
differential $\Omega$,
$$\sum_{l=1}\sp{g} U_l\oint_{\mathfrak{a}_l}\Omega=
\frac{1}{2\pi\imath}\sum_{l=1}\sp{g}\left(
\oint_{\mathfrak{a}_l}\Omega
\oint_{\mathfrak{b}_l}\gamma_{\infty}(P)-
\oint_{\mathfrak{b}_l}\Omega
\oint_{\mathfrak{a}_l}\gamma_{\infty}(P)
\right)=\sum_{i=1}^n\mathrm{Res}_{P\rightarrow\infty_i}
\gamma_{\infty}(P)\int_{P_0}^P \Omega.
$$
Houghton, Manton and Ram\~ao utilise this expression to express a
dual form of the Ercolani-Sinha constraints (\ref{ESC1}). Define
the 1-cycle
\begin{equation}\label{HMRcurve}
  \mathfrak{c}=\sum_{l=1}\sp{g}(n_l \mathfrak{a}_l +m_l \mathfrak{b}_l ).
\end{equation}
Then (upon recalling that $\tau_{lk}\oint_{\mathfrak{a}_k}\Omega
=\oint_{\mathfrak{b}_l}\Omega$, where $\tau$ is the period matrix)
we have the equivalent constraint:
\begin{equation}\label{ESC2}
\oint_{\mathfrak{c}}\Omega=2\sum_{i=1}^n\mathrm{Res}_{P\rightarrow\infty_i}
\gamma_{\infty}(P)\int_{P_0}^P \Omega.
\end{equation}
The right-hand side of this equation is readily evaluated. We may
express an arbitrary holomorphic differential $\Omega$ as,
\begin{align}
\Omega &=\frac{\beta_0\eta^{n-2}+\beta_1(\zeta)\eta^{n
-3}+\ldots+\beta_{n-2}(\zeta)}{\frac{\partial\mathcal{P}}{\partial
\eta}}\,d\zeta \label{holdefes} \\
&=\frac{\beta_0(\eta/\zeta^2)^{n-2}+\tilde\beta_1(1/\zeta)
(\eta/\zeta^2)^{n-3}+\ldots+\tilde\beta_{n-2}(1/\zeta)}{\sum_{i=1}\sp{n}
\prod_{\substack{j=1\\ j\ne
i}}\sp{n}\left(\eta/\zeta^2-\mu_j(1/\zeta)\right)}\,\frac{d\zeta}{\zeta^2},
\nonumber
\end{align}
where $\beta_j(\zeta)\equiv \zeta^{2j}\tilde \beta_j (1/\zeta)$ is
a polynomial of degree at most $2j$ in $\zeta$. Thus using
(\ref{rhodef}) we obtain
$$
\sum_{i=1}^n\mathrm{Res}_{P\rightarrow\infty_i}
\gamma_{\infty}(P)\int_{P_0}^P \Omega
=-\sum_{i=1}^n\frac{\beta_0\rho_i^{n-1}
+\tilde\beta_1(0)\rho_i^{n-2}+\ldots+\tilde\beta_{n-2}(0)\rho_i}
{\prod_{j\neq i}^n (\rho_i-\rho_j) }=-\beta_0,
$$
upon using Lagrange interpolation%
. At this stage we have from the condition H2,
\begin{lemma}[Ercolani-Sinha Constraints] The following are equivalent:
\begin{enumerate} \item $L\sp2$ is trivial on $\mathcal{C}$.

\item There exists a 1-cycle
$\mathfrak{c}=\boldsymbol{n}\cdot{\mathfrak{a}}+
\boldsymbol{m}\cdot{\mathfrak{b}}$ such that for every holomorphic
differential $\Omega$ (\ref{holdefes}),
\begin{equation}
\oint\limits_{\mathfrak{c}}\Omega=-2\beta_0,\label{HMREScond}
\end{equation}
\item $2\boldsymbol{U}\in \Lambda\Longleftrightarrow$
\begin{equation}\label{EScond}
\boldsymbol{U}=\frac{1}{2\pi\imath}\left(\oint_{\mathfrak{b}_1}\gamma_{\infty},
\ldots,\oint_{\mathfrak{b}_g}\gamma_{\infty}\right)\sp{T}= \frac12
\boldsymbol{n}+\frac12\tau\boldsymbol{m} .
\end{equation}
\end{enumerate}
\end{lemma}
Here (2) is the dual form of the Ercolani-Sinha constraints given
by Houghton, Manton and Ram\~ao. Their 1-cycle generalises a
similar constraint arising in the work of Corrigan and Goddard
\cite{CorGod}. The only difference between (3) and that of
Ercolani-Sinha Theorem II.2 is in the form of $\boldsymbol{U}$ in
which we disagree. We also know that $\boldsymbol{U}\ne 0$.

The Ercolani-Sinha constraints impose $g$ conditions on the period
matrix of our curve. We have seen that the coefficients $a_r(\zeta)$
each give $2r+1$ (real) parameters, thus the moduli space of charge
$n$ centred $SU(2)$ monopoles is
$$\sum_{r=2}\sp{n}(2r+1)-g=(n+3)(n-1)-(n-1)\sp2=4(n-1)$$
(real) dimensional.

The 1-cycle appearing in the work of Houghton, Manton and Ram\~ao
further satisfies
\begin{corollary}[Houghton, Manton and Ram\~ao,
 2000]\label{HMRinvc} $\tau_*\mathfrak{c}=-\mathfrak{c}$.
\end{corollary}
This result is the dual of Hitchin's remark \cite[p164]{hitchin83}
that the triviality of $L\sp2$ together with the antiholomorphic
isomorphism $L\cong L\sp*$ yields an imaginary lattice point with
respect to $H\sp1(\mathcal{C},\mathbb{Z})\subset
H\sp1(\mathcal{C},\mathcal{O})$.

The final constraint of Hitchin and various quantities needed for
our purposes are encoded in the expressions
\begin{equation}\label{newbafnch}
\Psi _{j }\left( z, P\right) =g_{j }(P)\, \frac{
\theta_{\frac{\boldsymbol{m}}{2},\frac{\boldsymbol{n}}{2}} \left(
\boldsymbol{\phi} (P)-\boldsymbol{\phi}(\infty_{j
})+z\,\boldsymbol{U}-\widetilde{\boldsymbol{K}}\right)
\theta_{\frac{\boldsymbol{m}}{2},\frac{\boldsymbol{n}}{2}}
\left(-\widetilde{\boldsymbol{K}} \right) } {
\theta_{\frac{\boldsymbol{m}}{2},\frac{\boldsymbol{n}}{2}}
\left(\boldsymbol{\phi} (P)-\boldsymbol{\phi}(\infty_{j })
-\widetilde{\boldsymbol{K}}\right)
\theta_{\frac{\boldsymbol{m}}{2},\frac{\boldsymbol{n}}{2}} \left(
z\,\boldsymbol{U}-\widetilde{\boldsymbol{K}}\right) }\,
e^{z\,\int\limits_{P_0}^{P}\gamma_{\infty}-z\,\nu_j}
\end{equation}
which we shall now unpack. Here
$\theta_{\frac{\boldsymbol{m}}{2},\frac{\boldsymbol{n}}{2}}$ are
theta functions with characteristics, $\phi$ is the Abel map,
$z\in(-1,1)$, and $P\in\mathcal{C}$. (Our conventions for theta
functions are given in the Appendix.) The vector
$\widetilde{\boldsymbol{K}}$ plays a special role in the monopole
construction. It is defined by
$$\widetilde{\boldsymbol{K}}=
\boldsymbol{K}+\boldsymbol{\phi}\left((n-2)
\sum_{k=1}\sp{n}\infty_k\right),$$ where $\boldsymbol{K}$ is the
vector of Riemann constants (our conventions regarding this are
also given in the Appendix). We have that
\begin{enumerate}\item $\widetilde{\boldsymbol{K}}$ is independent of the choice
of base point of the Abel map; \item $
\theta(\widetilde{\boldsymbol{K}})=0$;
\item $2\widetilde{\boldsymbol{K}} \in \Lambda$; \item for $n\ge3$ we
have $ \widetilde{\boldsymbol{K}}\in \Theta_{\rm singular}$.
\end{enumerate}
The point $\widetilde{\boldsymbol{K}}$ is the distinguished point
Hitchin uses to identify degree $g-1$ line bundles with
$\Jac(\mathcal{C})$. The proof of these properties together with
the following lemma further constraining the Ercolani-Sinha vector
may be found in \cite{be06}:
\begin{lemma}\label{ueventheta}$\boldsymbol{U}\pm
\widetilde{\boldsymbol{K}}$ is a non-singular even theta
characteristic.
\end{lemma}

The expressions (\ref{newbafnch}) are the components of a
Baker-Akhiezer function; they are sections of a line bundle
$L\sp{z+1}(n-1)$. The $g_j(P)$ which we have left unspecified form
a basis of the holomorphic sections of $L(n-1)$. (They can be
explicitly described in terms of theta functions and the
quantities introduced already, but we do not need this for our
present purposes.) Now the full condition H3 is that
$L\sp{z+1}(n-2)\in
\textrm{Jac}\sp{g_\mathcal{C}-1}(\mathcal{C})\setminus \Theta$ for
$z\in(-1,1)$. This constraint must be checked using knowledge of
the $\Theta$ divisor. The exact sequence
$\mathcal{O}(L\sp{s})\hookrightarrow \mathcal{O}(L\sp{s}(n-2))$
given by multiplication by a section of
$\pi\sp*\mathcal{O}(n-2)\vert_\mathcal{C}$ does however give us
the necessary condition
\begin{equation}\label{h0lz}
H\sp0\left(\mathcal{C},\mathcal{O}(L\sp{s}(n-2))\right)=0
\Longrightarrow
H\sp0\left(\mathcal{C},\mathcal{O}(L\sp{s})\right)=0,\qquad
s\in(0,2).
\end{equation}
If $L\sp{s}$ were trivial we would have a section, contradicting
this vanishing result. The same treatment given to the triviality
of $L\sp2$ shows that if $L\sp{s}$ were trivial then
$s\,\boldsymbol{U}\in \Lambda$. Therefore (\ref{h0lz}) shows that
$s\,\boldsymbol{U}\not\in \Lambda$ for $s\in(0,2)$. Thus
$2\boldsymbol{U}$ is a primitive vector in $\Lambda$ and we obtain
the final part of the Ercolani-Sinha constraints,
\begin{equation}\label{ESprim}
2\boldsymbol{U}\ \text{ is a primitive vector in} \ \Lambda
\Longleftrightarrow \mathfrak{c}\ \text{ is primitive in}\ H_1(
\mathcal{C},\mathbb{Z}).
\end{equation}

The Ercolani-Sinha constraints (\ref{EScond}) or (\ref{HMREScond})
place $g$ transcendental constraints on the spectral curve $
\mathcal{C}$ and a major difficulty in implementing this
construction has been in solving these, even in simple examples.
The case of $n=2$ has been treated by several authors and going
beyond these results we consider the case of $n=3$ in the
remaining portion of this work. Before doing this let us record
the remaining expressions needed for the integrable system
reconstruction which are to be found in an object $Q_0(z)$ used by
Ercolani-Sinha to reconstruct the Nahm data. The work \cite{be06}
gives
\begin{theorem} The matrix $Q_0(z)$ (which has poles of first order at
$z=\pm1$) may be written
\begin{equation}\label{ourq0}
Q_{0}(z)_{jl}  = \epsilon_{jl}\,\frac{\rho_{j}-
\rho_{l}}{\mathcal{E}(\infty_j,\infty_l)}\,e\sp{i\pi\boldsymbol{\tilde
q}\cdot(\boldsymbol{\phi}(\infty_l)-\boldsymbol{\phi}(\infty_j))}\,
    \,\frac{\theta(\boldsymbol{\phi}(\infty_{l}) -\boldsymbol{\phi}(
    \infty_{j}) + [z+1]\boldsymbol{U} - \widetilde{\boldsymbol{K}})}{\theta(
    [z+1]\boldsymbol{U}
    - \widetilde{\boldsymbol{K}})}\,e^{z(\nu_{l} - \nu_{j})}.
\end{equation}
Here $E(P,Q)=\mathcal{E}(P,Q)/\sqrt{dx(P)dx(Q)}$ is the
Schottky-Klein prime form, $\boldsymbol{U} -
\widetilde{\boldsymbol{K}}=\frac12\boldsymbol{\tilde
p}+\frac12\tau\boldsymbol{\tilde q}$ ($\boldsymbol{\tilde p}$,
$\boldsymbol{\tilde q}\in\mathbb{Z}\sp{g}$) is a non-singular even
theta characteristic, and $\epsilon_{jl}=\epsilon_{lj}=\pm1$ is
determined (for $j<l$)
 by
$\epsilon_{jl}=\epsilon_{jj+1}\epsilon_{j+1j+2}\dots\epsilon_{l-1l}$.
The $n-1$ signs $\epsilon_{jj+1}=\pm1$ are arbitrary.
\end{theorem}
In passing we note that a formula with similar features was
obtained by Dubrovin \cite{dub77} when giving a theta-functional
solution to the Euler equation describing motion of the
$n$-dimensional rigid body. (Dubrovin's curve was of course
different and was free of Hitchin's constraints.) This theorem
encodes the data needed to reconstruct the Nahm data and monopoles
given a suitable spectral curve. The whole construction is
predicated on the theta functions built from the spectral curve.
Thus we need
\begin{enumerate}
    \item To construct the period matrix $\tau$ associated to
    $\mathcal{C}$.
    \item To determine the half-period $\widetilde{\boldsymbol{K}}$.
    \item To determine the Ercolani-Sinha vector $\boldsymbol{U}$.
    \item For normalised holomorphic differentials $\boldsymbol{v}$ to
    calculate $\int\limits_{\infty_i}^{\infty_j}
   \boldsymbol{v}=\boldsymbol{\phi}(\infty_j)-\boldsymbol{\phi}(\infty_i)$.
   \item To determine $E(\infty_j,\infty_l)$.
   \item To determine $\gamma_{\infty}(P)$ and
   $\nu_i=\lim_{P\to\infty_i}\left(\int_{P_0}\sp{P}\gamma_{\infty}(P')+\frac{\eta}
{\zeta}(P)\right)$.
\end{enumerate}

\section{The trigonal curve}
We shall now introduce the class of curves that will be the focus
of our attention. These are
\begin{equation}
\eta^3+\hat\chi(\zeta-\lambda_1)(\zeta-\lambda_2)(\zeta-\lambda_3)
(\zeta-\lambda_4)(\zeta-\lambda_5)(\zeta-\lambda_6)=0.
\label{cubic}
\end{equation}
For suitable $\lambda_i$ they correspond\footnote{Here
$\{\lambda_i\}_{i=1}\sp{6}=\{\alpha_j,-{1}/{{\overline\alpha}_j}\}_{j=1}\sp{3}$
and $\hat\chi=\chi_3 \left[\prod_{l=1}\sp{3}\left(
\frac{\overline{\alpha}_l}{\alpha_l}\right)\sp{1/2}\right]$.} to
centred charge three monopoles restricted by $a_2(\zeta)=0$. Thus
the eight dimensional moduli space of centred monopoles has been
reduced to three dimensions. The asymptotic behaviour of the curve
gives us
\begin{equation}
\rho_k= -\hat\chi\sp{\frac13}\,e\sp{2\imath k\pi/3}.
\label{curverho}
\end{equation}
For notational convenience we will study (\ref{cubic}) in the form
($w=-\hat\chi\sp{-\frac{1}{3}}\eta$, $z=\zeta$)
\begin{equation}
w^3=\prod_{i=1}^6(z-\lambda_i). \label{curvegena}
\end{equation}
The moduli space of such curves with an homology marking can be
regarded as the configuration space of six distinct points on
$\mathbb{P}\sp1$. This class of curves has been studied by
Picard \cite{picard83}, Wellstein
\cite{wel99}, Shiga \cite{shiga88} and more
recently by Matsumoto \cite{matsu00}; we shall recall some of their
results. To make concrete the $\theta$-functions arising in the
Ercolani-Sinha construction we need to have the period matrix for
the curve, the vector of Riemann constants, and to understand the
special divisors. We shall now make these things explicit, beginning
first with our choice of homology basis.

\subsection{The curve and homologies}
Let $\mathcal{C}$ denote the curve (\ref{curvegena}) of genus four
where the six points $\lambda_i\in\mathbb{C}$ are assumed distinct
and ordered according to the rule $\mathrm{arg}(\lambda_1)<
\mathrm{arg}(\lambda_2)<\ldots< \mathrm{arg}(\lambda_6)$. Let
$\mathcal{R}$ be the automorphism of $\mathcal{C}$ defined by
\begin{equation}\label{curvesym}
    \mathcal{R}:(z,w)\rightarrow (z,\rho w),\quad
\rho=\mathrm{exp}\{2\imath\pi/3\}.
\end{equation}

The bilinear transformation $(z,w)\leftrightarrow (Z,W)$
\begin{align}
\begin{split}
Z&= \frac{(\lambda_2-\lambda_1)(z-\lambda_4)   }
{(\lambda_2-\lambda_4)(z-\lambda_1)},\\
W&=
-\frac{w}{(z-\lambda_1)^2}
\left(\prod_{k=2}^6(\lambda_1-\lambda_k)\right)^{-\frac13}
\left(\frac{(\lambda_1-\lambda_4)(\lambda_1-\lambda_2)}
{\lambda_2-\lambda_4}\right)^{\frac53}\end{split}\label{rotation1}\end{align}
and its inverse
\begin{align}
\begin{split}
z&=\frac{Z\lambda_1(\lambda_2-\lambda_4)
+\lambda_4(\lambda_1-\lambda_2)}{Z(\lambda_2-\lambda_4)
-(\lambda_2-\lambda_1)}\\
w&=-\frac{W}{(Z(\lambda_2-\lambda_4)-(\lambda_2-\lambda_1))^2}
\left(\prod_{k=2}^6(\lambda_1-\lambda_k)\right)^{\frac13}
\, (\lambda_1-\lambda_2)^{\frac13}(\lambda_1-\lambda_4)^{\frac13}
(\lambda_2-\lambda_4)^{\frac53}\end{split} \label{rotation2}
\end{align}
leads to the following normalization of the curve (\ref{curvegen})
\begin{equation}
W^3=Z(Z-1)(Z-\Lambda_1)(Z-\Lambda_2)(Z-\Lambda_3),
\label{curvegenb}
\end{equation}
where
\begin{equation}\Lambda_i=\frac{\lambda_2-\lambda_1}{\lambda_2-\lambda_4}
\frac{\lambda_{2+j(i)}-\lambda_4}{\lambda_{2+j(i)}-\lambda_1}, \quad
i=1,2,3;\ j(1)=1,\ j(2)=3,\ j(3)=4.
\end{equation}

Fix the following lexicographical ordering of independent canonical
holomorphic differentials of $\mathcal{C}$,
\begin{equation} {d}u_1= \frac{{d} z}{w},\quad
                 {d}u_2= \frac{{d} z}{w^2},\quad
                  {d}u_3= \frac{z{d} z}{w^2},\quad
                  {d}u_4= \frac{z^2{d} z}{w^2}.
\label{diffbasis}
 \end{equation}

To construct the symplectic basis
$(\mathfrak{a}_1,\ldots,\mathfrak{a}_4;
\mathfrak{b}_1,\ldots,\mathfrak{b}_4)$ of
$H_1(\mathcal{C},\mathbb{Z})$ we introduce oriented paths
$\gamma_k(z_i,z_j)$ going  from $P_i=(z_i,w_i)$ to $P_j=(z_j,w_j)$
in the  $k$-th sheet. Define 1-cycles
$\mathfrak{a}_i,\mathfrak{b}_i$ on $\mathcal{C}$ as
follows\footnote{This is the basis from \cite{matsu00}; another
but equivalent basis can be found in \cite{wel99}.}
\begin{align}\begin{split}
\mathfrak{a}_1&=\gamma_1(\lambda_1,\lambda_2)+\gamma_2(\lambda_2,\lambda_1),
\qquad
\mathfrak{b}_1=\gamma_1(\lambda_2,\lambda_1)+\gamma_3(\lambda_1,\lambda_2),\\
\mathfrak{a}_2&=\gamma_1(\lambda_3,\lambda_4)+\gamma_2(\lambda_4,\lambda_3)
,\qquad
\mathfrak{b}_2=\gamma_1(\lambda_4,\lambda_3)+\gamma_3(\lambda_3,\lambda_4),\\
\mathfrak{a}_3&=\gamma_1(\lambda_5,\lambda_6)
+\gamma_2(\lambda_6,\lambda_5),\qquad
\mathfrak{b}_3=\gamma_1(\lambda_6,\lambda_5)
+\gamma_3(\lambda_5,\lambda_6),\\
\mathfrak{a}_4&=\gamma_3(\lambda_1,\lambda_2)
+\gamma_1(\lambda_2,\lambda_6)
+\gamma_3(\lambda_6,\lambda_5)+\gamma_2(\lambda_5,\lambda_1),
\\
\mathfrak{b}_4&
=\gamma_2(\lambda_2,\lambda_1)+\gamma_3(\lambda_6,\lambda_2)
+\gamma_2(\lambda_5,\lambda_6)+\gamma_1(\lambda_1,\lambda_5).
\end{split}\label{homology}
\end{align}
The $\mathfrak{a}$-cycles of the homology basis are given in Figure
1, with the $\mathfrak{b}$-cycles shifted by one sheet.  We have the
pairings $\mathfrak{a}_k\circ \mathfrak{a}_l=\mathfrak{b}_k\circ
\mathfrak{b}_l=0$, $\mathfrak{a}_k\circ
\mathfrak{b}_l=-\mathfrak{b}_k\circ \mathfrak{a}_l= \delta_{k,l}$
and therefore
$(\mathfrak{a}_1,\ldots,\mathfrak{a}_4;\mathfrak{b}_1,\ldots,\mathfrak{b}_4)
$ is a symplectic basis of $ H_1(\mathcal{C},\mathbb{Z})$. In the
homology basis introduced we have
\begin{equation}
\mathcal{R}(\mathfrak{b}_i)=\mathfrak{a_i},\quad i=1,2,3, \quad
\mathcal{R}(\mathfrak{b}_4)=-\mathfrak{a}_4.
\end{equation}
As $(1+\mathcal{R}+\mathcal{R}\sp2)\mathfrak{c}=0$ for any cycle
$\mathfrak{c}$ we have, for example, that
$\mathcal{R}(\mathfrak{a}_i)=-\mathfrak{a}_i-\mathcal{R}\sp2(\mathfrak{a}_i)=
-\mathfrak{a}_i-\mathfrak{b}_i$ for $i=1,2,3$ and
$\mathcal{R}(\mathfrak{a}_4)= -\mathfrak{a}_4+\mathfrak{b}_4$, so
completing the $\mathcal{R}$ action on the homology basis.



\begin{figure}
\centering
\begin{minipage}[l]{0.6\textwidth}
\includegraphics[width=6cm]{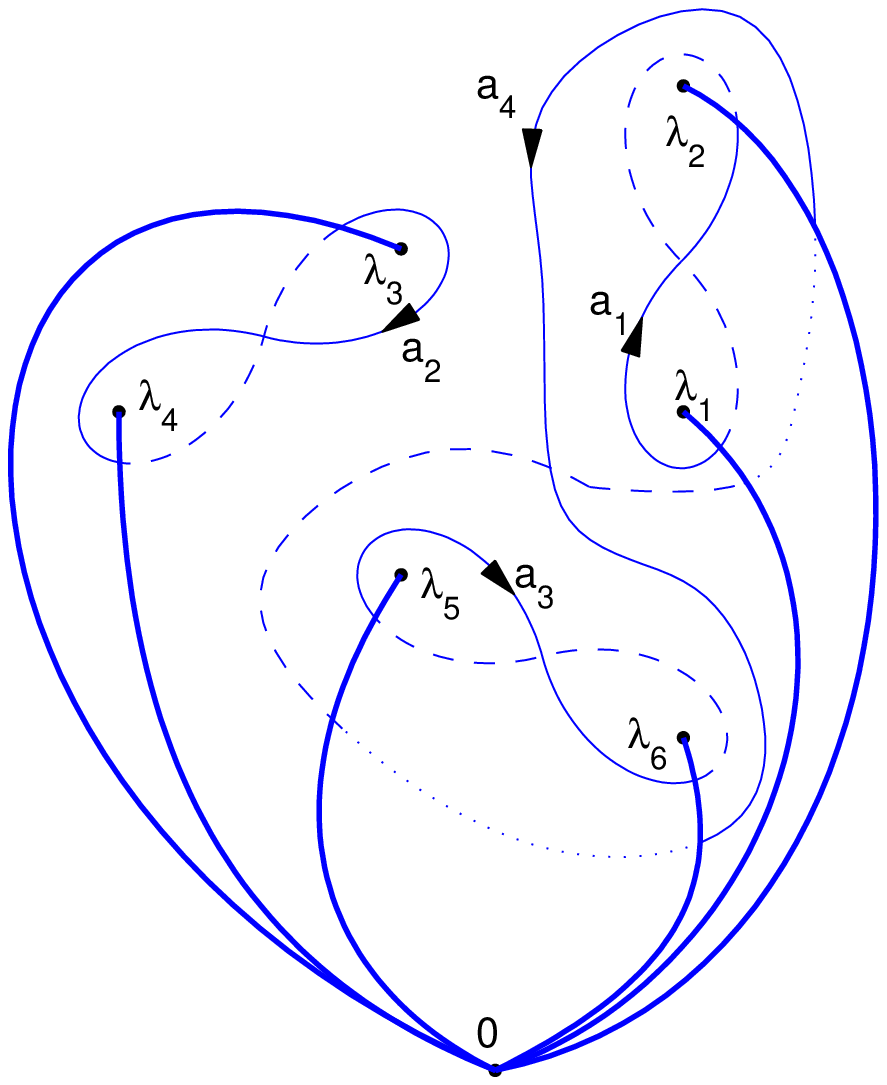} \caption{Homology basis: $\mathfrak{a}$-cycles }
\end{minipage}%
\begin{minipage}[l]{0.4\textwidth}
\includegraphics[width=3cm]{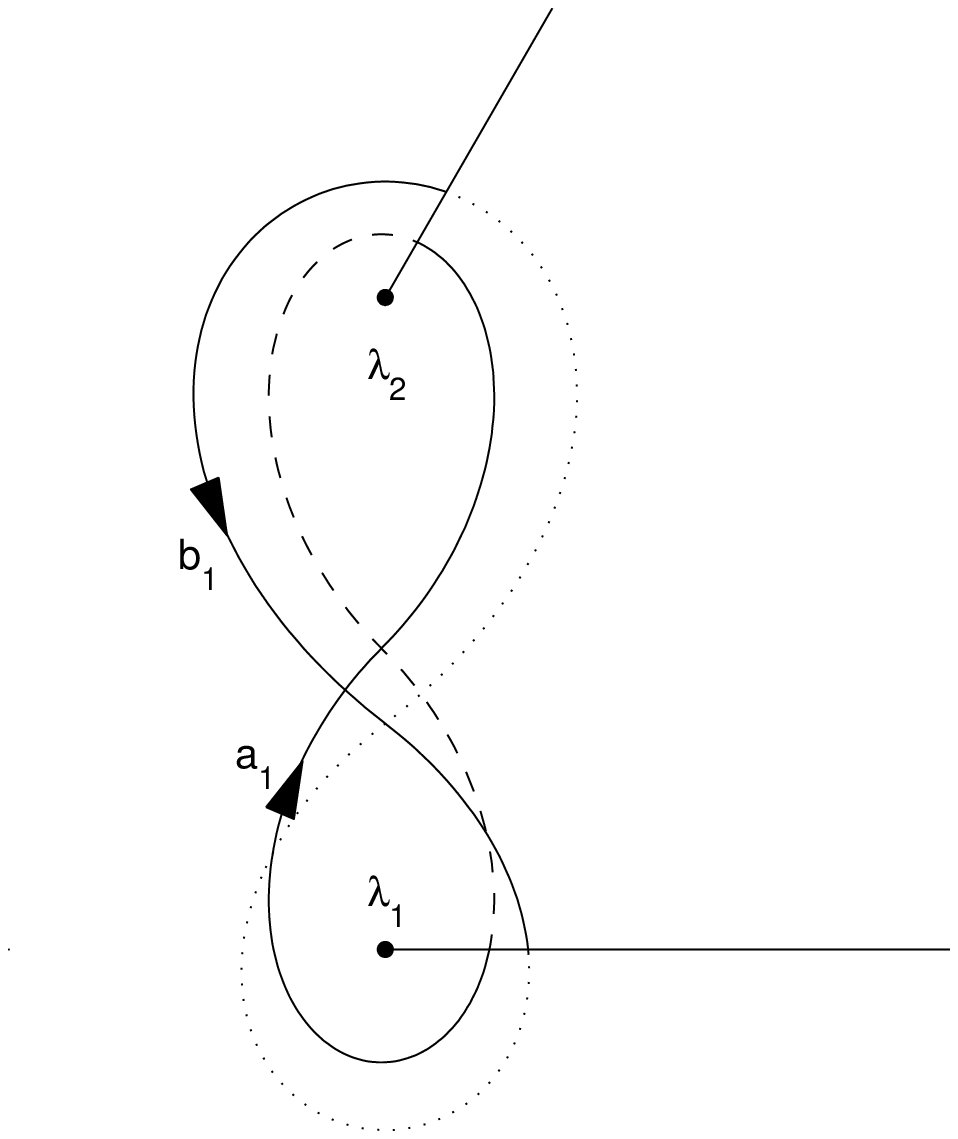} \caption{Cycles $\mathfrak{a}_1$ and
 $\mathfrak{b}_1$ }
\end{minipage}
\end{figure}

\subsection{The Riemann period matrix}
Denote vectors
\begin{align*}
\boldsymbol{x}&=(x_1,x_2,x_3,x_4)^T
=\left(
\oint_{\mathfrak{a}_1}{d}u_1,\ldots,\oint_{\mathfrak{a}_4}{d}u_1 \right)^T ,\\
\boldsymbol{b}&=(b_1,b_2,b_3,b_4)^T
=\left(
\oint_{\mathfrak{a}_1}{d}u_2,\ldots,\oint_{\mathfrak{a}_4}{d}u_2 \right)^T ,\\
\boldsymbol{c}&=(c_1,c_2,c_3,c_4)^T
=\left(
\oint_{\mathfrak{a}_1}{d}u_3,\ldots,\oint_{\mathfrak{a}_4}{d}u_3 \right)^T ,\\
\boldsymbol{d}&=(d_1,d_2,d_3,d_4)^T =\left(
\oint_{\mathfrak{a}_1}{d}u_4,\ldots,\oint_{\mathfrak{a}_4} {d}u_4
\right)^T .
\end{align*}
Crucial for us is the fact that the symmetry (\ref{curvesym})
allows us to relate the matrices of $\mathfrak{a}$ and
$\mathfrak{b}$-periods. For any contour $\Gamma$ and one form
$\omega$ we have that
$\oint\limits_{\mathcal{R}(\Gamma)}\omega=\oint\limits_{\Gamma}\mathcal{R}\sp*\omega$.
If $(\tilde z,\tilde w)=(z,\rho w)=R(z,w)$ then, for example,
$$\mathcal{R}\sp*\left(du_2\right)=\mathcal{R}\sp*\left(
\frac{d\tilde z}{{\tilde w}\sp2}\right)=\frac{d z}{{\tilde
w}\sp2}=\frac{d z}{\rho\sp2 {w}\sp2}=\rho\frac{d z}{{w}\sp2}$$
leading to
$$\oint\limits_{\mathfrak{a}_1} {d}u_2
=\oint\limits_{\mathcal{R}(\mathfrak{b}_1)} {d}u_2=
\oint\limits_{\mathfrak{b}_1}\mathcal{R}\sp*({d}u_2)=
\oint\limits_{\mathfrak{b}_1}\frac{dz}{\rho\sp{2}w\sp2}=
\rho\oint\limits_{\mathfrak{b}_1} {d}u_2.
$$
We find that
\begin{align}\begin{split}
\mathcal{A}&=\left(\mathcal{A}_{ki}\right)=\left(
  \oint\limits_{\mathfrak{a}_k} {d}u_i\right)_{i,k=1,\ldots,4}
=(\boldsymbol{x},\boldsymbol{b},\boldsymbol{c},\boldsymbol{d})
\\
 \mathcal{B}&=\left(\mathcal{B}_{ki}\right)=\left(
  \oint\limits_{\mathfrak{b}_k} {d}u_i\right)_{i,k=1,\ldots,4}
= (\rho H\boldsymbol{x},\rho^2 H\boldsymbol{b},\rho^2
H\boldsymbol{c}, \rho^2
H\boldsymbol{d})=H\mathcal{A}\Lambda,\end{split}\label{calba}
\end{align}
where $H=\mathrm{diag}(1,1,1,-1)$ and
$\Lambda=\mathrm{diag}(\rho,\rho\sp2,\rho\sp2,\rho\sp2)$. This
relationship between the $\mathfrak{a}$ and $\mathfrak{b}$-periods
leads to various simplifications of the Riemann identities,
$$\sum_i\left(
\oint\limits_{\mathfrak{a}_i} {d}u_k \oint\limits_{\mathfrak{b}_i}
{d}u_l - \oint\limits_{\mathfrak{b}_i} {d}u_k
\oint\limits_{\mathfrak{a}_i} {d}u_l \right)=0.
$$
For $k=1$ and $l=2,3,4$ we obtain (respectively) that
\begin{equation}
\boldsymbol{x}^TH\boldsymbol{b} =\boldsymbol{x}^TH\boldsymbol{c}
=\boldsymbol{x}^TH\boldsymbol{d}=0, \label{brr}
\end{equation}
relations we shall employ throughout the paper.

Given $\mathcal{A}$ and $\mathcal{B}$ we now construct the Riemann
period matrix which belongs to the Siegel upper half-space
$\mathbb{S}^4$ of degree 4. If one works with canonically
$\mathfrak{a}$-normalized differentials the period matrix (in our
conventions) is $\tau_\mathfrak{a}=\mathcal{B}\mathcal{A}^{-1}$
while for canonically $\mathfrak{b}$-normalized differentials it
is $\tau_{\mathfrak{b}}=\mathcal{A}\mathcal{B}^{-1}$. Clearly
$\tau_{\mathfrak{b}}=\tau_\mathfrak{a}\sp{-1}$ and we shall simply
denote the period matrix by $\tau$ if neither normalization is
necessary.

\begin{proposition}[Wellstein, 1899; Matsumoto, 2000]
\label{matsumoto1} Let $\mathcal{C}$ be the triple covering of
$\mathbb{P}^1$ with six distinct point $\lambda_1,
\ldots,\lambda_6$,
\begin{equation}
w^3=\prod_{i=1}^6(z-\lambda_i) .\label{curvegen}
\end{equation}
Then the Riemann period matrix is of the form
\begin{align}
\tau_{\mathfrak{b}}&=\rho\left(
H-(1-\rho)\frac{\boldsymbol{x}\boldsymbol{x}^T} {\boldsymbol{x}^T
H\boldsymbol{x}}   \right),\label{taumat}
\end{align}
where $H=\mathrm{diag}(1,1,1,-1)$. Then $\tau_{\mathfrak{b}}$ is
positive definite if and only if
\begin{align} \bar{\boldsymbol{x}}^T H \boldsymbol{x} <0.
\label{condition1}
\end{align}
\end{proposition}
Both Wellstein and Matsumoto give broadly similar proofs of
(\ref{taumat}) and we shall present another variant as we need to
use an identity established in the proof later in the text.

\begin{proof}From (\ref{brr}) we see that we have
$$\mathcal{A}\sp{T} H \boldsymbol{x}=(\Delta,0,0,0)\sp{T},\qquad
\Delta:={\boldsymbol{x}}^T H \boldsymbol{x}.$$ We know that
$\mathcal{A}$ is nonsingular and consequently $\boldsymbol{x}\ne0$
and $\Delta\ne0$. Now
$H\boldsymbol{x}=\mathcal{A}\sp{T\,-1}(\Delta,0,0,0)\sp{T}$ which
gives
\begin{equation}\label{invcalA}
    (H\boldsymbol{x})_{\mu}=\mathcal{A}\sp{-1}_{1\mu}\Delta.
\end{equation}
Now from (\ref{calba}) we see that
$$\mathcal{B}\mathcal{A}^{-1}=\rho\sp2H+(\rho-\rho\sp2)H
(\boldsymbol{x},0,0,0)\mathcal{A}^{-1}.$$ From (\ref{invcalA}) we
obtain
$$H(\boldsymbol{x},0,0,0)\mathcal{A}^{-1}=\frac{1}{\Delta}H
\boldsymbol{x}\boldsymbol{x}\sp{T}H$$ and therefore
$$\mathcal{B}\mathcal{A}^{-1}=\rho\sp2H+\frac{(\rho-\rho\sp2)}{\Delta}H
\boldsymbol{x}\boldsymbol{x}\sp{T}H.$$ Finally one sees that
$$\left[\rho\sp2H+\frac{(\rho-\rho\sp2)}{\Delta}H
\boldsymbol{x}\boldsymbol{x}\sp{T}H\right]\left[ \rho
H-\frac{(\rho-\rho\sp2)}{\Delta}
\boldsymbol{x}\boldsymbol{x}\sp{T}\right]=1,$$ whence the result
(\ref{taumat}) follows for
$\tau_{\mathfrak{b}}=\mathcal{A}\mathcal{B}^{-1}$. The remaining
constraint arises by requiring $\Im\tau$ to be positive definite.
We note that (\ref{condition1}) ensures that both
$\boldsymbol{x}\ne0$ and $\Delta\ne0$.
\end{proof}

The branch points can be expressed in terms of $\theta$-constants.
Following Matsumoto \cite{matsu00} we introduce the set of
characteristics
\begin{equation}  (\boldsymbol{a},\boldsymbol{b}),\qquad \boldsymbol{b}
=-\boldsymbol{a} H, \quad a_i\in \left\{ \frac16,\frac36, \frac
56\right\} \label{characteristics} \end{equation} and denote
$\theta_{\boldsymbol{a},-H\boldsymbol{b}}(\tau)
=\theta\{6\boldsymbol{a }\}(\tau)$ (see Appendix A for out theta
function conventions). The characteristics (\ref{characteristics})
are classified in \cite{matsu00} by the representations of the braid
group. Further, the period matrix determines the branch points as
follows.
\begin{proposition} [Diez 1991, Matsumoto 2000] \label{matsumoto2}
Let $\tau_{\mathfrak{b}}$ be the period matrix of
(\ref{curvegenb})
given in Proposition  \ref{matsumoto1}. Then
\begin{equation}
\Lambda_1=\left(\frac{\theta\{3,3,3,5\}} {\theta\{1,1,3,3\}
}\right)^3,\ \ \Lambda_2=-\left(\frac{\theta\{1,5,3,3\}}
{\theta\{1,1,5,5\} }\right)^3,\ \
 \Lambda_3
=-\left(\frac{\theta\{1,1,3,3\}} {\theta\{5,1,1,1\} }\right)^3.
\label{lambdas}
\end{equation}
\end{proposition}

These results have the following significance for our construction
of monopoles. First we observe that the period matrix is invariant
under $\mathbf{x}\rightarrow \lambda\mathbf{x}$. Thus to our
surface we may associate a point $[x_1:x_2:x_3:x_4]\in
\mathbb{B}\sp3=\{\mathbf{x}\in\mathbb{P}\sp3\,|\, \boldsymbol{x}^T
H \boldsymbol{x}<0\}\subset \mathbb{P}\sp3$ and from this point we
may obtain the normalized curve (\ref{curvegenb}). It is known
that a dense open subset of $\mathbb{B}\sp3$ arises in this way
from curves with distinct roots with the complement corresponding
to curves with multiple roots. Correspondingly, if we choose a
point $[x_1:x_2:x_3:x_4]\in \mathbb{B}\sp3$ we may construct a
period matrix and corresponding normalized curve.

We note that with $d\boldsymbol{u}=(du_1,\ldots,du_4)$ then
$$\tau\sp*(d\boldsymbol{u})=\overline{d\boldsymbol{u}}\cdot T,
\qquad T=\begin{pmatrix}
  -\kappa & 0& 0 & 0 \\
  0 & 0 & 0 & \kappa\sp2 \\
  0 & 0 & -\kappa\sp2 & 0 \\
  0 & \kappa\sp2 & 0 & 0
\end{pmatrix},\qquad
\kappa=\frac{\hat\chi\sp{\frac{1}{3}}}{\overline{\hat\chi}\sp{\frac{1}{3}}},
$$
and so we obtain
$$\oint_{\tau_*\mathfrak{c}}d\boldsymbol{u}=\oint_{\mathfrak{c}}
\tau\sp*(d\boldsymbol{u})=\oint_{\mathfrak{c}}\overline{d\boldsymbol{u}}\cdot
T=\overline{\left(
\oint_{\mathfrak{c}}{d\boldsymbol{u}}\right)}\cdot
T=\overline{\left(6\hat \chi\sp{\frac13}\left(
  1 \ 0\ 0 \ 0\right)\right)}\cdot
T=-6 \hat\chi\sp{\frac13}\left(
  1 \ 0\ 0 \ 0\right)=\oint_{-\mathfrak{c}}d\boldsymbol{u}$$
  and as a consequence corollary (\ref{HMRinvc}) of Houghton, Manton and
  Rom\~ao, $\tau_*\mathfrak{c}=-\mathfrak{c}$. More generally, let
  us write for an arbitrary cycle
  $$\gamma=\mathbf{p}\cdot \mathfrak{a}+\mathbf{q}\cdot \mathfrak{b}=
  \left(%
\begin{array}{cc}
  \mathbf{p} & \mathbf{q} \\
\end{array}%
\right)\left(%
\begin{array}{c}
  \mathfrak{a} \\
  \mathfrak{b} \\
\end{array}%
\right), \qquad \tau(\gamma)=  \left(%
\begin{array}{cc}
  \mathbf{p} & \mathbf{q} \\
\end{array}%
\right)\mathcal{M}\left(%
\begin{array}{c}
  \mathfrak{a} \\
  \mathfrak{b} \\
\end{array}%
\right).
$$
Then the equality
$$\oint_{\tau(\mathfrak{\gamma})}d\boldsymbol{u}=\oint_{\mathfrak{\gamma}}
\tau\sp*(d\boldsymbol{u})
=\oint_{\mathfrak{\gamma}}\overline{d\boldsymbol{u}}\cdot
T=\overline{\left(
\oint_{\mathfrak{\gamma}}{d\boldsymbol{u}}\right)}\cdot T$$ leads
to the equation
$$
 \left(%
\begin{array}{cc}
  \mathbf{p} & \mathbf{q} \\
\end{array}%
\right)\mathcal{M}\left(%
\begin{array}{c}
  \mathcal{A} \\
  \mathcal{B} \\
\end{array}%
\right) =\left(%
\begin{array}{cc}
  \mathbf{p} & \mathbf{q} \\
\end{array}%
\right)\overline{
\left(%
\begin{array}{c}
  \mathcal{A} \\
  \mathcal{B} \\
\end{array}%
\right)}\cdot T.
$$
We have then that the matrix $\mathcal{M}$ representing the
involution $\tau$ on homology and Ercolani-Sinha vector satisfy
\begin{equation}
\label{cplxMT} \mathcal{M}\sp2=\rm{Id},\quad
\mathcal{M}\left(%
\begin{array}{c}
  \mathcal{A} \\
  \mathcal{B} \\
\end{array}%
\right) =\overline{
\left(%
\begin{array}{c}
  \mathcal{A} \\
  \mathcal{B} \\
\end{array}%
\right)}\cdot T, \quad \mathbf{U}\mathcal{M}=
\left(%
\begin{array}{cc}
  \mathbf{n} & \mathbf{m} \\
\end{array}%
\right)\mathcal{M}=-\left(%
\begin{array}{cc}
  \mathbf{n} & \mathbf{m} \\
\end{array}%
\right).
\end{equation}
A calculation employing the algorithm of Tretkoff and Tretkoff
\cite{tret84} to describe the homology basis generators and
relations, together with some analytic continuation of the paths
associated to our chosen homology cycles (with the sheet
conventions described later in the text), yields that for our
curve
$$\mathcal{M}=\left( \begin {array}{cccccccc} 0&0&-1&0&0&-1&0&-1
\\\noalign{\medskip}0&0&1&1&-1&0&1&0\\\noalign{\medskip}-1&1&0&-1&0&1&0
&1\\\noalign{\medskip}0&-1&1&2&-1&0&1&0\\\noalign{\medskip}0&-1&1&1&0&0
&1&0\\\noalign{\medskip}-1&0&0&-1&0&0&-1&1\\\noalign{\medskip}1&0&0&0&
1&-1&0&-1\\\noalign{\medskip}1&-1&0&2&0&-1&1&-2\end {array}
\right).$$ The matrix $\mathcal{M}$ is not symplectic but
satisfies
$$\mathcal{M}J\mathcal{M}\sp{T}=-J,$$
where $J$ is the standard symplectic form. (The minus sign appears
here because of the reversal of orientation under the
antiholomorphic involution.)

\subsection{The vector $\widetilde{\boldsymbol{K}} $
and $\int\limits_{\infty_i}\sp{\infty_j}
\boldsymbol{v}$} We shall now describe the vector
$\widetilde{\boldsymbol{K}}$ and various related results,
including the quantity $\boldsymbol{\phi}(\infty_i)
-\boldsymbol{\phi}({\infty_j})$.

First let us record some elementary facts about our curve. For
ease in defining various divisors of the curve (\ref{curvegena})
let $\infty_{1,2,3}$ be the three points over infinity  and
$Q_i=(\lambda_i,0)$ ($i=1,\ldots 6$) be the branch points. Then
$$
\begin{array}{lll}
\div(z-\lambda_i)=\dfrac{Q_i\sp3}{\infty_1 \infty_2 \infty_3},&
\div(w)=\dfrac{\prod_{i=1}\sp6 Q_i}{(\infty_1 \infty_2
\infty_3)^2},&
\div(dz)=\dfrac{(\prod_{i=1}\sp6 Q_i)^2}{(\infty_1 \infty_2 \infty_3)^2},\\ \\
\div\left(\dfrac{dz}{w}\right)=\prod_{i=1}\sp6 Q_i,&
\div\left(\dfrac{dz}{w^2}\right)= (\infty_1 \infty_2 \infty_3)^2,&
\div\left(\dfrac{(z-\lambda_i)dz}{w^2}\right)= Q_i^3 \infty_1 \infty_2 \infty_3,\\
\\ \div\left(\dfrac{(z-\lambda_i)^2dz}{w^2}\right)=Q_i^6.
\end{array}
$$
Consideration of the function $(z-\lambda_i)/(z-\lambda_j)$ shows
that $3\int_{Q_j}\sp{Q_i}\boldsymbol{v}\in \Lambda$. The order of
vanishing of the differentials $d(z-\lambda_i)/w^2$,
$d(z-\lambda_i)/w$, $(z-\lambda_i)d(z-\lambda_i)/w^2$ and
$(z-\lambda_i)d(z-\lambda_i)/w^2$ at the point $Q_i$ are found to
be $0$, $1$, $3$ and $6$ respectively, which means that the gap
sequence at $Q_i$ is $1$, $2$, $4$ and $7$. From this we deduce
that the index of speciality of the divisor $Q_i^3$ is
$i(Q_i^3)=2$. Because the genus four curve $\mathcal{C}$ has the
function $w$ of degree $3$ then $\mathcal{C}$ is not
hyperelliptic. The function $1/(z-\lambda_i)$ has divisor
$\mathcal{U}/D$, with $\mathcal{U}=\infty_1 \infty_2 \infty_3$ and
$D=Q_i^3$ such that $D^2$ is canonical. This means that any other
function of degree $3$ on $\mathcal{C}$ is a fractional linear
transformation of $w$ and that $\Theta_{{\rm singular}}$ consists
of precisely one point which is of order $2$ in
$\Jac(\mathcal{C})$ \cite[III.8.7, VII.1.6]{fk80}. The vector of
Riemann constants $\boldsymbol{K}_{Q_i}$ is a point of order $2$
in $\Jac(\mathcal{C})$ because $Q_i\sp6$ is canonical
\cite[VI.3.6]{fk80}. Let us fix $Q_1$ to be our base point. Then
as $\boldsymbol{K}_{Q_1}=\boldsymbol{\phi}_{Q_1}(Q_1^3)+\boldsymbol{K}_{Q_1}$
we have that $\boldsymbol{K}_{Q_1}\in\Theta$. Because $i(Q_1^3)=2$
we may identify $\boldsymbol{K}_{Q_1}$ as the unique point in
$\Theta_{{\rm singular}}$. We may further identify
$\boldsymbol{K}_{Q_1}$ as the unique even theta characteristic
belonging to $\Theta$.

With $Q_1$ as our base point $\boldsymbol{\phi}\left(\sum_k \infty_k\right)$
corresponds to the image under the Abel map of the divisor of the
function $1/(z-\lambda_1)$, and so vanishes (modulo the period
lattice). Thus for our curve $\widetilde{\boldsymbol{K}}=
\boldsymbol{K}_{Q_1}+\boldsymbol{\phi}\left(\sum_k
\infty_k\right)=\boldsymbol{K}_{Q_1}=\Theta_{{\rm singular}}$ is
the unique even theta characteristic. The point
$\boldsymbol{K}_{Q_1}$ may be constructed several ways: directly,
using the formula (\ref{vecR}) of the Appendix (the evaluation of
the integrals of normalised holomorphic differentials between
branch points is described in Appendix B); by enumeration we may
find which of the $136$ even theta characteristics
$\left[\begin{matrix}\epsilon
\\ \epsilon'\end{matrix}\right]$ leads to the vanishing of $\theta
\left[\begin{matrix}\epsilon
\\ \epsilon'\end{matrix}\right](z;\tau)$; using a monodromy argument of Matsumoto
\cite{matsu00}. One finds that the relevant half period is
$\dfrac12\left[\begin{matrix}1&1&1&1\\1&1&1&1\end{matrix}\right]$.

The analysis of the previous paragraph, together with Lemma
\ref{ueventheta}, tells us that $\boldsymbol{U}$ must also be an
even theta characteristic.

Again using that $\sum_k \infty_k\sim_l 0$ we have that
$\infty_i-\infty_j\sim_l 2\infty_i+\infty_k$ (with $i$, $j$, $k$
distinct) and so
$\theta(\boldsymbol{\phi}(\infty_j)-\boldsymbol{\phi}(\infty_i)-\widetilde{\boldsymbol{K}})=
\theta(\boldsymbol{\phi}(2\infty_i+\infty_k)+ \boldsymbol{K})=0$. One sees from
the above divisors (in particular $\div(dz/w^2)$) that $\dim
H\sp0( \mathcal{C}, L_{2\infty_i+\infty_k})=
i(2\infty_i+\infty_k)=1$. Thus
$\theta(w+\boldsymbol{\phi}(\infty_j)-\boldsymbol{\phi}(\infty_i)-\widetilde{\boldsymbol{K}}$
and $\theta(w-\widetilde{\boldsymbol{K}})$ have order of vanishing
differing by one for (generic) $w\rightarrow0$.

\subsection{Calculating $\nu_i-\nu_j$}
From the results of the previous section we see that
$$\div\left(\dfrac{z^4dz}{w^2}\right)=\frac{(0_10_20_3)\sp4}{(\infty_1\infty_2\infty_3)\sp2}.
$$
This has precisely the same divisor of poles as $\gamma_\infty$
and we will use this to represent $\gamma_\infty$. It is
convenient to introduce the (meromorphic) differential
$$dr_1(P)=\frac{z^4 dz}{3w^2},$$
the factor of three here being introduced to give the pairing
\[ \sum_{s=1}^3  \mathrm{Res}_{P=\infty_s}d{r}_1(P)
  \int_{P_0}^P d{u}_1(P')
=1.
\]
We may therefore write \begin{equation}\label{gpdr}
\gamma_{\infty}(P)=-3 dr_1(P)+\sum_{i=1}^4 c_i {v}_i(P).
\end{equation} The constants $c_i$ are found from the condition of
normalisation
\[ \oint_{\mathfrak{a}_k } \gamma_{\infty}(P)=0\quad \Longleftrightarrow
c_k=3\oint_{\mathfrak{a}_k} dr_1(P)\equiv 3y_k,\quad k=1,\ldots,4,
\] where we have defined the vector of
$\mathfrak{a}$-periods
$\boldsymbol{y}\sp{T}=\left(\oint_{\mathfrak{a}_1} dr_1(P),\ldots,
\oint_{\mathfrak{a}_4} dr_1(P)     \right)$. The vector of
$\mathfrak{b}$-periods of $dr_1$ is found to be $\rho^2 H
\boldsymbol{y}$. The pairing with $du_1$ then yields the Legendre
relation
\begin{equation}\label{legndre}
\boldsymbol{y}{\cdot } H\boldsymbol{x}=-\frac{2\pi}{\sqrt{3}}.
\end{equation}
Now the $\mathfrak{b}$-periods of the differential
$\gamma_{\infty}$ give the Ercolani-Sinha vector. Using
(\ref{gpdr}) we then obtain the equality
\begin{equation}
-3(\rho^2 H-\tau_{\mathfrak{a}})\boldsymbol{y}=\pi\imath
\boldsymbol{n}+\pi\imath \tau \boldsymbol{m}.\label{relation}
\end{equation}

Finally, using (\ref{gpdr}), we may write
\begin{equation}\label{nuijgpgr}
\nu_i-\nu_j=3\boldsymbol{y}\cdot\int_{\infty_j}\sp{\infty_i}
\boldsymbol{v}+\int_{\infty_j}\sp{\infty_i}\left[
d\left(\frac{w}{z}\right)-3dr_1\right].
\end{equation}

\section{Solving the Ercolani-Sinha constraints}
Although we have now described how to evaluate most of the
quantities needed for the reconstruction of a monopole, at least
up to the evaluation of the integrals $\boldsymbol{x}$, we have
not described how to solve the Ercolani-Sinha constraints for the
spectral curve (\ref{cubic}) which encode one of Hitchin's
transcendental constraints. To this we now turn. As we shall see,
this reduces to constraints just on the four periods
$\boldsymbol{x}$. Later we shall restrict attention to the curves
(\ref{bren03}), which has the effect of reducing the number of
integrals to be evaluated to two and consequently simplifies our
present analysis.

We shall work with the Ercolani-Sinha constraints in the form
(\ref{HMREScond}). Let the holomorphic differentials be ordered as
in (\ref{diffbasis}). Then there exist two integer 4-vectors
$\boldsymbol{n}, \boldsymbol{m}\in \mathbb{Z}^4$ and values of the
parameters $\lambda_1,\ldots,\lambda_6 $ and $\chi$ such that
\begin{equation}\boldsymbol{n}^T \mathcal{A}+\boldsymbol{m}^T\mathcal{B}
=\nu(1,0,0,0).\label{esconda}
 \end{equation}
Here $\nu$ depends on normalizations.  For us this will be
\begin{equation*}
 \nu = 
 6 \hat\chi ^{\frac{1}{3}} .
\end{equation*}
To see this observe that (\ref{HMREScond}) requires that
$$-2\delta_{1k} = \oint_{\boldsymbol{n}\cdot\mathfrak{a}+
\boldsymbol{m}\cdot\mathfrak{b}} \Omega^{(k)}\ \textrm{ for the
differentials}\ \Omega^{(1)} = \frac{\eta^{n-2} d \zeta}
{\frac{\partial P}{\partial \eta}} = \frac{d \zeta}{n \eta},\\
\Omega^{(2)} = \frac{\eta^{n-3} d \zeta} {\frac{\partial P}
{\partial \eta}} ,\dots.
$$ In the parameterisation (\ref{curvegena}) we are using we have that
\begin{equation*}
    x_{i} = {\oint_{\mathfrak{a}_i}}\frac{dz}{w} =
   {\oint_{\mathfrak{a}_i}} \frac{d\zeta}{-\hat\chi^{-\frac{1}{3}}\eta} =
    -3\hat\chi^{\frac{1}{3}}{\oint_{\mathfrak{a}_i}} \Omega^{(1)}.
\end{equation*}
We wish
\begin{align}
    -2 = \oint_{\boldsymbol{n}\cdot\mathfrak{a}+
\boldsymbol{m}\cdot\mathfrak{b}} \Omega^{(1)} & =
    \frac{-1}{3 \hat\chi^{\frac{1}{3}}}(\boldsymbol{n}.\boldsymbol{ x} + \rho
    \boldsymbol{m}.H.\boldsymbol{x}) \nonumber \intertext{and so}
    \boldsymbol{n}.\boldsymbol{ x} + \rho
    \boldsymbol{m}.H.\boldsymbol{x}& = \nu,\label{esonx}
    \end{align}
with the value of $\nu$ stated. Consideration of the other
differentials then yields (\ref{esconda}), transcendental
constraints on the curve $\mathcal{C}$. These constraints may be
solved using the following result.
\begin{proposition}\label{ourh2}
The Ercolani-Sinha constraints (\ref{esconda}) are satisfied for
the curve (\ref{cubic}) if and only if
\begin{align}\label{expxx}
    \boldsymbol{x} & = \xi(H\boldsymbol{n} + \rho^{2}\boldsymbol{m}),
    \intertext{where }\xi &   = \frac{\nu}{[\boldsymbol{n}.H
    \boldsymbol{n}-\boldsymbol{m}.\boldsymbol{n}+\boldsymbol{m}.H
    \boldsymbol{m}]} = \frac{6
\hat\chi^{\frac{1}{3}}}{[\boldsymbol{n}.H
    \boldsymbol{n}-\boldsymbol{m}.\boldsymbol{n}+\boldsymbol{m}.H
    \boldsymbol{m}]}.\label{expxxx}
\end{align}
\end{proposition}

\begin{proof}
Rewriting (\ref{esconda}) we have that
\begin{align*}
    \boldsymbol{n}^{T} + \boldsymbol{m}^{T}\mathcal{B}\mathcal{A}^{-1}
    & = \nu (1,0,0,0)\mathcal{A}^{-1} = \nu
    \mathcal{A}^{-1}_{1\mu}. \intertext{Upon using (\ref{invcalA})
    we obtain}
    \boldsymbol{n}^{T} + \boldsymbol{m}^{T}\mathcal{B}\mathcal{A}^{-1}
    & = \frac{\nu}{\Delta}
    \boldsymbol{x}^{T}H.
    \end{align*}
  Therefore \begin{align*}  \boldsymbol{x}& = \frac{\Delta}{\nu}
  (H\boldsymbol{n} + H(\mathcal{B}\mathcal{A}^{-1})^{T}\boldsymbol{m}) \\
    & = \frac{\Delta}{\nu}(H\boldsymbol{n} + \rho^{2}\boldsymbol{m}
     + (\frac{\rho -
    \rho^{2}}{\Delta})\boldsymbol{x}\,\boldsymbol{x}^{T}H\boldsymbol{m})\end{align*}
  upon using that the period matrix is symmetric and our earlier expression
  for $\mathcal{B}\mathcal{A}^{-1}$. Rearranging now gives us that
  \begin{equation}\label{expx}
    (1 + \frac{\rho^{2}-\rho}{\nu}\boldsymbol{x}^{T}H\boldsymbol{m})
    \boldsymbol{x}  =
    \frac{\Delta}{\nu}(H\boldsymbol{n} + \rho^{2}\boldsymbol{m})
\end{equation}
and so we have established (\ref{expxx}) where
\begin{align}\label{expxxxx}
    \xi & = \frac{\Delta}{\nu}(1 +
    \frac{\rho^{2}-\rho}{\nu}\,\boldsymbol{x}.H\boldsymbol{m})^{-1}.
\end{align}
There are several constraints. First, the Ercolani-Sinha condition
(\ref{esonx}) is that
$$
    [\boldsymbol{n}^{T} + \rho \boldsymbol{m}^{T}H]
    \xi[H\boldsymbol{n} + \rho^{2}\boldsymbol{m}] = \nu $$
    and consequently
\begin{equation}\label{esnu}
[\boldsymbol{n}.H\boldsymbol{n} - \boldsymbol{m}.\boldsymbol{n} +
\boldsymbol{m}.H\boldsymbol{m}]\xi  = \nu =
6\hat\chi^{\frac{1}{3}},
\end{equation}
thus establishing (\ref{expxxx}). We remark that if $\hat\chi$ is
real, then $\hat\chi^{\frac{1}{3}}$ may be chosen real and hence
 $\xi$ is real. We observe that (\ref{expxxx}) and (\ref{expxxxx}) are consistent
with
\begin{align*}
    \Delta = \boldsymbol{x}^{T}H\boldsymbol{x} & =
     \xi^{2}(\boldsymbol{n}^{T}H + \rho^{2}\boldsymbol{m}\sp{T})H
     (H\boldsymbol{n} + \rho^{2}\boldsymbol{m})
     = \xi^{2}[\boldsymbol{n}.H\boldsymbol{n} + 2\rho^{2}
    \boldsymbol{m}.\boldsymbol{n} + \rho
    \boldsymbol{m}.H\boldsymbol{m}].
\end{align*}
A further consistency check is given by  (\ref{relation}). Using
the form of the period matrix, the Legendre relation
(\ref{legndre}) and the proposition (with $\nu=-6$) we obtain
(\ref{relation}).
\end{proof}

At this stage we have reduced the Ercolani-Sinha constraints to
one of imposing the four constraints (\ref{expxx}) on the periods
$x_k$. In particular this means we must solve
\begin{equation}
\frac{x_1}{n_1+\rho^2m_1} =\frac{x_2}{n_2+\rho^2m_2}
=\frac{x_3}{n_3+\rho^2m_3}
=\frac{x_4}{-n_4+\rho^2m_4}=\xi,\label{nmconditions1}
\end{equation}
which means $x_i/x_j\in\mathbb{Q}[\rho]$. Further we have from the
conditions (\ref{condition1}) that
\begin{align}
\frac{\bar{\boldsymbol{x}}\sp{T} H
\boldsymbol{x}}{|\xi|\sp2}&=[\boldsymbol{n}.H\boldsymbol{n} -
\boldsymbol{m}.\boldsymbol{n} +
\boldsymbol{m}.H\boldsymbol{m}]
= \sum_{i=1}^3 (n_i^2-n_im_i+m_i^2)-n_4^2-m_4^2-m_4n_4<0.
\label{nmconditions}
\end{align}

Our result admits another interpretation. Thus far we have assumed
we have been given an appropriate curve and sought to satisfy the
Ercolani-Sinha constraints. Alternatively we may start with a
curve satisfying (most of) the Ercolani-Sinha constraints and seek
one satisfying the reality constraints (and any remaining
Ercolani-Sinha constraints). How does this progress? First note
that the period matrix (\ref{taumat}) for a curve satisfying
(\ref{nmconditions1}) is independent of $\xi$: it is determined
wholly in terms of the Ercolani-Sinha vector. Let us then start
with a primitive vector $\mathbf{U}=(\mathbf{n},\mathbf{m})$
satisfying the hyperboloid condition (\ref{nmconditions}) and
lemma \ref{ueventheta}. From this we construct a period matrix and
then, via Proposition \ref{matsumoto2}, a normalized curve
(\ref{curvegenb}). Now we must address whether the curve has the
correct reality properties. For this we must show that there
exists a M\"obius transformation of the set $S=\{0,1,\infty,
\Lambda_1,\Lambda_2,\Lambda_3\}$ to one of the form
$H=\{\alpha_j,-{1}/{{\overline\alpha}_j}\}_{j=1}\sp{3}$. We will
show below that this question may be answered, with the roots
$\alpha_i$ being determined up to an overall rotation. At this
stage we have (using the rotational freedom) a curve of the form
$$W^3=Z(Z-a)(Z+\frac1{a})(Z-w)(Z+\frac{1}{\overline{w}}),\qquad
a\in\mathbb{R},\ w\in\mathbb{C}.$$ To reconstruct a monopole curve
we need a normalization $\hat\chi=\chi_3 \left[
\frac{\overline{w}}{w}\right]\sp{1/2}$. This is encoded in $\xi$,
which has not appeared thus far. To calculate the normalization we
must calculate a period. Then using (\ref{nmconditions1}) and
(\ref{expxxx}) we determine $\hat\chi$. This is a constraint. For
a consistent monopole curve we require $$\arg(\xi)=\arg\left[
\frac{\overline{w}}{w}\right]\sp{1/6}.$$ Of course, to complete
the construction we need to check there are no roots of the theta
function in $[-1,1]$. Although the procedure outlined involves
several transcendental calculations it is numerically feasible and
gives a means of constructing putative monopole curves.

To conclude we state when there exists a M\"obius transformation
of the set $S=\{0,1,  \infty,   \Lambda_1,  \Lambda_2,  \Lambda_3\}$ to
one of the form
$H=\{\alpha_j,-{1}/{{\overline\alpha}_j}\}_{j=1}\sp{3}$. For
simplicity we give the case of distinct roots:
\begin{theorem}The roots $S=\{0,1,\infty,
\Lambda_1,\Lambda_2,\Lambda_3\}$ are M\"obius equivalent to
$H=\{\alpha_j,-{1}/{{\overline\alpha}_j}\}_{j=1}\sp{3}$ if and
only if \begin{enumerate} \item If just one of the roots, say
$\Lambda_1$, is real and
\begin{itemize}
\item $\Lambda_1<0$ then
$\Lambda_2\overline{\Lambda}_3=\Lambda_1$, \item $0<\Lambda_1<1$
then
$\frac{\Lambda_2}{\Lambda_2-1}\overline{\frac{\Lambda_3}{\Lambda_3-1}}=\frac{\Lambda_1}{\Lambda_1-1}$,
\item $1<\Lambda_1$ then
$(1-\Lambda_2)\overline{(1-\Lambda_3)}=1-\Lambda_1$.
\end{itemize}
If all three roots are real then, up to relabelling, one of the
above must hold.

\item All three roots are complex and, up to relabelling,
$$0<\Lambda_1\overline{\Lambda}_2\in\mathbb{R},\qquad
1<\frac{\Lambda_1}{\Lambda_2},\qquad
\Lambda_3=\Lambda_2\,\frac{1-\overline{\Lambda}_1}{1-\Lambda_2}.$$
\end{enumerate}
\end{theorem}

\section{Symmetric 3-monopoles}
In this section we shall consider the curve $\mathcal{C}$
specialised to the form
\begin{equation}
\eta^3+\chi(\zeta^6+b\zeta^3-1)=0,\label{curve}
\end{equation}
where $b$ is a real parameter. In this case branch points are
\[ (\lambda_1,\lambda_2,\lambda_3,\lambda_4,\lambda_5,\lambda_6)
 =(\alpha,\;\rho^2 \beta,\; \rho \alpha,\; \beta,\; \rho^2 \alpha,\;
\rho \beta),   \] where $\alpha$ and $\beta$ are real,
\[ \alpha=\sqrt[3]{\frac{-b+\sqrt{b^2+4}}{2}}>0,\quad
\beta=\sqrt[3]{\frac{-b-\sqrt{b^2+4}}{2}}<0,\quad
\alpha^3\beta^3=-1. \] Here $\chi=\hat\chi$ is real and we choose
our branches so that $\hat\chi^{\frac{1}{3}}$ is also real.

The effect of choosing such a symmetric curve will be to reduce
the four period integrals $x_i$ to two independent integrals. The
tetrahedrally symmetric monopole is in the class (\ref{curve}). We
note that a general rotation will alter the form of $a_3(\zeta)$.
Thus the dimension of the moduli space is reduced from three by
the 3 degrees of freedom of the rotations yielding a discrete
space of solutions. We are seeking then a discrete family of
spectral curves.

We shall begin by calculating the period integrals, and then
imposing the Ercolani-Sinha constraints. We shall also consider
the geometry of the curves (\ref{curve}).

\subsection{The period integrals}
In terms of our Wellstein parameterisation we are working with
$$
w^3=z^6+bz^3-1=(z^3-\alpha\sp3)(z^3+\frac{1}{\alpha\sp3})
$$
($1/\alpha\sp3=-\beta\sp3=(b+\sqrt{b^2+4})/{2}$). We choose the
first sheet so that
$w=\sqrt[3]{(z^3-\alpha\sp3)(z^3+{1}/{\alpha\sp3})}$ is negative
and real on the real $z$-axis between the branch points
$(-1/\alpha,\alpha)$.

Introduce integrals computed on the first sheet
\begin{align}\begin{split}
\mathcal{I}_1(\alpha)&=\int\limits_{0}^{\alpha}\frac{{d}z}{w}
=-\frac{2\pi\sqrt{3}\alpha}{9} {_2F_1}\left(\frac13,\frac13;1;-\alpha^6\right),\\
\mathcal{J}_1(\alpha)&=\int\limits_{0}^{\beta}\frac{{d}z}{w} =
\frac{2\pi\sqrt{3}}{9\alpha} {_2F_1}\left(\frac13,\frac13;1;
-\alpha^{-6}\right).\end{split} \label{integralsij}\end{align}
Here $_2F_1(a,b;c;z)$ is the standard Gauss hypergeometric
function and we have, for example, evaluated the first integral
using the substitution $z=\alpha t\sp{1/3}$ and our specification
of the first sheet. We also have that
\[ \int\limits_{0}^{\rho^k\alpha}\frac{{d}z}{w}=\rho^k\mathcal{I}_1(\alpha),
\quad
\int\limits_{0}^{\rho^k\beta}\frac{{d}z}{w}=\rho^k\mathcal{J}_1(\alpha),\quad
k=1,2. \]

Our aim is to express the periods for our homology basis
(\ref{homology}) in terms of the integrals $\mathcal{I}_1(\alpha)$
and $\mathcal{J}_1(\alpha)$. Consider for example
\begin{align*}
x_1&=\oint_{\mathfrak{a}_1}{d}u_1=
\int_{\gamma_1(\lambda_1,\lambda_2)}\frac{{d}z}{w}+
\int_{\gamma_2(\lambda_2,\lambda_1)}\frac{{d}z}{w}
=\int_{\lambda_1}\sp{\lambda_2}\frac{{d}z}{w}- \rho\sp2\,
\int_{\lambda_1}\sp{\lambda_2}\frac{{d}z}{w}\\
&=(1-\rho\sp2) \int_{\alpha}\sp{\rho\sp2\beta}\frac{{d}z}{w}
=(1-\rho\sp2)\left[-\mathcal{I}(\alpha)+
\int\limits_{0}^{\rho^2\beta}\frac{{d}z}{w}\right]
=(1-\rho\sp2)\left[-\mathcal{I}_1(\alpha)+
\rho^2\mathcal{J}_1(\alpha)\right]\\
&=-2\mathcal{I}_1(\alpha)-\mathcal{J}_1(\alpha)-\rho\left[
\mathcal{I}_1(\alpha)+2\mathcal{J}_1(\alpha)\right].
\end{align*}
Here we have used that on the second sheet $w_2=\rho w_1$ to
obtain the last expression of the first line, and also that
$1+\rho+\rho\sp2=0$ to obtain the final expression. Similarly we
find (upon dropping the $\alpha$ dependence from $\mathcal{I}_1$
and $\mathcal{J}_1$ when no confusion arises) that
\begin{equation}
\begin{array}{rlrl}
x_{1}&=-(2\mathcal{J}_1+\mathcal{I}_1)\rho -2\mathcal{I}_1
-\mathcal{J}_1, &x_{2}&=(\mathcal{J}_1- \mathcal{I}_1 )\rho+
\mathcal{I}_1+2\mathcal{J}_1,
\\
x_{3}&=(\mathcal{J}_1+2\mathcal{I}_1)\rho-\mathcal{J}_1+\mathcal{I}_1,
&x_{4}&=3(\mathcal{J}_1-\mathcal{I}_1)\rho+3\mathcal{J}_1.
\end{array}
\label{ij}
\end{equation}
 Note that
\begin{equation} x_2=\rho x_1,\quad x_3=\rho^2 x_1 .\label{ij1} \end{equation}

\subsection{The Ercolani-Sinha constraints}
We next reduce the Ercolani-Sinha constraints to a number
theoretic one. Using (\ref{expxx}) and (\ref{ij}) we may rewrite
the constraints as
\begin{equation}
    x_{i} = \xi( \epsilon_{i}n_{i} + \rho^{2}m_{i}) = (\alpha_{i}\mathcal{I}_1 +
    \beta_{i} \mathcal{J}_1) + (\gamma_{i}\mathcal{I}_1 + \delta_{i}
    \mathcal{J}_1)\rho .\label{esred}
\end{equation}
We may solve for the various $n_{i},m_{i}$ in terms of
$n_{1},m_{1}$ as follows. Set
$$C_{i}=\begin{pmatrix} \epsilon_{i} & -1 \\
    0 & -1 \end{pmatrix},\
D_{i}=\begin{pmatrix} \alpha_{i} & \beta_{i} \\
    \gamma_{i} &\delta_{i} \end{pmatrix},\
\widehat{\mathcal{I}} = {\mathcal{I}_1}/{\xi},\ \widehat{\mathcal{J}} =
{\mathcal{J}_1}/{\xi}.$$ Then (\ref{esred}) may be rewritten as
\begin{align*}
    C_{i}\, \begin{pmatrix} n_{i} \\ m_{i}
    \end{pmatrix}  = D_i \begin{pmatrix} \widehat{\mathcal{I}} \\
    \widehat{\mathcal{J}} \end{pmatrix}
\end{align*}
giving
\begin{align*}
\begin{pmatrix} n_{i} \\ m_{i}
    \end{pmatrix} & = C^{-1}_{i} D_{i} \begin{pmatrix}\widehat{\mathcal{I}}
    \\ \widehat{\mathcal{J}} \end{pmatrix} = C^{-1}_{i} D_{i} D^{-1}_{1} C_{1}
    \begin{pmatrix} n_{1} \\m_{1} \end{pmatrix}.
\end{align*}
This yields that the vectors $\boldsymbol{n}$, $\boldsymbol{m}$
are of the form
\begin{equation}\label{solvn}\boldsymbol{n}=
\begin{pmatrix} n_{1} \\ n_{2} \\ n_{3} \\n_{4}\end{pmatrix} =
\begin{pmatrix} n_{1} \\ m_{1}-n_{1} \\ -m_{1}
\\2n_{1}-m_{1}\end{pmatrix}, \qquad
\boldsymbol{m}=\begin{pmatrix} m_{1} \\ m_{2} \\ m_{3}
\\m_{4}\end{pmatrix} =
\begin{pmatrix} m_{1} \\ -n_{1} \\ n_{1}-m_{1}
\\-3n_{1}\end{pmatrix}.
\end{equation}
One may verify that for vectors of this form then
$(\boldsymbol{n},\boldsymbol{m})\mathcal{M}=-(\boldsymbol{n},\boldsymbol{m})$
as required by (\ref{cplxMT}). Recall further that
$(\boldsymbol{n},\boldsymbol{m})$ is to be a primitive vector:
that is one for which the greatest common divisor of the
components is 1, and hence a generator of $\mathbb{Z}\sp8$. We see
that $(\boldsymbol{n},\boldsymbol{m})$ is primitive if and only if
\begin{equation}(n_1,m_1)=1.\label{primcond}\end{equation}

From
\begin{equation*}
    \begin{pmatrix} \widehat{\mathcal{I}} \\ \widehat{\mathcal{J}} \end{pmatrix} = D^{-1}_{i}
    C_{i} \begin{pmatrix}n_{i} \\m_{i}\end{pmatrix} =
    \frac{1}{3}\begin{pmatrix}-2 & 1 \\ 1 & 1 \end{pmatrix}
    \begin{pmatrix}n_{1} \\ m_{1}\end{pmatrix}
\end{equation*}
we obtain
\begin{equation*}
\begin{matrix}
    \dfrac{\widehat{\mathcal{I}}}{\widehat{\mathcal{J}}}
    & = &\dfrac{\mathcal{I}}{\mathcal{J}} = \dfrac{m_{1} - 2n_{1}}{m_{1} +
    n_{1}}, \\
    \mathcal{I}_1 & =& \dfrac{m_{1} - 2n_{1}}{3} \, \xi \quad & = &- \dfrac{2
    \pi}{3\sqrt{3}} \ \alpha\ {_2F_1}(\frac{1}{3}, \frac{1}{3}; 1,
    -\alpha^{6}), \\
    \mathcal{J}_1 & =& \dfrac{m_{1} + n_{1}}{3} \, \xi \quad & = &\dfrac{2
    \pi}{3\sqrt{3}} \ \dfrac{1}{\alpha}\ {_2F_1}(\frac{1}{3}, \frac{1}{3}; 1,
    -\alpha^{-6}).
\end{matrix}
\end{equation*}
Now given (\ref{solvn})  we find that
\begin{equation*}
    \boldsymbol{n}.H \boldsymbol{n} - \boldsymbol{m} .
\boldsymbol{n} + \boldsymbol{m}.H\boldsymbol{m} = 2(m_{1} +
n_{1})(m_{1} -
    2n_{1})
\end{equation*}
and so the constraint (\ref{condition1}) is satisfied if
$$
    \boldsymbol{\bar x}\sp{T}H\boldsymbol{x}  = \xi^{2}[
    \boldsymbol{n}.H \boldsymbol{n} - \boldsymbol{m} .
\boldsymbol{n} + \boldsymbol{m}.H\boldsymbol{m}
    ]= 2\xi^{2}(m_{1} +
n_{1})(m_{1} -
    2n_{1})< 0.
    $$
This requires
\begin{equation}\label{esccnm}
(m_{1} + n_{1})(m_{1} -
    2n_{1})< 0.
\end{equation}
In particular we have from (\ref{esnu}) that
$$\xi=\frac{3\chi^{\frac{1}{3}}}{(n_{1} + m_{1})(m_{1} -
2n_{1})}.$$

Thus we have to solve
\begin{equation}
\begin{split}
    \mathcal{I}_1 &= \frac{\chi^{\frac{1}{3}}}{n_{1} + m_{1}}
= - \frac{2 \pi}{3
    \sqrt{3}}\ \alpha\ {_2F_1}(\frac{1}{3}, \frac{1}{3}; 1,
    -\alpha^{6}), \\
    \mathcal{J}_1 &= \frac{\chi^{\frac{1}{3}}}{m_{1} - 2n_{1}}
= \frac{2 \pi}{3
    \sqrt{3}}\ \frac{1}{\alpha}\ {_2F_1}(\frac{1}{3}, \frac{1}{3}; 1,
    -\alpha^{-6}).
\end{split}\label{escj}
\end{equation}

Using the identity
\begin{equation*}
{_2F_1}(\frac{1}{3}, \frac{1}{3}; 1,x)=(1-x)\sp{-{1}/{3}}\,
{_2F_1}(\frac{1}{3}, \frac{2}{3}; 1,\frac{x}{x-1})
\end{equation*}
we then seek solutions of
\begin{equation*}
\dfrac{\mathcal{I}_1}{\mathcal{J}_1} = \dfrac{m_{1} -
2n_{1}}{m_{1} +    n_{1}}=- \frac{{_2F_1}(\frac{1}{3},
\frac{2}{3}; 1,t)}{{_2F_1}(\frac{1}{3}, \frac{2}{3};
1,1-t)},\qquad t=\frac{\alpha\sp6}{1+\alpha\sp6}=
\frac{-b+\sqrt{b^2+4}}{2\sqrt{b^2+4}}.
\end{equation*}
From (\ref{esccnm}) the ratio of ${\mathcal{I}_1}/{\mathcal{J}_1}$
is negative. Consideration of the function
\[f(t)=\frac{{_2F_1}\left(\frac13,\frac23;1;t\right)}
{ {_2F_1}\left(\frac13,\frac23;1;1-t\right)} .
\]
(see Figure 2 for its plot) shows that there exists unique root
$t\in(0,1)$ for each value $f(t)\in(0,\infty)$ and correspondingly
a unique real positive $\alpha=\sqrt[6]{t/(1-t)}$.
\begin{figure}\label{figratio}
\epsfxsize=12cm \epsfysize=10cm \epsffile{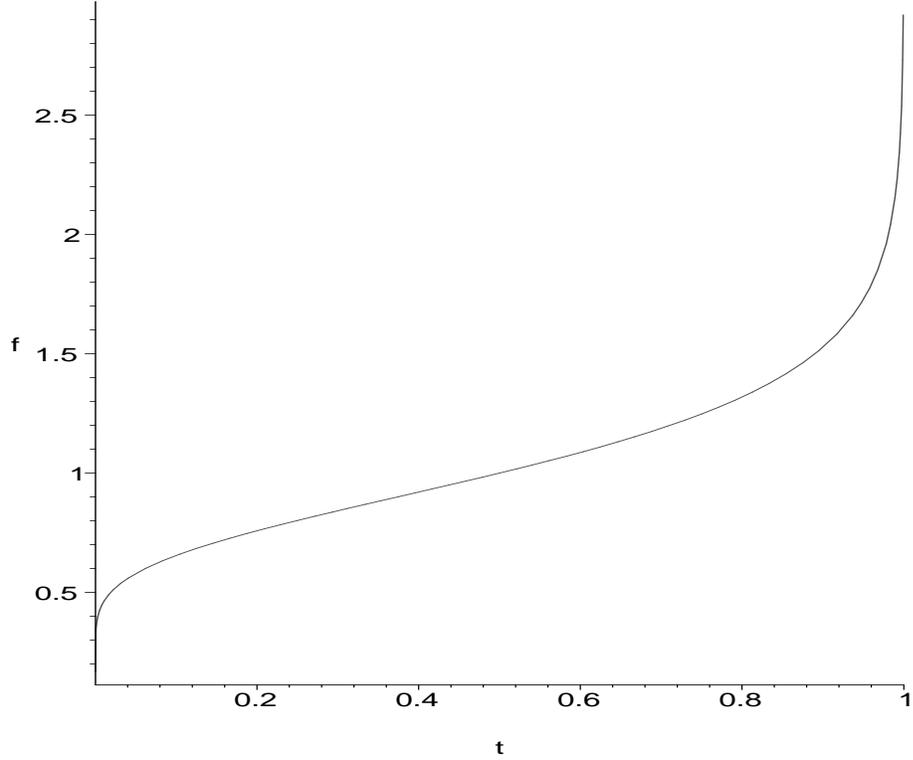} \caption{The
function $f(t)= \frac{{_2F_1}(\frac{1}{3}, \frac{2}{3};
1,t)}{{_2F_1}(\frac{1}{3}, \frac{2}{3}; 1,1-t)}$.}
\end{figure}

Bringing these results together we have established:
\begin{proposition}\label{propesnt}
To each pair of relatively prime integers
$(n_1,m_1)=1$ for which
$$(m_{1} + n_{1})(m_{1} -2n_{1})<0$$
we obtain a solution to the Ercolani-Sinha constraints for a curve
of the form (\ref{curve}) as follows. First we solve for $t$,
where
\begin{equation}\label{esct}
\dfrac{2n_{1}-m_{1}}{m_{1} + n_{1}}=\frac{{_2F_1}(\frac{1}{3},
\frac{2}{3}; 1,t)}{{_2F_1}(\frac{1}{3}, \frac{2}{3}; 1,1-t)}.
\end{equation}
Then
\begin{equation}
b=\frac{1-2t}{\sqrt{t(1-t)}},\qquad t=
\frac{-b+\sqrt{b^2+4}}{2\sqrt{b^2+4}},
\end{equation}
and we obtain $\chi$ from
\begin{equation}\label{esevch}
\chi^{\frac{1}{3}} = -(n_{1} + m_{1} )\, \frac{2 \pi}{3
    \sqrt{3}}\ \frac{\alpha}{(1+\alpha\sp6)\sp\frac13}\ {_2F_1}(\frac{1}{3}, \frac{2}{3}; 1,
    t)
\end{equation}
with $\alpha\sp6=t/(1-t)$.
\end{proposition}

\subsection{Ramanujan}
Thus far we have reduced the problem of finding an appropriate
monopole curve within the class (\ref{curve}) to that of solving
the transcendental equation (\ref{esct}) for which a unique
solution exists. Can this ever be solved apart from numerically?
Here we shall recount how a (recently proved) result of Ramanujan
enables us to find solutions.

Let $n$ be a natural number. A modular equation of degree $n$ and
signature $r$ ($r=2,3,4,6$) is a relation between $\alpha$,
$\beta$ of the form
\begin{equation}\label{modnr}
n\,\frac{_2F_1(\frac{1}{r},\frac{r-1}{r};1;1-\alpha)}
{_2F_1(\frac{1}{r},\frac{r-1}{r};1;\alpha)}=
\frac{_2F_1(\frac{1}{r},\frac{r-1}{r};1;1-\beta)}
{_2F_1(\frac{1}{r},\frac{r-1}{r};1;\beta)}. \end{equation} When
$r=2$ we have the complete elliptic integral
$\mathbf{K}(k)=\frac{\pi}{2}\, _2F_1(\frac12,\frac12;1;k\sp2)$ and
(\ref{modnr}) yields the usual modular relations. By interchanging
$\alpha\leftrightarrow\beta$ we may interchange $n\leftrightarrow
1/n$. This, together with iteration of these modular equations,
means we may obtain relations with $n$ being an arbitrary rational
number. Our equation (\ref{esct}) is precisely of this form for
signature $r=3$ and starting with say $\alpha=1/2$.

Ramanujan in his second notebook presents results pertaining to
these generalised modular equations and various theta function
identities. For example, if $n=2$ in signature $r=3$ then $\alpha$
and $\beta$ are related by
\begin{equation}\label{mod23}
    (\alpha\beta)\sp\frac{1}{3}+\left(
    (1-\alpha)(1-\beta)\right)\sp\frac{1}{3}=1.
\end{equation}
He also states that (for $0\le p<1$)
\begin{equation}\label{ramf23}
    (1+p+p^2)\, _2F_1\left(\frac12,\frac12;1,\frac{p^3(2+p)}{1+2p}\right)=
    \sqrt{1+2p}\;
    _2F_1\left(\frac13,\frac23;1,\frac{27p^2(1+p)^2}{4(1+p+p^2)^3}\right).
\end{equation}
Ramanujan's results were derived in \cite{bbg95} (see also
\cite{h98}), though some related to expansions of $1/\pi$ had been
obtained earlier by J.M. and P.B. Borwein \cite{BB87}. An account
of the history and the associated theory of these equations may be
found in the last volume dedicated to Ramanujan's notebooks
\cite{bv98}. The associated theory of these modular equations
presented in the accounts just cited is largely based on direct
verification that appropriate expressions of hypergeometric
functions satisfy the same differential equations and initial
conditions and so are equal: we shall present a more geometric
picture in due course.

Analogous expressions to (\ref{mod23}) are known for $n=3$, $5$,
$7$ and $11$ \cite[7.13, 7.17, 7.24, 2.28 respectively]{bv98}.
Thus by iteration we may solve (\ref{esct}) for rational numbers
whose numerator and denominator have these as their only factors.
We include some examples of these in the table below. Thus to get
the value $2$ for the ratio $(2n_1-m_1)/(m_1+n_1)$ we set
$\alpha=\frac12$ in (\ref{mod23}) and solve for
$$t\sp\frac13+(1-t)\sp\frac13=2\sp\frac13,$$
taking the larger value $t=\frac{1}{2}+\frac{5\sqrt{3}}{18}$ (the
smaller value yielding the ratio $\frac12$).
$$
\begin{array}{|c|c|c|c|c|} \hline
n_1&m_1&(2n_1-m_1)/(m_1+n_1)&t&b\\
\hline 2&1&1&\frac{1}{2}&0\\
\hline
1&0&{2}&\frac{1}{2}+\frac{5\sqrt{3}}{18}&5\sqrt{2}\\
\hline1&1&\frac{1}{2}&\frac{1}{2}-\frac{5\sqrt{3}}{18}&5\sqrt{2}\\
 \hline 4&-1&3&(63+171\sqrt[3]{2}-18\sqrt[3]{4})/250&
(44+38\sqrt[3]{2}+26\sqrt[3]{4})/3\\
\hline 5&-2&4&\frac{1}{2}+\frac{153\sqrt{3}-99\sqrt{2}}{250}&
9\sqrt{458+187\sqrt{6}}\\
\hline
\end{array}
$$

A theory exists then for solving (\ref{modnr}) and this has been
worked out for various low primes. These results enable us to
reduce the Ercolani-Sinha conditions (\ref{esct}) to solving an
algebraic equation.

\subsection{Covers of the sextic} We shall now describe some
geometry underlying our curves (\ref{curve}) which will lead to an
understanding of the results of the last section. We shall first
present a more computational approach, useful in actual
calculations, and then follow this with a more invariant
discussion. We begin with the observation that our curves each
cover four elliptic curves.

\begin{lemma}\label{coverlem}
The curve $\mathcal{C}:= \{(x,y)| y^3+x^6+bx^3-1=0\}$ with
arbitrary value of the parameter $b$ is a simultaneous covering of
the four elliptic curves $\mathcal{E}_{\pm}$, $\mathcal{E}_{1,2}$
as indicated in the diagram, where $\mathcal{C}\sp*$ is an
intermediate genus two curve:

\vskip 0.5cm
\begin{center}
\unitlength=1mm
\begin{picture}(60,30)(0,0)
\put(9,23){\makebox(0,0){${\mathcal{C}}=(x,y)$}}
\put(6,20){\vector(-4,-1){20}} \put(-8,20){\makebox(0,0){$\pi^*$}}
\put(-18,15){\makebox(0,0){$\mathcal{C}^*$}}
\put(-26,0){\makebox(0,0){$\mathcal{E}_+=(z_+,w_+)$}}
\put(-20,12){\vector(-1,-1){9}} \put(-16,12){\vector(1,-1){9}}
\put(0,0){\makebox(0,0){$\mathcal{E}_-=(z_-,w_-)$}}
\put(-9,10){\makebox(0,0){$\pi_-$}}
\put(-26,10){\makebox(0,0){$\pi_+$}}
\put(10,19){\vector(1,-1){16}} \put(16,10){\makebox(0,0){$\pi_1$}}
\put(28,0){\makebox(0,0){$\mathcal{E}_1=(z_1,w_1)$}}
\put(12,20){\vector(2,-1){34}} \put(38,10){\makebox(0,0){$\pi_2$}}
\put(55,0){\makebox(0,0){$\mathcal{E}_2=(z_2,w_2)$}}
\end{picture}
\end{center}
\vskip 1cm

The equations of the elliptic curves are
\begin{align}
\mathcal{E}_{\pm}&= \{ (z_{\pm},w_{\pm})|
w_{\pm}^2=z_{\pm}(1-z_{\pm})
(1-k_{\pm}^2z_{\pm})   \},\label{curvepm}\\
\mathcal{E}_1&= \{ (z_1,w_1)
|z_1^3+{w_1}^3+3z_1+b=0\},\label{curve1}\\
\mathcal{E}_2&= \{ (z_2,w_2) | {w_2}^3+{z_2}^2+bz_2-1 =0  \},
\label{curve2}
\end{align}
where the Jacobi moduli, $k_{\pm}$ are given by
\begin{equation}
k_{\pm}^2=-\frac{\rho(\rho M \pm 1)(\rho M \mp 1)^3  } {(M\pm
1)(M\mp 1 )^3}
\end{equation}
with
\begin{equation} M=\frac{K}{L},\quad K=(2\imath -b)^{\frac13},\quad
L=(b^2+4)^{\frac16}. \label{KLM}  \end{equation}

The covers $\pi_{\pm},\pi_{1,2}$ are given by
\begin{align}
\pi_{\pm}:\quad\begin{split} z_{\pm}&=-\frac{K^2-L^2}{K^2-\rho
L^2}\,
\frac{Kx-y}{\rho Kx-y}\,\frac{L^2x-Ky}{L^2x-K\rho y},\\
w_{\pm}&=\imath\sqrt{2+\rho}\sqrt{\frac{L\pm K}{L\mp K}}
\frac{K^2}{L} \frac{L^2-\rho K^2}{\rho L^2-K^2}\frac{(Lx\mp
y)(x^6+1)} {(\rho K x- y)^2(L^2x-\rho K y)^2  }
\end{split}\label{pipm}
\end{align}
and
\begin{align*}
\pi_1:&\quad z_1=x-\frac{1}{x},\quad w_1=\frac{y}{x},\\
\pi_2:&\quad z_2=x^3,\quad w_2=y.
\end{align*}

The elliptic curves $\mathcal{E}_{1,2}$ are equianharmonic
($g_2=0$) and consequently have vanishing $j$-invariant,
$j\left(\mathcal{E}_{1,2}\right)=0$.
\end{lemma}

\begin{proof}
The derivation of the covers $\pi_{1,2}$ and the underlying curves
is straightforward. The pullbacks $\pi_{1,2}\sp{-1}$ of these
covers are
\begin{align*}
\pi_1^{-1}:= \begin{cases}   x=(z_1
\pm\sqrt{{z_1}^2+4})/2\\
                            y=w_1(z_1
\pm\sqrt{{z_1}^2+4})/2\end{cases}\qquad
 \pi_2^{-1}:= \begin{cases}
x=\rho\sqrt[3]{z_2}\\
                               y=w_2\end{cases}
\end{align*}
showing that the degrees of the cover are 2 and 3 respectively. A
direct calculation putting these elliptic curves into Weierstrass
form shows $g_2=0$ and hence the  elliptic curves
$\mathcal{E}_{1,2}$ are equianharmonic. Their $j$-invariants are
therefore vanishing and $\mathcal{E}_{1,2}$ are birationally
equivalent.

To derive the covers $\pi_{\pm}$ we first note that the curve
$\mathcal{C}$ is a covering of the hyperelliptic  curve
$\mathcal{C}^{\ast}$ of genus two,
\begin{equation}
\mathcal{C}^{\ast}= \{ (\mu,\nu)| \nu^2=(\mu^3+b)^2+4   \}.
\label{hcurve}
\end{equation}
The cover of this curve is given by the formulae
\begin{equation}
\pi^{\ast}:\quad \mu=\frac{y}{x},\quad
\nu=-x^3-\frac{1}{x^3}.\label{piast}
\end{equation}

The curve $\mathcal{C}^{\ast}$ covers two-sheetedly the two
elliptic curves $\mathcal{E}_{\pm}$ given in (\ref{curvepm})
\begin{align}\begin{split}
  z_{\pm}&=\frac{K^2-L^2}{K^2-\rho L^2}\,
\frac{K-\mu}{\rho K-\mu}\,\frac{L^2-K\mu}{L^2-K\rho \mu},\\
w_{\pm}&=-\imath\sqrt{2+\rho}\sqrt{\frac{L\pm K}{L\mp K}}
\frac{K^2}{L} \frac{L^2-\rho K^2}{\rho L^2-K^2}\frac{\nu(L\mp
\mu)} {(\mu-\rho K)^2(L^2-\rho K \mu)^2
}.\end{split}\label{cover2a}
\end{align}
Composition of (\ref{piast})  and (\ref{cover2a}) leads to
(\ref{pipm}).
\end{proof}

Using these formulae direct calculation then yields
\begin{corollary}
The holomorphic differentials of $\mathcal{C}$ are mapped to
holomorphic differentials of $\mathcal{E}_{\pm}$,
$\mathcal{E}_{1,2}$ as follows
\begin{align}\begin{split}
\frac{{d} z_{\pm}}{w_{\pm}}&=\sqrt{1+2\rho}\frac{L}{K}
\sqrt{(L\pm K)(L\mp K)^3}\,\frac{Lx\pm y}{y^2}\,{d}x,\end{split}
\label{invdiffpma}\\
&=\sqrt{1+2\rho}\frac{L}{K} \sqrt{(L\pm K)(L\mp K)^3}\,
(L\pm \mu)\,\frac{{d} \mu}{\nu}  \nonumber\\
\frac{{d} z_1}
{w_1^2}&=\frac{x^2+1}{y^2}{d}x, \label{invdiff1}\\
\frac{{d}z_2 }{w_2^2}&=\frac{3x^2}{y^2}{d}x, \label{invdiff2}
\end{align}
where $L,K$ are given in (\ref{KLM}).
\end{corollary}

The absolute invariants $j_{\pm}$ of the curves
$\mathcal{E}_{\pm}$ are
\begin{equation}
j_{\pm}=108\,{\frac {{L}^{3} \left( 5\,{L}^{3}\mp 4\,b \right)
^{3}}{ \left( {L}^ {3}\pm b \right) ^{2} }} .\label{invariants}
\end{equation}
Evidently $j_{\pm}\neq 0$ in general, as well $j_+\neq j_-$;
therefore these elliptic curves are not birationally equivalent to
that one appearing in Hitchin's theory of the tetrahedral monopole
which is equianharmonic \cite{hmm95}. We observe that the
substitution
\[ M=\frac{1+2\rho+p}{1+2\rho-p}  \]
leads to the parameterisation of Jacobi moduli being
\begin{equation} k_+^2=\frac{(p+1)^3(3-p)}{16p},\quad
                  k_-^2=\frac{(p+1)(3-p)^3}{16p^3},
\end{equation}
which Ramanujan used in his hypergeometric relations of signature
3, see e.g. \cite{bbg95}. The $\theta$-functional representation
of the moduli $k_{\pm}$ and parameter $p$ can be found in
\cite[Section 9.7]{lawd89},
\[  k_+=\frac{\vartheta_2^2(0|\tau) }{\vartheta_3^2(0|\tau)},\quad
   k_-=\frac{\vartheta_2^2(0|3\tau) }{\vartheta_3^2(0|3\tau)},\quad
  p=\frac{3\vartheta_3^2(0|3\tau) }{\vartheta_3^2(0|\tau)}.
  \]

We shall now describe the geometry of the covers we have just
presented explicitly. Our curve has several explicit symmetries
which lie behind the covers described. We will first describe
these symmetries acting on the field of functions $\mathfrak{k}$
of our curve as this field does not depend on whether we have a
singular or nonsingular model of the curve; we will subsequently
give a projective model for these, typically working in weighted
projective spaces where the curves will be nonsingular.

Viewing $\bar y=y/x$ and $x$ as functions on $\mathcal{C}$ we see
that
$${\bar y}^3=x^3+b-\frac{1}{x^3}$$
has symmetries ($\rho=e\sp{2\imath\pi /3}$)
\begin{align*}\mathrm{a}:\ &
x\rightarrow  x,\quad {\bar y}\rightarrow \rho{\bar y},\\
\mathrm{b}:\ &
x\rightarrow \rho x,\quad {\bar y}\rightarrow {\bar y},\\
\mathrm{c}:\ & x\rightarrow  -1/x,\quad {\bar y}\rightarrow {\bar
y}.
\end{align*}
Together these yield the group $G=C_3\times S_3$, with
$C_3=<\mathrm{a}|\,\mathrm{a}^3=1>$ and
$S_3=<\mathrm{b},\mathrm{c}|\,\mathrm{b}^3=1,\mathrm{c}^2=1,\mathrm{cbc}=\mathrm{b}^2>$.
When $b=5\sqrt{2}$, the dihedral symmetry $S_3$ is enlarged to
tetrahedral symmetry by
$$ \mathrm{t}: x\rightarrow \frac{\sqrt{2}-x}{1+\sqrt{2}x},\quad \bar y\rightarrow
\frac{3 x \bar y}{(1+\sqrt{2}x)(x-\sqrt{2})},\qquad
\mathrm{t}^2=1,$$ with $A_4$ being generated by $\mathrm{b}$ and
$\mathrm{t}$. Now to each subgroup $H\leq G$ we have the fixed
field $\mathfrak{k}\sp{H}$ associated to the quotient curve
$\mathcal{C}/H$.

The canonical curve of a non-hyperelliptic curve of genus 4 is
given by the intersection of an irreducible quadric and cubic
surface in $\mathbb{P}\sp3$. In our case the quadric is in fact a
cone and we may represent our curve $\mathcal{C}$  as the
nonsingular curve\footnote{Had we represented
$\mathcal{C}\subset\mathbb{P}\sp2$ as the plane curve given by the
vanishing of $z^6+b\, z^3 t^3 -t^6-w^3 t^3$  the curve is
singular. When $b$ is real the point $[z,t,w]=[0,0,1]$ is the only
singular point of $\mathcal{C}$ with delta invariant $6$ and
multiplicity $3$ yielding $g_\mathcal{C}=4$.} in the weighted
projective space $\mathbb{P}\sp{1,1,2}=\{[z,t,w]\,|\,[z,t,w]\sim
[\lambda z,\lambda t, \lambda\sp2 w]\}$ given by the vanishing of
$$f(z,t,w)=z^6+b\, z^3 t^3 -t^6-w^3 .$$
The group $G$ acts on this as ($x=z/t$, $\bar y=w/(zt)$)
\begin{align*}\mathrm{a}:\ &
[z,t,w]\rightarrow  [z,t,\rho w]\sim [\rho z,\rho t,w],\\
\mathrm{b}:\ & [z,t,w]\rightarrow [\rho z,t,\rho w]\sim [\rho\sp2 z,\rho t, w] ,\\
\mathrm{c}:\ & [z,t,w]\rightarrow [t,-z,-w]\sim [\imath t, -\imath
z,w].
\end{align*}
The fixed points of these actions on $\mathcal{C}$ and quotient
curves are as follows:
\begin{description}\item[$\mathrm{a}$]
There are 6 fixed points, $[1,\rho\sp{k}\,\alpha_\pm,0]$, where
$\alpha_\pm$ are the two roots of $\alpha\sp2-b\alpha-1=0$. For
other points we have a $3:1$ map
$\mathcal{C}\rightarrow\mathcal{C}/<\mathrm{a}>$. An application
of the Riemann-Hurwitz theorem shows the genus of
$\mathcal{C}/<\mathrm{a}>$ to be $g_{\mathcal{C}/<\mathrm{a}>}=0$.
\item[$\mathrm{b}$] This has no fixed points and an application of
the Riemann-Hurwitz theorem shows the genus of
$\mathcal{C}/<\mathrm{b}>$ to be $g_{\mathcal{C}/<\mathrm{b}>}=2$.
\item[$\mathrm{c}$]There are 6 fixed points, $[1,\pm\imath,
\rho\sp{k}\,\beta_\pm]$, where $\beta_\pm$ is a root of
$\beta_\pm\sp3=2\pm\imath b$. Here the Riemann-Hurwitz theorem
shows the genus of $\mathcal{C}/<\mathrm{c}>$ to be
$g_{\mathcal{C}/<\mathrm{c}>}=1$.
\end{description}

By using the invariants of $H$ we may obtain nonsingular
projective models of $\mathfrak{k}\sp{H}$. Take for example
$H=<\mathrm{c}>$ with invariants $u=zt$, $v=z^2-t^2$ and $w$ (in
degree $2$). Then we obtain the quotient curve $w^3=v^3+3u^2 v+b
u^3$ in
$\mathbb{P}\sp{2,2,2}\sim\mathbb{P}\sp{1,1,1}=\{[u,v,w]\}$. The
genus of the quotient is seen to be $1$. We recognise this as the
curve $\mathcal{E}_1$. One verifies that
$$\mathrm{c}\sp*\left(\frac{x^2+1}{y^2}\,dx\right)=\frac{x^2+1}{y^2}\,dx$$
giving us the invariant differential (\ref{invdiff1}). Similarly,
by taking $H=<\mathrm{b}\mathrm{ c}>$ and
$H=<\mathrm{b}^2\mathrm{c}>$, we also obtain equianharmonic
elliptic curves. The invariants of the involution
$\mathrm{b}^2\mathrm{c}$ are again all in degree $2$ and now are
$u=zt$, $v=\rho\sp{1/2}z^2-\rho\sp{-1/2}\,t^2$ and $w$.

By taking $H=<\mathrm{a}^2\mathrm{b}>$ we may identify
$\mathcal{E}_2$. The invariant of
$<\mathrm{a}^2\mathrm{b}>:[z,t,w]\rightarrow[\rho z,t, w]$ is
$u=z^3$ and the curve $w^3=u^2-but^3+t^6$ in
$\mathbb{P}\sp{3,1,2}=\{[u,t,w]\}$.Using the formula for the genus
of a smooth curve of degree $d$ in $\mathbb{P}\sp{a_0,a_1,a_2}$,
$$g=\frac12\left(\frac{d\sp2}{a_0a_1a_2}-d\sum_{i<j}\frac{\textrm{gcd}(a_i,a_j)}{a_i a_j}+
\sum_{i=0}\sp{2}\frac{\textrm{gcd}(a_i,d)}{a_i}-1 \right),$$ the
genus is seen to be $1$. Now (\ref{invdiff2}) is the invariant
differential for this action. If we had taken $H=<\mathrm{a}>$
with invariants $u=z^3$, $v=t^3$ and $w$ we obtain the curve
$w^3=u^2+buv-v^2$ in $\mathbb{P}\sp{3,3,2}$ (which is equivalent
to $W=u^2+buv-v^2$ in $\mathbb{P}\sp{1,1,2}$). The genus of this
quotient is seen to be $0$.

We obtain the genus $2$ curve $\mathcal{C}^{\ast}$ as follows. The
invariants of $H=<\mathrm{b}>$ are $U=zt$, $V=z^3$, $T=t^3$ and
$w$, subject to the relation $U^3=VT$. The curve $\mathcal{C}$ may
be written $T^2=-w^3+bU^3+V^2$, and hence
$U^6=V^2T^2=V^2(-w^3+bU^3+V^2)$. This curve has genus 2 in
$\mathbb{P}\sp{2,3,2}=\{[U,V,w]\}$ and may be identified with
$\mathcal{C}^{\ast}$. Be setting $\nu=2V^2-(w^3-bU^3)$ this curve
takes the form $$\nu^2=(w^3-bU^3)^2 +4U^6$$ in
$\mathbb{P}\sp{1,3,1}=\{[U,\nu,w]\}$ and the identification with
$\mathcal{C}^{\ast}$ in the affine chart of earlier is given by
$\mu=-w$, $U=1$. In this latter form we find that the action of
$\mathrm{c}$ is given by $[U,\nu,w]\rightarrow [-U,\nu,-w]\sim
[U,-\nu,w]$ which is the hyperelliptic involution; further
quotienting yields a genus 0 curve.

The remaining genus 1 curves $\mathcal{E}_\pm$ are identified with
the quotients of $\mathcal{C}^{\ast}$ by $U\rightarrow \pm
w/\sqrt[6]{4+b^2}$, $w\rightarrow\pm\sqrt[6]{4+b^2}U$,
$\nu\rightarrow\nu$. This action has invariants $A=Uw$ (in degree
2), $B=w\pm\sqrt[6]{4+b^2}U$ (in degree 1), and $\nu$ (in degree
3). The resulting degree 6 curve is
$$\nu^2=B^6\mp 6 L A B^4+9 L^2 A^2 B^2\mp2L^3A^3-2b A^3,$$
where, as previously, $L=\sqrt[6]{4+b^2}$. These curves have genus
$1$ in $\mathbb{P}\sp{2,1,3}=\{[A,B,\nu]\}$. To complete the
identification with $\mathcal{E}_\pm$ we compute the
$j$-invariants of these curves. In the affine patch with $B\ne0$
which looks like $\mathbb{C}\sp2$ (the other affine patches have
orbifold singularities and hence this choice) the curve takes the
form
$$Y^2=1\mp 6 L X+9 L^2 X^2-2(b\pm L^3) X^3.$$
The $j$-invariants of these curves agree with (\ref{invariants})
and hence the identifications as stated.

Both the differentials ${d}x/y$ and $x\,{d}x/y^2$ are invariant
under $\mathrm{b}$. These may be obtained by linear combinations
of ${d}z_\pm/w_\pm$ (\ref{invdiffpma}). The latter differentials
are those invariant under the symmetry of (\ref{hcurve})
$$\mu\rightarrow \frac{L\sp2}{\mu},\qquad \nu\rightarrow
\pm \frac{L\sp3 \nu}{\mu\sp3},
$$
which yield the quotients $\mathcal{E}_\pm$. A birational
transformation makes this symmetry more manifest\footnote{We thank
Chris Eilbeck for this observation.}. Let
$$T=\frac{L+\mu}{L-\mu},\ S=\frac{8\nu}{(L-\mu)\sp3},\qquad
\mu=L\,\frac{T-1}{T+1},\ \nu=\frac{L^3 S}{(T+1)^3}.$$ Then
(\ref{hcurve}) transforms to
$$S^2=(T-1)^6+2\frac{b}{L^3}(T^2-1)^3+(T+1)^6$$
which is manifestly invariant under $T\rightarrow-T$,
$S\rightarrow\mp S$. The substitution $W=T^2$ reduces the
canonical differentials ${d}T/S$ and $T\,{d}T/S^2$ to the
canonical differentials the elliptic curves
\begin{align*}
\mathcal{E}_+:&\quad
S^2=2(1+\frac{b}{L^3})W^3+6(5-\frac{b}{L^3})W^2+6(5+\frac{b}{L^3})W+
2(1-\frac{b}{L^3})
,\\
\mathcal{E}_-:&\quad
S^2=2(1+\frac{b}{L^3})W^4+6(5-\frac{b}{L^3})W^3+6(5+\frac{b}{L^3})W^2+
2(1-\frac{b}{L^3})W,
\end{align*}
which correspond to our earlier parameterisations.

\subsection{Role of the higher Goursat hypergeometric identities}
We have seen that complete Abelian integrals of the curve
$\mathcal{C}$ (\ref{bren03}) are given by hypergeometric
functions. The same is true for the various curves given in lemma
\ref{coverlem} covered by $\mathcal{C}$. Relating the periods of
$\mathcal{C}$ and the curves it covers leads to various relations
between hypergeometric functions, and this underlies the higher
hypergeometric identities of Goursat \cite{goursat81}. Goursat
gave detailed tables of transformations of hypergeometric
functions up to order four that will be enough for our purposes.

The simplest example of this is the cover
$\pi:\mathcal{C}\rightarrow\mathcal{C}\sp*$ for which
$\pi\sp*(\mu\,d\mu/\nu)=dx/y$ and $\pi\sp*(d\mu/\nu)=x\,dx/y^2$.
One then finds for example that
\begin{equation}\label{gour1}
\int_0\sp\alpha\frac{dx}{y}=\int_0\sp\infty\frac{\mu\,d\mu}{\nu},
\end{equation}
where both $y$ and $\nu$ are evaluated on the first sheet. A
change of variable shows that
$$\int_0\sp\infty\frac{\mu\,d\mu}{\nu}=\frac{2\pi}{3\sqrt{3}}\,
(b-2\imath)\sp{-1/3}\,{_2F_1}\left(\frac12,\frac13;1;\frac{4\imath}{2\imath-b}\right).$$
Now the left-hand side of equation (\ref{gour1}) is
$-\mathcal{I}_1$ (the minus sign arising when we go to Wellstein
variables $y\rightarrow -w$) and this has been evaluated in
(\ref{integralsij}). Comparison of these two representations
yields the hypergeometric equality
\[ F\left(\frac12,\frac13;1;\frac{4\imath}{2\imath-b}\right)=
      \left( \frac{2(b-2\imath)}{b+\sqrt{b^2+4}}   \right)^{\frac13}
F\left( \frac13,\frac13;1;\frac{b-\sqrt{b^2+4}}{b+\sqrt{b^2+4}}
\right) ,  \] which is one of Goursat's quadratic equalities
\cite{goursat81}; see also \cite[Sect. 2.11, Eq. (31)]{ba55}.
Further identities ensue from the coverings
$\mathcal{C}\rightarrow\mathcal{E}_\pm$ and we shall describe
these as needed below.

We remark that the curve (\ref{hcurve}) already appeared in
Hutchinson's study \cite{hut02} of automorphic functions
associated with singular, genus two, trigonal curves in which he
developed earlier investigations of Burkhardt \cite{bur93}.  These
results were employed by Grava and one of the authors \cite{eg04}
to solve the Riemann-Hilbert problem and associated Schlesinger
system for certain class of curves with $Z_N$-symmetry.

\subsection{Weierstrass reduction}
It is possible for the theta functions associated to a period
matrix $\tau$ to simplify (or admit reduction) and be expressible
in terms of lower dimensional theta functions. Such happens when
the curve covers a curve of lower genus, but it may also occur
without there being a covering. Reduction may be described purely
in terms of the Riemann matrix of periods (see \cite{ma92a}; for
more recent expositions and applications see
\cite{be02a},\cite{be02b}).
A $2g\times g$ Riemann matrix $\Pi=\left(%
\begin{array}{c}
  \mathcal{A} \\
  \mathcal{B} \\
\end{array}%
\right)$ is said to admit \emph{reduction} if there exists a
$g\times g_1$ matrix of complex numbers $\lambda$ of maximal rank,
a $2g_1\times g_1$ matrix of complex numbers $\Pi_1$ and a
$2g\times 2g_1$ matrix of integers $M$ also of maximal rank such
that
\begin{equation} \Pi \lambda  = M \Pi_1, \end{equation}
where  $1\leq g_1 < g$. When a Riemann matrix admits reduction the
corresponding period matrix may be put in the form
\begin{equation}\label{redpergen}
    \tau=\left(%
\begin{array}{cc}
  \tau_1 & Q\\
  Q\sp{T} & \tau\sp{\#} \\
\end{array}%
\right),
\end{equation}
where $Q$ is a $g_1\times(g-g_1)$ matrix with rational entries and
the matrices $\tau_1$ and $\tau\sp{\#}$ have the properties of
period matrices. Because $Q$ here has rational entries there
exists a diagonal $(g-g_1)\times(g-g_1)$ matrix
$D=\diag(d_1,\ldots,d_{g-g_1})$ with positive integer entries for
which $(QD)_{jk}\in\mathbb{Z}$. With
$(z,w)=(z_1,\ldots,z_{g_1},w_1,\ldots,w_{g-g_1})$ the theta
function associated with $\tau$ may then be expressed in terms of
lower dimensional theta functions as
\begin{equation}\label{redthetagen}
\theta((z,w);\tau)=\sum_{\substack{
\mathbf{m}=(m_1,\ldots,m_{g-g_1})\\
0\le m_i\le d_i-1}}
\theta(z+Q\mathbf{m};\tau_1)\,\theta\left[%
\begin{array}{c}
  D\sp{-1}\mathbf{m} \\
  0 \\
\end{array}%
\right](Dw;D\tau\sp{\#}D).
\end{equation}

Our curve admits many reductions. Of itself this just means that
the theta functions may be reduced to theta functions of fewer
variables. It is only when the Ercolani-Sinha vector
correspondingly reduces that we obtain real simplification. In the
remainder of this section we shall describe these reductions and
later see how dramatic simplifications occur.

First let us describe the Riemann matrix of periods. We may
evaluate the remaining period integrals as follows. Let
\begin{align*}
\int\limits_0^{\alpha}{d}u_i=\mathcal{I}_i(\alpha),\quad
\int\limits_0^{\beta}{d}u_i=\mathcal{J}_i(\alpha),\quad
i=1,\ldots,4.
\end{align*}
Then for $k=1,2$ we have that
\begin{align*}
\int\limits_0^{\rho^k\alpha}{d}u_{1,2}&=\rho^k\mathcal{I}_{1,2}(\alpha),\quad
\int\limits_0^{\rho^k\beta}{d}u_{1,2}=\rho^k\mathcal{J}_{1,2}
(\alpha),\\
\int\limits_0^{\rho^k\alpha}{d}u_{3}&=\rho^{2k}\mathcal{I}_{3}(\alpha),\quad
\int\limits_0^{\rho^k\beta}{d}u_{3}=\rho^{2k}\mathcal{J}_{3}
(\alpha),\\
 \int\limits_0^{\rho^k\alpha}{d}u_{4}&=\mathcal{I}_{4}(\alpha),\quad\quad
\int\limits_0^{\rho^k\beta}{d}u_{4}=\mathcal{J}_{4} (\alpha),
\end{align*}
where it is again supposed that the integrals  $\mathcal{I}_*$ and
$\mathcal{J}_*$ are computed on the first sheet. We have already
computed $\mathcal{I}_1(\alpha)$ and $\mathcal{J}_1(\alpha)$. The
integrals $\mathcal{I}_*$ and $\mathcal{J}_*$ are found to be
\begin{align*}
\mathcal{I}_1(\alpha)&=
-\frac{2\pi\alpha}{3\sqrt{3}}\,{_2F_1}\left(\frac13,\frac13;1;-\alpha^6\right)
=-\frac{2\pi}{3\sqrt{3}}\,\frac{\alpha}{(1+\alpha^6)\sp\frac13}
\ {_2F_1}(\frac{1}{3}, \frac{2}{3}; 1,t),\\
\mathcal{J}_1(\alpha)&=
\frac{2\pi}{3\sqrt{3}\alpha}\,{_2F_1}\left(\frac13,\frac13;1;-\frac{1}{\alpha^{6}}
\right)=\frac{2\pi}{3\sqrt{3}}\,\frac{\alpha}{(1+\alpha^6)\sp\frac13}
\ {_2F_1}(\frac{1}{3}, \frac{2}{3}; 1,1-t),\\
\mathcal{I}_2(\alpha)&= \frac{4\pi^2}{9\Gamma\left(
\frac23\right)^3}\frac{\alpha}{(1+\alpha^6)^{\frac13}},\\
\mathcal{J}_2(\alpha)&=-\frac{4\pi^2}{9\Gamma\left(
\frac23\right)^3}\frac{\alpha}{(1+\alpha^6)^{\frac13}},\\
\mathcal{I}_3(\alpha)&=
\frac{2\pi\alpha^2}{3\sqrt{3}}\,{_2F_1}\left(\frac23,\frac23;1;-\alpha^6\right)
=\frac{2\pi}{3\sqrt{3}}\,\frac{\alpha^2}{(1+\alpha^6)\sp\frac23}
\ {_2F_1}(\frac{1}{3}, \frac{2}{3}; 1,t),\\
\mathcal{J}_3(\alpha)&=
\frac{2\pi}{3\sqrt{3}\alpha^2}{_2F_1}\left(\frac23,\frac23;1;-\frac{1}{\alpha^{6}}
\right)=\frac{2\pi}{3\sqrt{3}}\,\frac{\alpha^2}{(1+\alpha^6)\sp\frac23}
\ {_2F_1}(\frac{1}{3}, \frac{2}{3}; 1,1-t),\\
\mathcal{I}_4(\alpha)&= \alpha^3\, {_2F_1}\left(\frac23,1;\frac43;
-\alpha^6\right),\\
\mathcal{J}_4(\alpha)&=-\frac{1}{\alpha^3}\,
{_2F_1}\left(\frac23,1;\frac43;-\frac{1}{\alpha^{6}} \right),
\end{align*}
with $t=\alpha^6/(1+\alpha^6)$.

We observe that the relations
\begin{equation}\mathcal{R} \equiv\frac{\mathcal{I}_1(\alpha)}{\mathcal{J}_1(\alpha) }
=-\frac{\mathcal{I}_3(\alpha)}{\mathcal{J}_3(\alpha) }, \qquad
\mathcal{I}_2(\alpha)+\mathcal{J}_2(\alpha)=0, \qquad
\mathcal{I}_4(\alpha)-\mathcal{J}_4(\alpha)=\mathcal{I}_2(\alpha),
\label{relation13}\end{equation} follow from the above formulae.

The vectors $\boldsymbol{x},\ldots,\boldsymbol{d}$ are
\begin{equation}
\begin{split}
\boldsymbol{x}&=\left(\begin{array}{c}
 - (2\mathcal{J}_1+\mathcal{I}_1)\rho-2\mathcal{I}_1-\mathcal{J}_1\\
  (\mathcal{J}_1-\mathcal{I}_1)\rho+\mathcal{I}_1+2\mathcal{J}_1\\
 (\mathcal{J}_1+2\mathcal{I}_1)\rho+\mathcal{I}_1-\mathcal{J}_1\\
  3(\mathcal{J}_1-\mathcal{I}_1)\rho+3\mathcal{J}_1
\end{array}\right),\quad \boldsymbol{b}=\mathcal{I}_2
\left(\begin{array}{c} 1+2\rho\\-2-\rho\\1-\rho\\0
\end{array}\right),\\
\boldsymbol{c}&=\left(\begin{array}{c}
 (\mathcal{I}_3+2\mathcal{J}_3)\rho+\mathcal{J}_3-\mathcal{I}_3\\
 (\mathcal{I}_3-\mathcal{J}_3)\rho+\mathcal{J}_3+2\mathcal{I}_3\\
 - (2\mathcal{I}_3+\mathcal{J}_3)\rho-2\mathcal{J}_3-\mathcal{I}_3\\
 3(\mathcal{I}_3-\mathcal{J}_3)\rho+3\mathcal{I}_3
\end{array}    \right),\quad \boldsymbol{d}
=(\rho-1)\mathcal{I}_2 \left(\begin{array}{c} 1\\1\\1\\0
\end{array}\right).
\end{split}
\end{equation}

One may can easily check that
\[ \boldsymbol{x}\sp{T}H\boldsymbol{b}= \boldsymbol{x}\sp{T} H\boldsymbol{c}=
   \boldsymbol{x}\sp{T}H\boldsymbol{d}=0 .  \]
We then have that
\begin{align}\begin{split}
\mathcal{A}&= \left(\begin{array}{cccc}
-1-2\rho-(2+\rho)\mathcal{R}&1+2\rho&1+2\rho+(1-\rho)\mathcal{R}&-1+\rho\\
2+\rho+(1-\rho)\mathcal{R}&-2-\rho&1-\rho-(2+\rho)\mathcal{R}&-1+\rho\\
-1+\rho+(1+2\rho)\mathcal{R}&1-\rho&-2-\rho+(1+2\rho)\mathcal{R}&-1+\rho\\
3+3\rho-3\rho\mathcal{R}&0&-3\rho-3(1+\rho)\mathcal{R}&0
\end{array}\right)\left(\begin{array}{cccc} \mathcal{J}_1&&&\\
                           &\mathcal{I}_2\\
                           &&\mathcal{J}_3\\
                           &&&\mathcal{I}_2
\end{array}\right),\\
\mathcal{B}&=\left(\begin{array}{cccc}
2+\rho+(1-\rho)\mathcal{R}&1-\rho&1-\rho-(2+\rho)\mathcal{R}&2+\rho\\
-1+\rho+(1+2\rho)\mathcal{R}&1+2\rho&-2-\rho+(1+2\rho)\mathcal{R}&2+\rho\\
-1-2\rho-(2+\rho)\mathcal{R}&-2-\rho&1+2\rho+(1-\rho)\mathcal{R}&2+\rho\\
3-3(1+\rho)\mathcal{R}&0&3-3\rho\mathcal{R}&0
\end{array}\right)\left(\begin{array}{cccc} \mathcal{J}_1&&&\\
                           &\mathcal{I}_2\\
                           &&\mathcal{J}_3\\
                           &&&\mathcal{I}_2
\end{array}\right). \end{split}\label{ABsym}
\end{align}
The Ercolani-Sinha conditions,
$\boldsymbol{n}^T\mathcal{A}+\boldsymbol{m}\sp{T}\mathcal{B} =6
\chi\sp{\frac13} (1,0,0,0)$ written for the vectors
\begin{equation}
\boldsymbol{n}=\left(\begin{array}{c}n_1\\m_1-n_1\\-m_1\\2n_1-m_1
\end{array}\right), \qquad
\boldsymbol{m}=\left(\begin{array}{c}m_1\\-n_1\\n_1-m_1\\-3n_1
\end{array}\right)
\end{equation}
lead to the equations
\begin{equation}
\mathcal{R}=-\frac{2n_1-m_1}{m_1+n_1},\qquad
\mathcal{J}_1=\frac{\chi\sp{\frac13}}{m_1-2n_1},
\end{equation}
which were obtained earlier. A calculation also shows that the
relation (\ref{cplxMT})
$$\mathcal{M}\left(%
\begin{array}{c}
  \mathcal{A} \\
  \mathcal{B} \\
\end{array}%
\right) =\overline{
\left(%
\begin{array}{c}
  \mathcal{A} \\
  \mathcal{B} \\
\end{array}%
\right)}\cdot T$$ yielding a nontrivial check of our procedure.

The integrals between infinities may be reduced to our standard
integrals by writing
$$\int_{\infty_i}^{\infty_j} d\boldsymbol{u}
=\int_{\tau(0_{\tau(i)})}^{\tau(0_{\tau(j)})} d\boldsymbol{u}
=\int_{0_{\tau(i)}}^{0_{\tau(j)}} \tau\sp*(d\boldsymbol{u})
=\overline{\int_{0_{\tau(i)}}^{0_{\tau(j)}}d\boldsymbol{u}}\cdot T
=\left( \overline{\int_{0_{\tau(i)}}^{\lambda_*}d\boldsymbol{u}}-
\overline{\int_{0_{\tau(j)}}^{\lambda_*}d\boldsymbol{u}}\right)\cdot
T,
$$
where we write $\tau(\infty_i)=0_{\tau(i)}$ and $\lambda_*$ is any
of the branch points. These are then calculated to be
\begin{equation}
\int_{\infty_1}^{\infty_2} d\boldsymbol{u}=\left(
\begin{array}{c}
(\rho-1)\mathcal{J}_1\\
-(\rho^2-1)\mathcal{J}_4\\
(\rho^2-1)\mathcal{J}_3\\
-(\rho^2-1)\mathcal{J}_2
\end{array} \right),\
\int_{\infty_1}^{\infty_3} d\boldsymbol{u} =\left(
\begin{array}{c}
(\rho^2-1)\mathcal{J}_1\\
-(\rho-1)\mathcal{J}_4\\
(\rho-1)\mathcal{J}_3\\
-(\rho-1)\mathcal{J}_2
\end{array} \right),\
\int_{\infty_2}^{\infty_3}
 d\boldsymbol{u}=\left(
\begin{array}{c}
(\rho^2-\rho)\mathcal{J}_1\\
-(\rho-\rho^2)\mathcal{J}_4\\
(\rho-\rho^2)\mathcal{J}_3\\
-(\rho-\rho^2)\mathcal{J}_2
\end{array} \right).\label{qintegrals}
\end{equation}

Our Riemann matrix admits a reduction with respect to any of it
columns. We will exemplify this with the first column, a result we
will use next; similar considerations apply to the other columns.
Now from the above and (\ref{expxx}) it follows that
\begin{align}
\Pi\lambda=\left(%
\begin{array}{c}
  \mathcal{A} \\
  \mathcal{B} \\
\end{array}%
\right) \,\left(%
\begin{array}{c}
  1 \\
  0 \\
  0 \\
  0 \\
\end{array}%
\right)=\left(\begin{array}{c}\boldsymbol{x}\\\boldsymbol{y}
\end{array}\right)=
\left(\begin{array}{c}\oint_{\mathfrak{a}_i}{d} u_1
\\ \oint_{\mathfrak{b}_i}{d} u_1
\end{array}\right)=
\left(\begin{array}{c}\xi(H\boldsymbol{n}+\rho^2\boldsymbol{m})\\
\xi(\rho \boldsymbol{n}+H\boldsymbol{m})
\end{array}\right)=\xi\,M\left(\begin{array}{c}1\\\rho
\end{array}\right),\label{escond2}
\end{align}
where $M$ is the $2g\times2$ integral matrix
\begin{equation} \label{escond3}
M^T=\left(\begin{array}{cccccccc}n_1-m_1&n_2-m_2&n_3-m_3&-n_4-m_4
&m_1&m_2&m_3&-m_4      \\ -m_1&-m_2&-m_3&-m_4&n_1&n_2&n_3&n_4
\end{array}\right).
\end{equation}
Then to every two Ercolani-Sinha vectors $\boldsymbol{n}$,
$\boldsymbol{m}$ we have that
\begin{equation}
M^TJM=d\left(\begin{array}{cc}0&1\\-1&0\end{array}\right),\quad
d=\boldsymbol{n}.H\boldsymbol{n}- \boldsymbol{m}. \boldsymbol{n}
+\boldsymbol{m}.H\boldsymbol{m}= \sum_{j=1}^4(\varepsilon_j
n_j^2-n_jm_j+\varepsilon_j m_j^2).
\end{equation}
The number $d$ here is often called the Hopf number. In particular
for $d\ne0$ then $M$ is of maximal rank and consequently our
Riemann matrix admits reduction.

Let us now focus on the consequences of reduction for symmetric
monopoles.

\begin{theorem}\label{symwp} For the
symmetric monopole we may reduce by the first column using the
vector (\ref{escond2}) whose elements are related by
(\ref{solvn}), with $(n_1,m_1)=1$. Then
$$d=2(n_1+m_1)(m_1-2n_1)$$ and for $d\ne0$ there exists an element $\sigma$ of the
symplectic group $\mathrm{Sp}_{2g}(\mathbb{Z})$ such that
\begin{align}
\tau'_{\mathfrak{b}}=\sigma\circ\tau_{\mathfrak{b}}=
\left(\begin{array}{ccccc}
(\rho+2)/d&{\alpha}/{d}&0&\ldots&0\\
{\alpha}/{d}&&&&\\
0&&{\tau}^{\#}&&\\
\vdots&&&&\\ 0&&&&
\end{array}\right).\label{taureduced}
\end{align}
Letting $p\, m_1+q\, n_1=1$ then
\begin{equation}\label{alphareduced}
\alpha=\gcd(m_1+4n_1-q\,[m_1-2n_1],n_1-2m_1-p\,[m_1-2n_1]).
\end{equation}
When $\alpha=1$ a further symplectic transformation allows the
simplification $\tau'_{11}=\rho/d$.

Under $\sigma$ the Ercolani-Sinha vector transforms as
\begin{equation}\label{essymptrans}
  \sigma\circ{\boldsymbol U}=  \sigma\circ({\boldsymbol m}\sp{T}+{\boldsymbol
    n}\sp{T}\tau_{\mathfrak{b}})=(1/2,0,0,0).
\end{equation}
\end{theorem}
The proof of the theorem is constructive using work of Krazer,
Weierstrass and Kowalewski. Martens \cite{ma92,ma92a} has given an
algorithm for constructing $\sigma$ which we have implemented
using $Maple$. Because $\sigma$ depends on number theoretic
properties of $n_1$ and $m_1$ the form is rather unilluminating
and we simply record the result (though an explicit example will
be given in the following section). What is remarkable however is
the simple universal form the Ercolani-Sinha vector takes under
this transformation. This has great significance for us as we next
describe.

Using (\ref{redthetagen}),(\ref{taureduced}) and say
$D=\diag(d,1,1)$ we have that\footnote{When $\gcd(\alpha,d)\ne1$ a
smaller multiple than $d_1=d$ would suffice here with
correspondingly fewer terms in the sums $0\le m\le d_1-1$.}
\begin{align*}\theta((z,w);\tau'_\mathfrak{b})&=\sum_{m=0}\sp{d-1}
\theta(z+\frac{m\alpha}{d};\frac{\rho+2}{d})\,\theta\left[%
\begin{array}{ccc}
  \frac{m}{d} & 0 & 0 \\
  0 & 0 & 0 \\
\end{array}%
\right](Dw;D{\tau}^{\#}D)\\
&=\sum_{m=0}\sp{d-1} \theta \left[%
\begin{array}{c}\frac{m}{d}\\0\end{array}%
\right](dz;d(\rho+2))\,\theta((w_1+\frac{m\alpha}{d},w_2,w_3);{\tau}^{\#}),
\end{align*}
where we have genus one and three theta functions on the right
hand-side here. Comparison of (\ref{newbafnch}) and
(\ref{essymptrans}) then reveals that the theta function dependence
of the Baker-Akhiezer is given wholly by the genus one theta
functions. Further simplifications ensue from the identity
$$\theta \left[%
\begin{array}{c}\frac{\epsilon}{d}\\ \epsilon'\end{array}%
\right](dz;d\tau)=\mu(\tau)\,\prod_{l=0}\sp{d-1}
\theta \left[%
\begin{array}{c}\frac{\epsilon}{d}\\
\frac{ \epsilon'}{d}+\frac{d-(1+2l)}{2d}\end{array}%
\right](z;\tau),
$$
where $\mu(\tau)$ is a constant. We then have
\begin{theorem}\label{symel} For symmetric monopoles the theta function $z$-dependence of
(\ref{newbafnch}) is expressible in terms of elliptic functions.
\end{theorem}

Thus far we have not discussed the final Hitchin constraint for
symmetric monopoles. This theorem reduces the problem to one of
the zeros of elliptic functions.
\begin{figure}
\centering
\begin{minipage}[c]{0.5\textwidth}
\includegraphics[width=7cm]{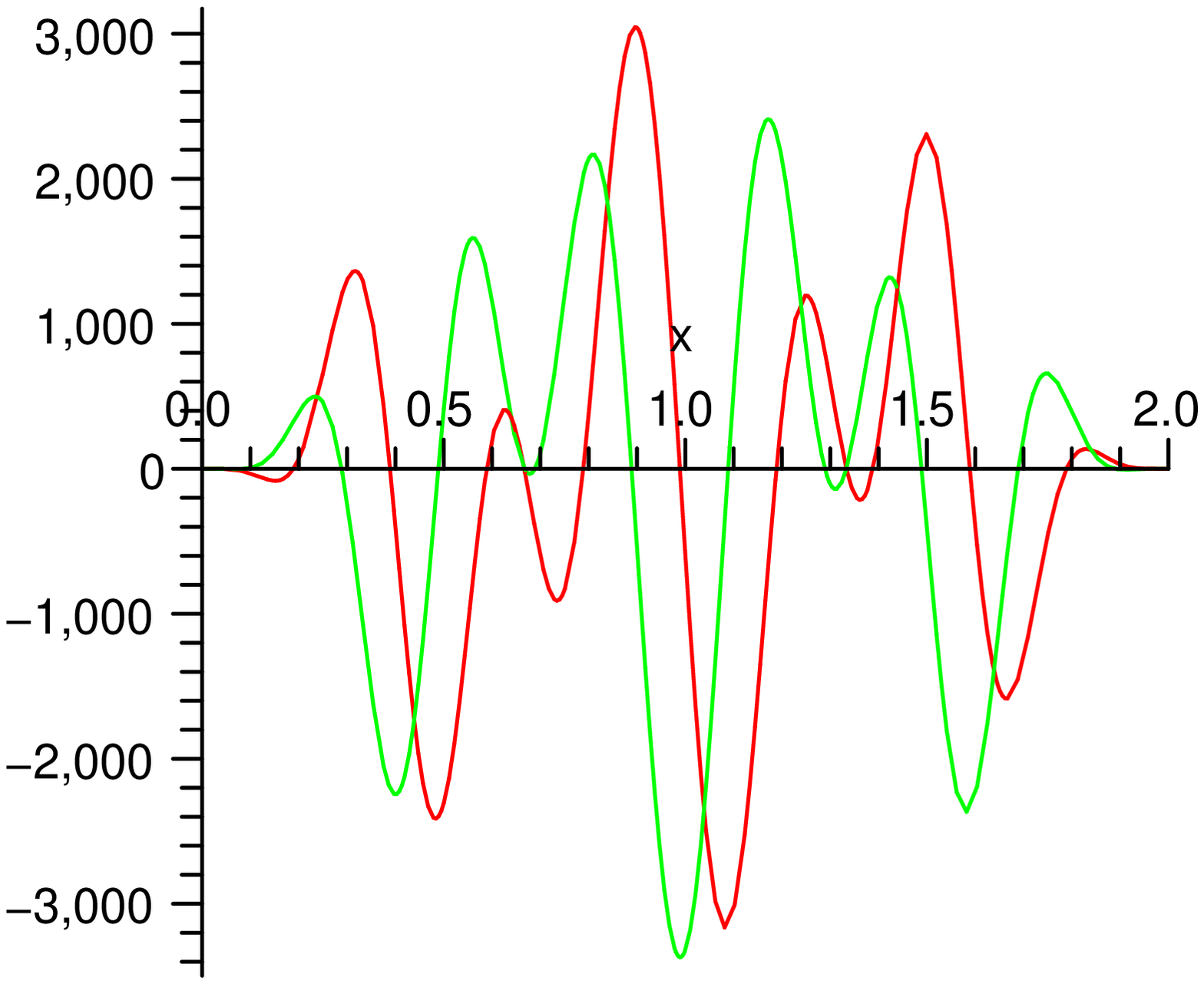} \caption{$n_1=2$, $m_1=1$. }
\label{fig:21}
\end{minipage}%
\begin{minipage}[c]{0.5\textwidth}
\includegraphics[width=7cm]{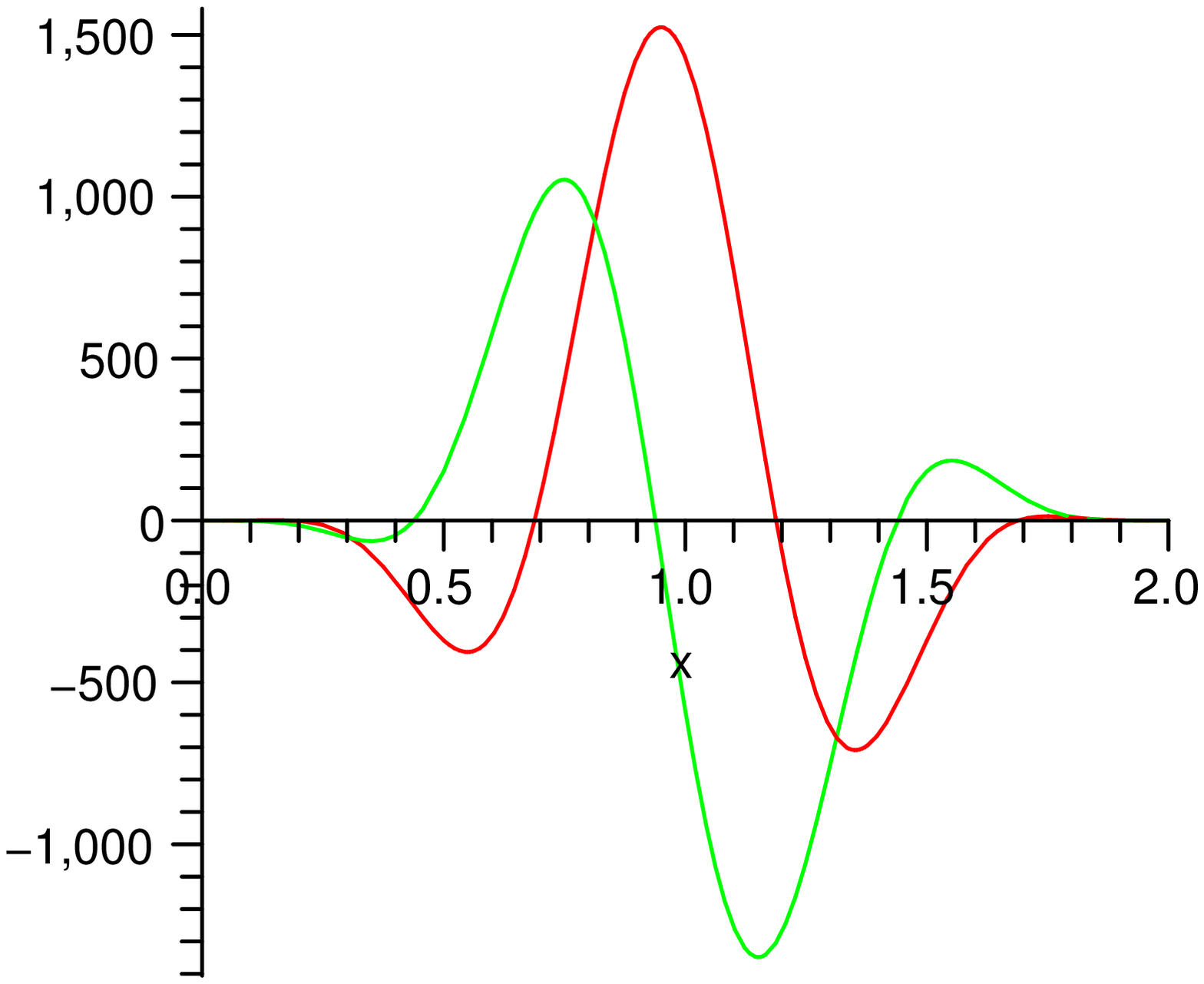} \caption{$n_1=1$, $m_1=1$. }
\label{fig:11}
\end{minipage}
\end{figure}
The graph in Figure~\ref{fig:21} shows the real and imaginary
parts of the theta function denominator of $Q_0(z)$ for the
$n_1=2$, $m_1=1$ symmetric monopole, the $b=0$ Ramanjuan case.
These vanish at $z=0$ and $z=2$ as desired, but additionally one
finds vanishing at $z=2/3$ and $z=4/3$. Calculating the theta
function with shifted argument in the numerator shows that there
is no corresponding vanishing and consequently $Q_0(z)$ yields
unwanted poles in $z\in(0,2)$. Thus the $n_1=2$, $m_1=1$ curve
does not yield a monopole.

A similar evaluation of the relevant $n_1=4$, $m_1=-1$ and
$n_1=5$, $m_1=-2$ theta functions also reveals unwanted zeros and
of the those cases from our table of symmetric 3-monopoles only
the tetrahedrally symmetric case has the required vanishing.
Indeed extensive numerical calculations suggests:

\begin{conjecture}For a symmetric monopole the denominator of $Q_0(z)$
 has $2(|n_1|-1)$ zeros and consequently the tetrahedrally
symmetric monopole is the only monopole in this class.
\end{conjecture}

Although other techniques exist for the study of monopoles which
might allow one to prove that the tetrahedrally symmetric monopole
is the only monopole in the class of symmetric monopoles we don't
know of a suitable theory that counts the number of times a
\emph{real} line (interval) intersects the theta divisor. We would
welcome such a theory. Our results do however allow us to say more
about the curve of the tetrahedrally symmetric monopole. Before
turning to a more detailed examination of this case in our next
section we first describe how to calculate the remaining
quantities appearing in the formula for $Q_0(z)$.

\subsection{Calculating $\nu_i-\nu_j$} Here we follow section
\S5.4. We calculate the $\mathfrak{a}$-periods of the differential
$dr_1$ in a manner similar to the period integrals already
calculated. Introduce integrals on the first sheet
\begin{align}
\mathcal{K}_1(\alpha)=\int_{0}^{\alpha}\frac{z^4 dz}{3w^2},\qquad
\mathcal{L}_1(\beta)&=\int_{0}^{\beta}\frac{z^4 dz}{3w^2}, \qquad
\beta=-\frac{1}{\alpha}.
\end{align}
Evidently
$\mathcal{K}_1(\rho^k\alpha)=\rho^{2k}\mathcal{K}(\alpha) $ and
$\mathcal{L}_1(\rho^k\beta)=\rho^{2k}\mathcal{L}(\beta)$ and one
finds that
\begin{equation}
\mathcal{K}_1=-\frac{4\sqrt{3}\pi}{27} \alpha^5\,
_2F_1\left(\frac23,\frac53;2; -\alpha^6   \right),\qquad
\mathcal{L}_1=\frac{4\sqrt{3}\pi}{27} \frac{1}{\alpha^5}\,
_2F_1\left(\frac23,\frac53;2; -\frac{1}{\alpha^6}
\right).\label{kl}
\end{equation}

We find, as before in the case of holomorphic differentials, that
\begin{align*}
y_1&=(\mathcal{K}_1+2\mathcal{L}_1)\rho-\mathcal{K}_1+\mathcal{L}_1,
&y_2&=(\mathcal{K}_1-\mathcal{L}_1)\rho+2\mathcal{K}_1+\mathcal{L}_1\\
y_3&=-(2\mathcal{K}_1+\mathcal{L}_1
)\rho-\mathcal{K}_1-2\mathcal{L}_1,
&y_4&=3(\mathcal{K}_1-\mathcal{L}_1)\rho +3\mathcal{K}_1.
\end{align*}
The Legendre relation (\ref{legndre}) gives a non trivial
consistency check of our calculations. This may be written in the
form of the following hypergeometric equality
\begin{align*}\frac{27}{4\sqrt{3}\pi}=
\alpha^4\,
_2F_1\left(\frac13,\frac13;1;-\frac{1}{\alpha^6}\right)\, _2F_1
\left(\frac23,\frac53;2;-\alpha^6 \right)+\frac{1}{\alpha^4}\,
_2F_1\left(\frac13,\frac13;1;-\alpha^6\right)\, _2F_1
\left(\frac23,\frac53;2;-\frac{1}{\alpha^6} \right)  \label{new}
\end{align*}
and this may be established by standard means.

To calculating $\nu_i-\nu_j$ using (\ref{nuijgpgr}) introduce the
differential of the second kind,
\begin{equation}s= d\left( \frac{w}{z}\right)(P)  -3 d r_1(P)\equiv
\frac{dz}{z^2 w^2},
\end{equation}
with second order pole at $0$ on all sheets,
\begin{align*}
\left.\frac{dz}{z^2 w^2}\right|_{P=0_k}&= \left\{
\frac{1}{w(0_k)^2}
 \frac{1}{\xi^2}  +\frac{2b}{3}\xi+\ldots  \right\} d\xi
= \left\{-
 \frac{w(0_k)}{\xi^2}  +\frac{2b}{3}\xi+\ldots  \right\} d\xi.
 \end{align*}
(Here we took into account $w(0_k)^3=-1$ for $k=1,2,3$.) Then
\begin{align}
\nu_i-\nu_j&=3 \boldsymbol{y}.\int_{\infty_j}^{\infty_j}
\boldsymbol{v}+\int_{\infty_j}^{\infty_j} \frac{dz}{z^2 w^2}
\label{nuij1s}
\end{align}
The last integral in (\ref{nuij1s}) may also be expressed in terms
of hypergeometric functions as follows. First we remark that
\[ \int_{\infty_i}^{\infty_j} \frac{dz}{z^2 w^2}
= (\rho_i-\rho_j)\int_{\alpha}^{\infty_1} \frac{dz}{z^2 w^2}, \]
where $\rho_i=\rho^{i-1}$. Next, for the integrals on the first
sheet we have
\begin{align*} \int_{\alpha}^{\infty} \frac{dz}{z^2 w^2}
=\frac{4\sqrt{3}\pi}{27} \frac{1}{\alpha^5}
F\left(\frac23,\frac53;2;
-\frac{1}{\alpha^6}   \right)=\mathcal{L}_1,\\
\int_{-\frac{1}{\alpha}}^{\infty} \frac{dz}{z^2 w^2}
=-\frac{4\sqrt{3}\pi}{27} \alpha^5 F\left(\frac23,\frac53;2;
-\alpha^6   \right)=\mathcal{K}_1.
\end{align*}

\section{The tetrahedral 3-monopole}
The curve of the tetrahedrally symmetric monopole is of the form
\begin{equation} \eta^3+\chi(\zeta^6+5\sqrt{2} \zeta^3-1)=0.
\label{thetracurve}
\end{equation}
In this case we may take
$$
t=\frac12-\frac{5\sqrt{3}}{18},\ \alpha
=\frac{\sqrt{3}-1}{\sqrt{2}},\
\mathcal{J}_1(\alpha)=-2\mathcal{I}_1(\alpha).
$$
For these values we may explicitly evaluate the various
hypergeometric functions. Using Ramanujan's identity
(\ref{ramf23}) together with the standard quadratic transformation
of the hypergeometric function
$$_2F_1\left(\frac12,\frac12,1,z\right)=(1+\sqrt{z})\sp{-1}\,
 _2F_1\left(\frac12,\frac12,1,\frac{4\sqrt{z}}{(1+\sqrt{z})\sp2}\right),
$$
(valid for $|z|<1$, $\arg{z}<\pi$) we find that
$$
_2F_1\left(\frac13,\frac23,1,t\right)=\frac{3\sp{\frac{5}{4}}}{4}\,
 _2F_1\left(\frac12,\frac12,1,\frac{2-\sqrt{3}}{4}\right).
$$
(In verifying this we note that
$p=4+3\sqrt{3}-2\sqrt{6}-3\sqrt{2}$ is the relevant value leading
to our $t$ in (\ref{ramf23}).) Now this last hypergeometric
function is related to an elliptic integral we may evaluate
\cite[p 86]{lawd89},
$$
K\left(\frac{\sqrt{3}-1}{2\sqrt{2}}\right)=\frac{\pi}{2}\,
_2F_1\left(\frac12,\frac12,1,\frac{2-\sqrt{3}}{4}\right)
=\frac{\Gamma(\frac16)\Gamma(\frac13)}{3\sp{\frac14}\, 4
\sqrt{\pi}}.
$$
Bringing these results together we finally obtain
\begin{equation}\label{evalhyp}
_2F_1\left(\frac13,\frac23,1,t\right)=
\frac{3\,\Gamma(\frac16)\Gamma(\frac13)}{8\pi\sp{\frac32}}\, .
\end{equation}
Then from (\ref{esevch}) we obtain that
\begin{equation}\label{tetchi}
\chi^{\frac{1}{3}} = -2\, \frac{2 \pi}{3
    \sqrt{3}}\ \frac{\alpha}{(1-\alpha\sp6)\sp\frac13}\ {_2F_1}(\frac{1}{3}, \frac{2}{3}; 1,
    t)=-\frac{1}{2\sp\frac16 \,\sqrt{3}}\,
    \frac{\Gamma(\frac16)\Gamma(\frac13)}{2\sqrt{3}\pi\sp{\frac12}}.
\end{equation}
This agrees with the result of \cite{hmr99}. We also note that
upon using Goursat's identity \cite[(39)]{goursat81}
$$
_2F_1\left(\frac13,\frac23,1,x\right)= (1-2x)\sp{\frac{-1}{3}}\,
_2F_1\left(\frac16,\frac23,1,\frac{4x(x-1)}{(2x-1)\sp2}\right),
$$
we may establish the result of \cite{hmr99} based on numerical
evaluation, that
$$
_2F_1\left(\frac16,\frac23,1,-\frac{2}{25}\right)=
\frac{5\sp{\frac13}\,3}{8}\,
\frac{\Gamma(\frac16)\Gamma(\frac13)}{\sqrt{3}\,\pi\sp{\frac32}}.
$$

Using these results and those of the previous section we have,
\begin{theorem}
The tetrahedral 3-monopole for which $b=5\sqrt{2}$ admits the
$\tau$-matrix of the form
\begin{equation}\label{tetratau}
\begin{split}\tau&=
{\frac {1}{98}}\, \left( \begin {array}{cccc} -73+51\,\imath\sqrt
{3}&9-13
\,i\sqrt {3}&15+11\,\imath\sqrt {3}&42-28\,\imath\sqrt {3}\\
\noalign{\medskip}9- 13\,\imath\sqrt {3}&-34+60\,\imath\sqrt
{3}&2\,\imath\sqrt {3}-24&21+35\, \imath\sqrt {3}
\\\noalign{\medskip}15+11\,\imath\sqrt {3}&2\,\imath\sqrt {3}-24&-40+36\,
\imath\sqrt {3}&-63-7\,\imath\sqrt
{3}\\\noalign{\medskip}42-28\,\imath\sqrt {3}&21+35\, \imath \sqrt
{3}&-63-7\,\imath\sqrt {3}&49+49\,\imath\sqrt {3}\end {array}
\right) \\
&=\left( \begin {array}{cccc} -{\frac {11}{49}}+{\frac
{51}{49}}\,\rho& -{\frac {2}{49}}-{\frac {13}{49}}\,\rho&{\frac
{13}{49}}+{\frac {11}{
49}}\,\rho&\frac17-\frac47\,\rho\\\noalign{\medskip}-{\frac
{2}{49}}-{\frac { 13}{49}}\,\rho&{\frac {13}{49}}+{\frac
{60}{49}}\,\rho&-{\frac {11}{49 }}+{\frac
{2}{49}}\,\rho&\frac47+\frac57\,\rho\\\noalign{\medskip}{\frac
{13}{ 49}}+{\frac {11}{49}}\,\rho&-{\frac {11}{49}}+{\frac
{2}{49}}\,\rho&-{ \frac {2}{49}}+{\frac
{36}{49}}\,\rho&-\frac57-\frac17\,\rho
\\\noalign{\medskip}\frac17-\frac47\,\rho&\frac47+\frac57\,\rho&-\frac57-\frac17\,\rho&1+\rho
\end {array} \right).
\end{split}
\end{equation}
\end{theorem}

We have already seen that the symmetric monopole curve
$\mathcal{C}$ covers two equianharmonic torii $\mathcal{E}_{1,2}$.
For the value of the parameter $b=5\sqrt{2}$ the curve covers
three further equianharmonic elliptic curves. These may be
described as follows.  For $i=3$, $4$, $5$ let $\pi_i:\,
\mathcal{C}\rightarrow\mathcal{E}_i$ be defined by the formulae
\begin{align}
\mu_3&=-\frac{\imath}{2^{\frac43}}\frac{(1+z\alpha)^4
+(z-\alpha)^4}{\alpha^2w^2},& \nu_3&=\frac{1+\imath}{\sqrt{2}}
\frac{(1+\alpha^2)(z^2+1)}{(z-\alpha)(z\alpha+1)},\nonumber \\
\mu_4&=-2^{\frac23}\frac{(1+z\alpha)^4 +(z-\alpha)^4}{2\alpha w
(1+z\alpha)(z-\alpha) },& \nu_4&=2^{\frac23}\alpha w \frac{
(1+z\alpha)^4 -(z-\alpha)^4
}{(z-\alpha)^3(z\alpha+1)^3},\label{cover345}\\
\mu_5&=-\sqrt{3}\imath\frac{(z^2+1)(z^2-2\sqrt{2}z-1)}
{(z^2+\sqrt{2}z-1)^2},& \nu_5&=-4\sqrt{6}\imath \frac{w
(z^4-\sqrt{2}z^3+3z^2 +\sqrt{2}z+1 )}
{(z^2+\sqrt{2}z-1)^3}.\nonumber
\end{align}
Then
\begin{align}
\mathcal{E}_3:&\quad \{(\nu_3,\mu_3)\,|\, {\nu_3}^2-{\mu_3}^3-2\imath=0 \},\nonumber\\
 \mathcal{E}_4:&\quad \{(\nu_4,\mu_4)\,|\,
  {\nu_4}^2-  \mu_4({\mu_4}^3+4)=0\}, \label{curves345} \\
\mathcal{E}_5:&\quad \{(\nu_5,\mu_5)\,|\,
{\nu_5}^3+24\sqrt{6}\imath ({\mu_5}^2-1)^2=0 \},\nonumber
 \end{align}
and we have the following relations between holomorphic
differentials
\begin{align}
{d}u_2&=\frac{z\,{d}z}{w^2}=\frac{1}{2^{\frac53}\sqrt{3}}\left\{
 (\imath-1)\pi_3\sp*\left(\frac{{d}\mu_3 }{\nu_3}\right) +
\pi_4\sp*\left(\frac{{d}\mu_4 }{\nu_4}\right) \right\} \label{differentials34} \\
{d}u_1&=\frac{{d} z}{w}
=\pi_5\sp*\left(\frac{{d}\mu_5}{\nu_5}\right).
\label{differentials5}
 \end{align}
The final of these rational maps was introduced by \cite{hmr99}
and has the following significance.

\begin{proposition}
Let $\boldsymbol{x}$ and $\boldsymbol{y}$ be the $\mathfrak{a}$
and $\mathfrak{b}$-periods of the differential ${d}u_1$ and denote
by $X$, $Y$ the $\mathfrak{a}$ and $\mathfrak{b}$-periods of the
elliptic differential ${d}\mu_5/\nu_5$. Then
\begin{equation}
\left(
\begin{array}{c}
  \boldsymbol{x} \\
  \boldsymbol{y} \\
\end{array}
\right) = M_5\left(%
\begin{array}{c}
  X \\
  Y \\
\end{array}
\right),\label{transfhom}
\end{equation}
where $M_5$ is the matrix
\begin{align}
M_5\sp{T}= \left(\begin{array}{rrrrrrrr}-1 &1 &0  &3 &1  &0 &-1  &1\\
                                 0 &1 &-1 &2 &1 & -1 &0  &3
  \end{array}\right)\label{mmatrix}
\end{align}
satisfying the condition
\begin{equation}
M_5\sp{T}\left(\begin{array}{cc}0_4&1_4\\
                        -1_4&0_4 \end{array}\right)M_5=
4\left(\begin{array}{cc}
 0&1\\-1&0  \end{array}\right).
\label{hopf1}\end{equation}
\end{proposition}

\begin{proof}
Introduce the homology basis for the elliptic curve as shown in
Figure~\ref{fig:Elldiffbasis} and
\begin{figure}
\centering
\begin{minipage}[c]{0.8\textwidth}
\includegraphics[width=6cm]{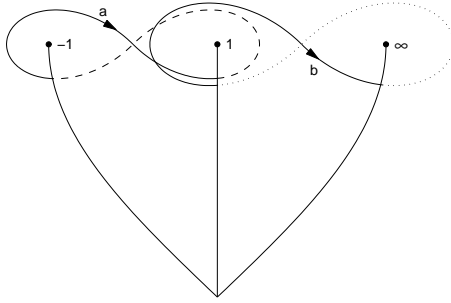} \caption{The elliptic curve homology basis }
\label{fig:Elldiffbasis}
\end{minipage}%
\end{figure}
set
\[ \mathcal{K}(\alpha)
=\int\limits_{-1}^{1}\frac{{d}\mu_5}{\nu_5}.\] Then
\begin{equation}
X=(2+\rho)\mathcal{K}(\alpha), \quad
Y=-(2\rho+1)\mathcal{K}(\alpha).\label{XY}
\end{equation}
From the reduction formula (\ref{differentials5}) we next conclude
that
\begin{equation}
\int\limits_{\alpha\rho}^{\alpha\rho^2}\frac{{d}z}{w}
=\mathcal{K}(\alpha)
\end{equation}
and therefore have that
\begin{align}\begin{split}
&2\mathcal{I}_1(\alpha)+\rho\mathcal{I}_1(\alpha)=\rho\mathcal{K}(\alpha),\\
&-2\rho\mathcal{I}_1(\alpha)-\mathcal{I}_1(\alpha)=\mathcal{K}(\alpha).
\end{split}\label{XY1}
\end{align}
Equations (\ref{XY}) and (\ref{XY1}) permit us to express
\begin{equation} \mathcal{I}_1(\alpha)=-\frac{Y}{3}   ,\qquad
\rho\mathcal{I}_1(\alpha)=-\frac{X}{3}
\end{equation}
and comparison with (\ref{ABsym}) yields the given $M_5$. The
condition (\ref{hopf1}) is checked directly. The number 4
appearing in (\ref{hopf1}) means that the cover $\pi_5$ given in
(\ref{cover345}) be of degree 4.

\end{proof}
We remark that the matrix $M_5$ of the proposition is obtained
from the $M$ of (\ref{escond3}) by 
$M_5=-\left(%
\begin{array}{cc}
  0 & 1 \\
  1 & 0 \\
\end{array}%
\right)M$, which simply reflects our choice of homology basis.
Thus we are discussing the reduction of the previous section.
Indeed with
$$\sigma=
\left[ \begin {array}{cccccccc}
1&0&0&0&0&0&0&0\\\noalign{\medskip}0&0
&0&0&0&1&0&0\\\noalign{\medskip}0&0&-1&0&0&1&0&0\\\noalign{\medskip}2&0
&-1&1&0&0&0&-1\\\noalign{\medskip}1&0&-1&1&1&-1&0&-3
\\\noalign{\medskip}5&-1&0&3&0&-1&1&-2\\\noalign{\medskip}-6&0&0&-3&0&0
&-1&2\\\noalign{\medskip}7&0&0&3&0&0&0&-2\end {array} \right]
\equiv \left(%
\begin{array}{cc}
  a & b \\
  c & d \\
\end{array}%
\right)
$$
we find that the $\tau$ matrix (\ref{tetratau}) transforms to
$$
\tau'=\sigma\circ\tau=(a\tau+b)(c\tau+d)\sp{-1}=\left[ \begin
{array}{cccc} \rho/4&1/4&0&0\\\noalign{\medskip}1/4&
5\rho/4&\rho&0\\\noalign{\medskip}0&\rho&2\,\rho&\rho
\\\noalign{\medskip}0&0&\rho&2/7+6\rho/7 \end {array} \right].
$$

Combined with Theorem \ref{symwp} we may reduce our expression for
the tetrahedral monopoles $Q_0(z)$ to one built out of Jacobi
elliptic theta functions.

\section{Conclusions}
Although monopoles have been studied now for many years and from
various perspectives, relatively few analytic solutions are known.
The outstanding problem is constructing curves satisfying the
transcendental constraints of Hitchin placed upon it. Using the
connection with integrable systems this article has sought to make
effective such solutions. It is nevertheless only early steps upon
this road. Here we have seen how one class of transcendental
constraints may be replaced by a number theoretic problem.

In applying our techniques beyond the known case of charge two we
considered the restricted class of charge three monopoles
(\ref{welstein99}) which includes the tetrahedrally symmetric
monopole. This family of curves has many arithmetic properties
that facilitates analytic integration. In particular the period
matrix may be explicitly expressed in terms of just four
integrals. Using this we were able to explicitly solve the
Ercolani-Sinha constraints that are equivalent to Hitchin's
transcendental condition (H2) of the triviality of a certain line
bundle over the spectral curve (Proposition \ref{ourh2}). Our
approach reduces the problem to that of determining certain
rationality properties of the (four) relevant periods. (Our result
also admits another approach to seeking monopole curves: we may
solve the Ercolani-Sinha constraints and then seek to impose
Hitchin's reality conditions on the resulting curves.) To proceed
further in this rather uncharted territory we further restricted
our attention to what we have referred to as ``symmetric
3-monopoles" whose spectral curve has the form (\ref{bren03}).
This reduced the required independent integrals from four to two,
each of which were hypergeometric in form, and the rationality
requirement is now for the ratio of these (Proposition
\ref{propesnt}). Extensions of work by Ramanujan mean this latter
question may be replaced by number theory and of seeking solutions
of various algebraic equations (depending on the primes involved
in the rational ratio). Examples of such solutions were given
(again including the tetrahedral case). We further examined the
symmetries and coverings of these symmetric curves and their
relation to higher Goursat hypergeometric identities. Having at
hand now many putative spectral curves we proceeded to evaluate
the remaining integrals needed in our construction. Remarkably we
discovered that application of Weierstrass reduction theory showed
that the Ercolani-Sinha vector transformed to a universal form and
that all of the theta function $z$-dependence for symmetric
3-monopoles was expressible in terms of elliptic functions
(Theorems \ref{symwp}, \ref{symel}). The final selection of
permissible spectral curves at last reduced to the question of
zeros of these elliptic functions. We are hampered by not knowing
a suitable theory for the calculation of the number of times a
real line interval intersects the theta divisor. Numerical
calculation of these zeros led to a universal conjecture (5.6)
suggesting that of the symmetric 3-monopoles only the tetrahedral
monopole has the required zeros.

Our final section then was devoted to the charge three
tetrahedrally symmetric monopole. Here we were able to
substantially simplify known expressions for the period matrix of
the spectral curve as well as prove a conjectured identity of
earlier workers. Again an explicit map was given and we have been
able to reduce entirely to elliptic functions. The final
comparison with the Nahm data of \cite{hmm95} will be left for a
subsequent work.

\section*{Acknowledgements}
This paper was conceived at the EPSRC funded Newton Institute
programme ``Integrable Systems'' in 2001 and had early encouragement
from Hermann Flaschka and Nick Ercolani. The many technical hurdles
encountered in this work meant growth has come in spurts rather than
continuous progress and it has occupied our thoughts for much of
this time. We are grateful to an EPSRC small grant and most recently
a MISGAM small grant enabling the authors to come together to draw
the work into its present form. Over the intervening years we have
benefited from discussions and correspondence with many colleagues
and we wish to thank: Nigel Hitchin, Conor Houghton and Paul
Sutcliffe for their remarks on monopoles; David Calderbank, Miles
Reid, Richard Thomas and Armando Treibich for their geometric and
algebro-geometric advice; Raimundas Vidunas and Adri Olde Daalhuis
for references on matters hypergeometric; Mike Eggar and Keiji
Matsumoto for help with topological aspects of our curve; and Chris
Eilbeck, Tamara Grava, Yuri Fedorov, John McKay and Stan Richardson
for discussions pertaining to Riemann surfaces. We have enjoyed a
problem touching on so many aspects of mathematics.

\appendix
\section{Theta Functions}
 For $r\in \mathbb{N}$ the canonical Riemann
 $\theta$-function is given by
\begin{equation}
\theta(\boldsymbol{z};\tau) =\sum_{\boldsymbol{n}\in \mathbb{Z}^r}
\exp(\imath\pi \boldsymbol{n}^T \tau\boldsymbol{n}+2 \imath\pi
\boldsymbol{z}^T \boldsymbol{n}).
\end{equation}
The $\theta$-function is holomorphic on
$\mathbb{C}^r\times\mathbb{S}^r$ and satisfies
\begin{equation}\theta(\boldsymbol{z}+\boldsymbol{p}\, ;\tau)=
\theta(\boldsymbol{z};\tau),\quad
\theta(\boldsymbol{z}+\boldsymbol{p}\tau;\tau)=
\mathrm{exp}\{-\imath\pi(\boldsymbol{p}^T\tau \boldsymbol{p}
+2\boldsymbol{z}^T\boldsymbol{p})\}\, \theta(\boldsymbol{z};\tau),
\label{transformation}
\end{equation}
where $\boldsymbol{p}\in\mathbb{Z}^r$.

The Riemann $\theta$-function
$\theta_{\boldsymbol{a},\boldsymbol{b}}(\boldsymbol{z};\tau)$ with
characteristics $\boldsymbol{a},\boldsymbol{b}\in\mathbb{Q}$ is
defined by
\begin{align*}
\theta_{\boldsymbol{a},\boldsymbol{b}}(\boldsymbol{z};\tau)
&=\mathrm{exp} \left\{ \imath\pi
(\boldsymbol{a}^T\tau\boldsymbol{a}
+2\boldsymbol{a}^T(\boldsymbol{z}+\boldsymbol{b})))\right\}
\theta(\boldsymbol{z}+\tau\boldsymbol{a}+\boldsymbol{b};\tau)\\
& =\sum_{\boldsymbol{n}\in\mathbb{Z}^r}\mathrm{exp}
\left\{\imath\pi(\boldsymbol{n}+\boldsymbol{a})^T\tau
            (\boldsymbol{n}+\boldsymbol{a})
+2\imath\pi
(\boldsymbol{n}+\boldsymbol{a})^T(\boldsymbol{z}+\boldsymbol{b})
\right\},
\end{align*}
where $\boldsymbol{a},\boldsymbol{b}\in \mathbb{Q}^r$. This is also
written as
$$\theta_{\boldsymbol{a},\boldsymbol{b}}(\boldsymbol{z};\tau)=
\theta\left[\begin{matrix}\boldsymbol{a}
\\ \boldsymbol{b}\end{matrix}\right](\boldsymbol{z};\tau).
$$
For arbitrary $\boldsymbol{a},\boldsymbol{b}\in \mathbb{Q}^r$ and
$\boldsymbol{a}',\boldsymbol{b}'\in \mathbb{Q}^r$ the following
formula is valid
\begin{align}
\theta_{\boldsymbol{a},\boldsymbol{b}}
(\boldsymbol{z}+\boldsymbol{a}'\tau+\boldsymbol{b}';\tau)&=\mathrm{exp}\left\{
-\imath\pi
{\boldsymbol{a}'}^T\tau{\boldsymbol{a}'}-2\imath\pi{\boldsymbol{a}'}^T\boldsymbol{z}
 -2\imath\pi (\boldsymbol{b}+\boldsymbol{b}')^T{\boldsymbol{a}'}   \right\}
 \times
\theta_{\boldsymbol{a}+\boldsymbol{a}',\boldsymbol{b}+\boldsymbol{b}'}
(\boldsymbol{z};\tau).\label{thetatransf}
\end{align}

The function $\theta_{\boldsymbol{a},\boldsymbol{b}}(\tau)=
\theta_{\boldsymbol{a},\boldsymbol{b}}(\boldsymbol{0};\tau) $ is
called the $\theta$-constant with characteristic
$\boldsymbol{a},\boldsymbol{b}$. We have
\begin{align*} \
&\theta_{-\boldsymbol{a},-\boldsymbol{b}}(\boldsymbol{z};\tau)=
\theta_{\boldsymbol{a},\boldsymbol{b}}(-\boldsymbol{z};\tau)\\
&\theta_{\boldsymbol{a}+\boldsymbol{p},\boldsymbol{b}+\boldsymbol{q}}
(\boldsymbol{z};\tau)= \mathrm{exp}(2\pi\imath
\boldsymbol{a}^T\boldsymbol{q})
\theta_{\boldsymbol{a},\boldsymbol{b}}(\boldsymbol{z};\tau)
\end{align*}

The following transformation formula is given in \cite[p85,
p176]{igusa72}.
\begin{proposition}
For any $\mathfrak{g}=\left(\begin{array}{cc}A&B\\C&D
\end{array}\right)\in\mathrm{Sp}(2g,\mathbb{Z})$ and
$(\boldsymbol{a},\boldsymbol{b})\in\mathbb{Q}^{2g}$ we put
\begin{align*}
\mathfrak{g}\cdot(\boldsymbol{a},\boldsymbol{b})&=
(\boldsymbol{a},\boldsymbol{b})\mathfrak{g}^{-1}
+\frac12(\mathrm{diag}(CD^T),\mathrm{diag}(AB^T) )\\
\boldsymbol{\phi}_{\boldsymbol{a},\boldsymbol{b}}(\mathfrak{g})&=-\frac12
(\boldsymbol{a}D^TB\boldsymbol{a}^T
-2\boldsymbol{a}B^TC\boldsymbol{b}^T+
\boldsymbol{b}C^TA\boldsymbol{b}^T)
+\frac12(\boldsymbol{a}D^T-\boldsymbol{b}C^T)^T\mathrm{diag}(AB^T)
,\end{align*} where $\mathrm{diag}(A)$ is the row vector
consisting of the diagonal components of $A$. Then for every
$\mathfrak{g}\in\mathrm{Sp}(2g,\mathbb{Z})$ we have
\begin{align}\begin{split}
&\theta_{\mathfrak{g}\cdot(\boldsymbol{a},
\boldsymbol{b})}(0;(A\tau_\mathfrak{b}+B)(C\tau_\mathfrak{b}+D)^{-1})
=\kappa(\mathfrak{g})\mathrm{exp}(2\pi\imath
\boldsymbol{\phi}_{\boldsymbol{a},\boldsymbol{b}}(\mathfrak{g})   )\,
\mathrm{det}(C\tau_\mathfrak{b}+D)^{\frac12}
\theta_{(\boldsymbol{a},\boldsymbol{b})}(0;\tau_\mathfrak{b})
\end{split}\label{igusa}
\end{align}
in which $\kappa(\mathfrak{g})^2$ is a $4$-th root of unity
depending only on $\mathfrak{g}$ while
\begin{equation} \begin{split}
\theta_{\mathfrak{g}\cdot(\boldsymbol{a},
\boldsymbol{b})}(z(C\tau_\mathfrak{b}+D)^{-1};(A\tau_\mathfrak{b}+B)(C\tau_\mathfrak{b}+D)^{-1})
&=\mu\,\exp\left(i\pi z(C\tau_\mathfrak{b}+D)^{-1}Cz\sp{T}\right)
\,
\mathrm{det}(C\tau_\mathfrak{b}+D)^{\frac12}\\
&\qquad
\times\theta_{(\boldsymbol{a},\boldsymbol{b})}(z;\tau_\mathfrak{b})
\end{split} \label{igusab}
\end{equation}
and $\mu$ is a complex number independent of $\tau$ and $z$ such
that $|\mu|=1$.
\end{proposition}

\noindent{\bf The Vector of Riemann Constants} The convention we
adopt for our vector of Riemann constants is
$$\theta\left(\boldsymbol{\phi}(P)-\boldsymbol{\phi}(\sum_{i=1}\sp{g}Q_i) -K\right)=0$$
in the Jacobi inversion. This is the convention used by Farkas and
Kra and the negative of that of Mumford; the choice of signs
appears in the actual construction of $K$, such as (2.4.1) of
Farkas and Kra. Then \begin{align}\label{vecR}\begin{split} (K_Q)_j&=\frac12
\tau_{jj}
-\sum_{k}\oint_{\mathfrak{a}_k}\omega_k(P)\int_Q\sp{P}\omega_j,\\
 & = \frac{1}{2}\left( \tau_{jj} + 1\right) -
    \sum_{k\ne j}\oint_{\mathbf{\mathfrak{a}}_{k}}
    \omega_{k}(P)\int^{P}_{Q}\omega_{j}.\end{split}
\end{align}
The vector of Riemann constants depends on the homology basis and
base point $Q$. If we change base points of the Abel map
$\boldsymbol{\phi}_{Q}\rightarrow \boldsymbol{\phi}_{ Q'}$ then $K_{Q}=K_{ Q'}+\boldsymbol{\phi}_{
Q'}({Q}\sp{g-1})$. With this convention
\begin{equation}\label{KKC}
\boldsymbol{\phi}_Q(\div(K_\mathcal{C}))=-2K_Q.
\end{equation}

\noindent{\bf Theta Characteristics}
\def\Pic{\mathop{\rm Pic}\nolimits}
\def\div{\mathop{\rm div}\nolimits}
The set $\Sigma$ of divisor classes $D$ such that
$2D=K_\mathcal{C}$, the canonical class, is called the set of
\emph{theta characteristics} of $\mathcal{C}$. The set $\Sigma$ is a
principal homogeneous space for the group $J_2$, the group of
$2$-torsion points of the group $\Pic\sp0(\mathcal{C})$ of degree
zero line bundles on $\mathcal{C}$. Equivalently this may be viewed
as the $2$-torsion points of the Jacobian, $J_2=\frac12
\Lambda/\Lambda$. Geometrically if $\xi$ is a holomorphic line
bundle on $\mathcal{C}$ such that $\xi\sp2$ is holomorphically
equivalent to $K_\mathcal{C}$ then the divisor of $\xi$ is a theta
characteristic. If $L$ is a holomorphic line bundle of order $2$,
that is $L\sp2$ is holomorphically trivial, then the divisor of
$\xi\otimes L$ is also a theta characteristic. Thus there are
$|J_2|=2\sp{2g}$ theta characteristics.

We may view $J_2=\{v\in \Pic\sp0(\mathcal{C})|2v=0\}$ as a vector
space of dimension $2g$ over $\mathbb{F}_2$. This vector space has a
nondegenerate symplectic (and hence symmetric as the field is
$\mathbb{F}_2$) form defined by the Weil pairing. If $D$ and $E$ are
divisors with disjoint support in the classes of $u$ and $v$
respectively, and $2D=\div(f)$, $2E=\div(g)$ then the Weil Pairing
is
$$\lambda_2:J_2\times J_2\rightarrow \mathbb{F}_2,\qquad
\lambda_2(u,v)=\frac{g(D)}{f(E)},$$ where if $D=\sum_j n_j x_j$ then
$g(D)=\prod_j g(x_j)\sp{n_j}$. Mumford identifies $\mathbb{F}_2$
with $\pm1$ by sending $0$ to $1$ and $1$ to $-1$. (In general we
may consider $J_r$, the $r$-torsion points of
$\Pic\sp0(\mathcal{C})$, and the Weil pairing gives us a
nondegenerate antisymmetric map $\lambda_2:J_r\times J_r\rightarrow
\mu_r$ where $\mu_r$ are $r$-th roots of unity.) The $\mathbb{F}_2$
vector space $J_2$ may be identified with
$H\sp1(\mathcal{C},\mathbb{F}_2)$ and with this identification
$\lambda_2$ is simply the cup product.

Define $\omega_\xi:J_2\rightarrow \mathbb{F}_2$ by
\begin{equation}\label{quadform}\omega_\xi(u)=\dim H\sp0(\mathcal{C},\xi\otimes u)- \dim
H\sp0(\mathcal{C},\xi)\quad (\mod 2),\end{equation}
 where $u=L_D$ is
the line bundle with divisor $D$. Then
$$\lambda_2(u,v)=\omega_\xi(u\otimes v)-\omega_\xi(u)-
\omega_\xi(v).$$ Any function $\omega_\xi$ satisfying this identity
is known as an Arf function, and any Arf function is given by
$\omega_\xi$ for some theta characteristic with corresponding line
bundle $\xi$. Thus the space of theta characteristics may be
identified with the space of quadratic forms (\ref{quadform}).

\section{Integrals between branch points}
We shall now describe how to integrate holomorphic differentials
between branch points.  We use the fact that for non-invariant
holomorphic differentials (as we have)
\begin{equation*}
    \sum^{3}_{i=1} \int_{\gamma_{i}(\lambda_{A},\lambda_{B})} \omega =
    \int^{\lambda_{B}}_{\lambda_{A}}(\omega + R_{*}\omega + R^{2}_{*}\omega) =
    0.
\end{equation*}
Indeed, if $\omega$ is any holomorphic differential on a compact
Riemann surface which is an $N$-fold branched cover of
$\mathbb{CP}\sp1$ then $\sum_{j=1}\sp{N}\omega(P\sp{(j)})=0$,
where $P\sp{(j)}$ are the preimages of $P\in\mathbb{CP}\sp1$.
 Then
\begin{align*}
   \oint_{ \mathfrak{a}_{1} - \mathfrak{b}_{1} }\omega= 3
   \int_{ \gamma_{1}(\lambda_{1},\lambda_{2})}\omega,\qquad
    \oint_{\mathfrak{a}_{2} - \mathfrak{b}_{2}}\omega = 3
    \int_{\gamma_{1}(\lambda_{3},\lambda_{4})}\omega,\qquad
    \oint_{\mathfrak{a}_{3} - \mathfrak{b}_{3}}\omega = 3
    \int_{\gamma_{1}(\lambda_{5},\lambda_{6})}\omega,
\end{align*}
and consequently
\begin{align*}
    \int_{\gamma_{1}(\lambda_{1}, \lambda_{2})}\omega & = \frac{1}{3}
    \oint_{\mathfrak{a}_{1} - \mathfrak{b}_{1}} \omega, \\
    \int_{\gamma_{2}(\lambda_{1}, \lambda_{2})}\omega & =
    \int_{\gamma_{1}(\lambda_{1}, \lambda_{2})-\mathfrak{a}_{1}}\omega =
    \frac{1}{3} \oint_{-2\mathfrak{a}_{1} - \mathfrak{b}_{1}} \omega, \\
    \int_{\gamma_{3}(\lambda_{1}, \lambda_{2})}\omega & = \frac{1}{3}
    \oint_{2\mathfrak{b}_{1} + \mathfrak{a}_{1}} \omega, \\
\end{align*}
with similar expressions obtained for $\gamma_{i}(\lambda_{3},
\lambda_{4})$ and $\gamma_{i}(\lambda_{5},\lambda_{6})$.

Further utilising $
    \gamma_{1}(\lambda_{2},\lambda_{6})  = \gamma_{1}(\lambda_{2},\lambda_{1}) +
    \gamma_{1}(\lambda_{1},\lambda_{6})$ and $\gamma_{2}(\lambda_{5},\lambda_{1})
    = \gamma_{2}(\lambda_{5},\lambda_{6}) +
    \gamma_{2}(\lambda_{6},\lambda_{1})$ we may write
\begin{align*}
    \mathfrak{a}_{4} & =  \mathfrak{b}_{1}
    -\mathfrak{b}_{3}-\mathfrak{a}_{3}+
    \gamma_{1}(\lambda_{1},\lambda_{6})+
     \gamma_{2}(\lambda_{6},\lambda_{1}), \\
    \mathfrak{b}_{4} & = \mathfrak{a}_{1} +
    \mathfrak{b}_{1}-\mathfrak{a}_{3}+ \gamma_{1}(\lambda_{1},\lambda_{6})+
     \gamma_{3}(\lambda_{6},\lambda_{1}) .
\end{align*}
Appropriate linear combinations of these yield
$\int_{\gamma_{i}(\lambda_{1},\lambda_{6})}\omega$ for $i=1,2,3$.
For example
$$\int_{\gamma_{1}(\lambda_{1},\lambda_{6})}\omega=
\frac{1}{3} \oint_{2\mathfrak{a}_{3} -
2\mathfrak{b}_{1}-\mathfrak{a}_{1}+\mathfrak{b}_{3}+\mathfrak{a}_{4}+\mathfrak{b}_{4}}
\omega .$$

In order to be able to integrate a holomorphic differential
between any branch point we must show how we may integrate such
between $\lambda_4$ and $\lambda_5$ on any branch. Now we use that
there exist meromorphic functions $f=w/{(z - \lambda_{1})^{2}}$
and $g=(z - \lambda_{i})/(z - \lambda_{j})$ (for each $i$, $j$)
with (respective) divisors
\begin{align*}
    (f) = \lambda_{2} + \lambda_{3} + \lambda_{4} + \lambda_{5} +
    \lambda_{6} - 5\lambda_{1},\qquad (g)=3(\lambda_{i} - \lambda_{j}).
\end{align*}
Thus for any normalized holomorphic differential $\boldsymbol{v}$
\begin{align*}
    \Lambda & \ni \int^{\lambda_{2}}_{\lambda_{1}}\boldsymbol{v}+
    \int^{\lambda_{3}}_{\lambda_{1}}\boldsymbol{v} +
    \int^{\lambda_{4}}_{\lambda_{1}}\boldsymbol{v} +
    \int^{\lambda_{5}}_{\lambda_{1}}\boldsymbol{v} +
    \int^{\lambda_{6}}_{\lambda_{1}}\boldsymbol{v}
     = 4\int^{\lambda_{2}}_{\lambda_{1}}\boldsymbol{v} +
    3\int^{\lambda_{3}}_{\lambda_{2}}\boldsymbol{v} +
    2\int^{\lambda_{4}}_{\lambda_{3}}\boldsymbol{v} +
    \int^{\lambda_{5}}_{\lambda_{4}}\boldsymbol{v} +
    \int^{\lambda_{6}}_{\lambda_{1}}\boldsymbol{v},
\end{align*}
and $  3\int^{\lambda_{i}}_{\lambda_{j}}\boldsymbol{v}
\in\Lambda$, where $\Lambda$ is the period lattice. These
equalities hold (modulo a lattice vector) for a path of
integration on any branch and so, for example,
$$\int_{\gamma_{1}(\lambda_{4},\lambda_{5})}\boldsymbol{v}\equiv\int_{\gamma_{1}(\lambda_{3},\lambda_{4})}\boldsymbol{v}
-\int_{\gamma_{1}(\lambda_{1},\lambda_{2})}\boldsymbol{v}
-\int_{\gamma_{1}(\lambda_{1},\lambda_{6})}\boldsymbol{v}\qquad\mod\Lambda.$$

\section{M\"obius Transformations}
We wish to determine when there is a M\"obius transformation
between the sets $H = \{\alpha_{1}, -{1}/{\overline{\alpha}_{1}},
\alpha_{2}, -{1}/{\overline{\alpha}_{2}}, \alpha_{3},
-{1}/{\overline{\alpha}_{3}}\}$ and $S = \{0, 1, \infty,
\Lambda_{1}, \Lambda_{2}, \Lambda_{3}\}$.  The former corresponds
to reality constraints on our data arising from $(H1)$ while the
latter may be constructed from the period matrix of the curve in
terms of various theta constants.  If we have a period matrix
satisfying
$(H2)$ then we must satisfy $(H1)$. \\
At the outset we note that the M\"obius transformation $M$ sending
$a \rightarrow 0, b \rightarrow 1, c \rightarrow \infty$ and its
inverse $M^{-1}$
\begin{equation*}
    \begin{matrix}
        M(a) = 0 & & M^{-1}(0) = a \\
        M(b) = 1 & & M^{-1}(1) = b \\
        M(c) = \infty & & M^{-1}(\infty) = c
    \end{matrix}
\end{equation*}
are given by
\begin{equation}\label{mob1}
    M(z) = \frac{b - c}{b - a} \frac{z-a}{z-c} \qquad M^{-1}(z) =
    \frac{z\,c(b - a) - a(b - c)}{z(b - a) - (b - c)}.
\end{equation}
The transformation
\begin{equation*}
    M(z) = \lambda \cdot \frac{z - a}{z - c} \qquad = \frac{\alpha z
    + \beta}{\gamma z + \delta}
\end{equation*}
may be represented by the $SL(2, \mathbb{C})$ matrix
\begin{equation}\label{mob2}
    \begin{pmatrix}
        \alpha & \beta \\ \gamma & \delta
    \end{pmatrix} =
    \begin{pmatrix}
        \frac{i \sqrt{\lambda}}{\sqrt{a + c}}
        & - \frac{ia \sqrt{\lambda}}{\sqrt{a + c}} \\
        \frac{i}{\sqrt{\lambda}\sqrt{a + c}} &
        -\frac{ic}{\sqrt{\lambda}\sqrt{a + c}}
    \end{pmatrix}
\end{equation}
and upon setting $\lambda = {(b - c)}/{(b - a)}$ we may determine
a $SL(2, \mathbb{C})$ representation of (\ref{mob1}).

A M\"obius transformation is conjugate to a rotation if and only
if it is of the  form $M(z) = \dfrac{(\alpha z +
\beta)}{(-\overline{\beta}Z + \overline{\alpha})}$.  In terms of
(\ref{mob2}) this means
\begin{equation*}
    a \overline{c} = -1 \qquad \text{and} \qquad \lambda
    \overline{\lambda} = \frac{1}{a \overline{a}}.
\end{equation*}
Then $M(0) \overline{M(\infty)} = -1$.

The rotation $\begin{pmatrix}\dfrac{
\overline{\alpha}_{1}}{\sqrt{1
+ |\alpha_{1}|^{2}}} & \dfrac{1}{\sqrt{1 + | \alpha_{1}|^{2}}} \\
\dfrac{-1}{\sqrt{1
+ |\alpha_{1}|^{2}}} & \dfrac{\alpha_{1}}{\sqrt{1 + | \alpha_{1}|^{2}}} \\
\end{pmatrix}$ transforms the set $H$ to one of the form $\{0, \infty,
\tilde{\alpha}_{2}, -{1}/{\tilde{\alpha}_{2}}, \tilde{\alpha}_{3},
-{1}/{\tilde{\alpha}_{3}} \}$ where $\tilde{\alpha}_{r} =
M(\alpha_{r}) = ({1 + \overline{\alpha}_{1}
\alpha_{r}})/({\alpha_{1} - \alpha_{r}})$ ($r = 2,3$).  Upon
setting $\tilde{\alpha}_2 = ae^{i \theta}$, $ a =
|\tilde{\alpha}_{2}|$ the rotation $\begin{pmatrix} e^{i
{\theta}/{2}} & 0 \\ 0 & e^{-i {\theta}/{2}}\end{pmatrix}$ will
transform the latter set to one of the form $\{0, \infty, a,
{-1}/{a}, w, {-1}/{\overline{w}}\}$.  Finally the scaling $z
\rightarrow {z}/{a}$ given by $\begin{pmatrix}\dfrac{1}{\sqrt{a}}
& 0 \\ 0 & \sqrt{a}\end{pmatrix}$ transforms $H$ to $H_{s} = \{0,
1, \infty,{-1}/{a^{2}}, {w}/{a}, {-1}/{(a \overline{w})}\}$ . Such
a set is of the desired form $S$ and is characterised by 3 (real)
parameters.  With $\Lambda_{1} = {-1}/{a^{2}}$, $\Lambda_{2} =
{w}/{a}$, $\Lambda_{3} = {-1}/{a \overline{w}}$ we see we have
$\Lambda_{1} \in \mathbb{R}$, $\Lambda_{1} < 0$,
$\Lambda_{2}\overline{\Lambda}_{3} = \Lambda_{1}$. From a set
$H_{s}$ and a choice of $\theta$ and $\alpha_{1}$ (equivalently, a
rotation) we may reconstruct $H$.

More generally, let us consider images $M(H)$ under M\"obius
transformations.  Up to a relabelling of roots we have four
possibilities of those roots we map to $\{0, 1, \infty \}$:
\begin{equation*}
\begin{matrix}
    \text{a.} & \alpha_{1} \rightarrow 0 & \alpha_{2} \rightarrow 1
    & {-1}/{\overline{\alpha}_{1}} \rightarrow \infty, \\
    \text{b.} & \alpha_{1} \rightarrow 0 & {-1}/{\overline{\alpha}_{1}}
\rightarrow 1
    & \alpha_{2} \rightarrow \infty, \\
    \text{c.} & \alpha_{1} \rightarrow 0 & {-1}/{\overline{\alpha}_{2}}
\rightarrow 1
    & \alpha_{2} \rightarrow \infty, \\
    \text{d.} & \alpha_{1} \rightarrow 0 & \alpha_{3} \rightarrow 1
    & \alpha_{2} \rightarrow \infty.
\end{matrix}
\end{equation*}
We have already considered $(a)$ in the previous paragraph.  For
completeness let us give $\Lambda_{1}, \Lambda_{2}, \Lambda_{3}$
for
the various cases and the various restrictions arising\\
a.
\begin{align}
    \Lambda_{1} = M(\frac{-1}{\overline{\alpha}_{2}}) & = - \frac{(1
    + \overline{\alpha}_{1} \alpha_{2})(1 +
    \overline{\alpha}_{2}\alpha_{1} )}{(\alpha_{1} -
    \alpha_{2}){(\overline{\alpha}_{1} - \overline{\alpha}_{2})}} <
    0 ,\nonumber \\
    \Lambda_{2} = M(\alpha_{3}) & = \frac{\alpha_{1} - \alpha_{3}}{\alpha_{1} -
    \alpha_{2}}\frac{1 + \overline{\alpha}_{1} \alpha_{2}}{1 + \overline{\alpha}_{1}
    \alpha_{3}}, \nonumber\\
    \Lambda_{3} = M(\frac{-1}{\overline{\alpha}_{3}}) & = - \frac{1
    + \alpha_{1} \overline{\alpha}_{3}}{\overline{\alpha}_{1}- \overline{\alpha}_{3}}\frac{1 + \alpha_{2}
    \overline{\alpha}_{1}}{\alpha_{1} - \alpha_{2}},\nonumber \\
    \Lambda_{2} \overline{\Lambda}_{3} & = \Lambda_{1};
    \label{mobca}
\end{align}
b.
\begin{align}
    \Lambda_{1} = M(\frac{-1}{\overline{\alpha}_{2}}) & = - \frac{1
    + \overline{\alpha}_{1} \alpha_{2}}{1 + \alpha_{1}
\overline{\alpha}_{1}}\cdot
    \frac{1 + \alpha_{1} \overline{\alpha}_{2}}
    {1 + \alpha_{2}\overline{\alpha}_{2}} \in \mathbb{R}
\quad 0 <\Lambda_{1} < 1 ,\nonumber\\
    \Lambda_{2} = M(\alpha_{3}) & = \frac{1
+ \overline{\alpha}_{1} \alpha_{2}}{1 + \alpha_{1}
    \overline{\alpha}_{1}}\frac{\alpha_{3} - \alpha_{1}}{\alpha_{3}
- \alpha_{2}}, \nonumber\\
    \Lambda_{3} = M(\frac{-1}{\overline{\alpha}_{3}}) & = \frac{1
    + \overline{\alpha}_{1} \alpha_{2}}{1
+ \alpha_{1} \overline{\alpha}_{1}} \frac{1 + \alpha_{1}
    \overline{\alpha}_{3}}{1 + \alpha_{2} \overline{\alpha}_{3}}, \nonumber\\
    \frac{\Lambda_{2}}{\Lambda_{2} - 1}
\overline{\left(\frac{\Lambda_{3}}{\Lambda_{3} - 1}\right)} & =
    \frac{\Lambda_{1}}{\Lambda_{1} - 1}; \label{mobcb}
\end{align}
c.
\begin{align}
    \Lambda_{1} = M(\frac{-1}{\overline{\alpha}_{1}}) & = - \frac{1
    + \alpha_{2} \overline{\alpha}_{2}}{1 + \alpha_{1}
\overline{\alpha}_{2}}\cdot
    \frac{1 + \alpha_{1} \overline{\alpha}_{1}}
    {1 + \alpha_{2}\overline{\alpha}_{1}} \in \mathbb{R} \quad 1 <\Lambda_{1} < \infty\nonumber, \\
    \Lambda_{2} = M(\alpha_{3}) & = \frac{1 + \alpha_{2} \overline{\alpha}_{2}}{1 + \alpha_{1}
    \overline{\alpha}_{2}}\frac{\alpha_{3} - \alpha_{1}}{\alpha_{3} - \alpha_{2}} \nonumber,\\
    \Lambda_{3} = M(\frac{-1}{\overline{\alpha}_{3}}) & = \frac{1
    + \alpha_{2} \overline{\alpha}_{2}}{1 + \alpha_{1} \overline{\alpha}_{2}} \frac{1 + \alpha_{1}
    \overline{\alpha}_{3}}{1 + \alpha_{2} \overline{\alpha}_{3}}, \nonumber\\
   (1 - \Lambda_{2}) \overline{(1 - \Lambda_{3})} & = 1 - \Lambda_{1}
 \label{mobcc};
\end{align}
d.
\begin{align}
    \Lambda_{r} = M(-\frac{1}{\overline{\alpha}_{r}}) & =
    \frac{\alpha_{3} - \alpha_{2}}{\alpha_{3} - \alpha_{1}} \frac{1 +
    \alpha_{1}\overline{\alpha}_{r}}{1 + \alpha_{2}
    \overline{\alpha}_{r}}, \qquad r = 1, 2, 3, \nonumber\\
    0 < \Lambda_{1} \overline{\Lambda}_{2} &\in \mathbb{R},\
    1 < \frac{\Lambda_{1}}{\Lambda_{2}} \in \mathbb{R}, \qquad
    \Lambda_{3} = \Lambda_{2} \frac{(1 -
    \overline{\Lambda}_{1})}{1 - \overline{\Lambda}_{2}}.
   \label{mobcd}
\end{align}

The constraints $(\ref{mobcb})$ for case $(b)$ may be obtained as
follows. Further composing the M\"obius transformation leading to
$(b)$ with that giving $0 \rightarrow 0$, $1 \rightarrow \infty$,
 $\infty \rightarrow 1$ gives us case $(a)$ for which we know the
constraint.  This second M\"obius transformation is given by $M(z)
= M^{-1}(z) = {z}/{(z - 1)}$ and we may transfer the constraint of
$(a)$ to $(b)$.  Similarly composing $(c)$ with $M(z) = 1-z$
yields case $(a)$ up to a relabelling of roots. Geometrically
cases $(a)$, $(b)$, $(c)$ consist of the following.  A circle
passes through $\{ \alpha_{1}, {-1}/{\overline{\alpha}_{1}},
\alpha_{2},{-1}/{\overline{\alpha}_{2}} \}$.  Under a M\"obius
transformation to the set $\{ 0, 1, \infty, \mu \}$ the circle
becomes the real axis and so $\mu \in \mathbb{R}$. This is the
real parameter appearing in each of these cases.  A similar
argument composing $(d)$ with $M(z) = {z}/{(z - \Lambda_{1})}$
will give the constraints (\ref{mobcd}).

In each case, given $\alpha$, and a choice of $\theta$ (a
rotation)
we can construct $S$ from $H$.

\providecommand{\bysame}{\leavevmode\hbox
to3em{\hrulefill}\thinspace}
\bibliographystyle{amsalpha}

\end{document}